\documentclass[
  twocolumn,english,aps,pra,
  superscriptaddress,amsmath,amssymb,floatfix,nofootinbib,longbibliography
]{revtex4-2}

\usepackage{amsthm}
\usepackage{amsfonts}
\usepackage{siunitx}
\usepackage{amsmath}
\usepackage{amssymb}
\usepackage{graphicx}
\usepackage{verbatim}
\usepackage[colorlinks]{hyperref}
\usepackage{tikz}
\usepackage{pgfplots}
\usepackage{adjustbox}
\usepackage{braket}
\usepackage{xcolor}
\usepackage{physics}
\usepackage{amssymb}
\usepackage{bm}
\usepackage{bbm}
\usepackage{mathtools}
\usepackage{mathrsfs}
\usepackage{dashrule}
\usepackage{caption}
\usepackage{braket}
\usepackage{enumitem}
\usepackage{amsthm}
\usepackage{amsfonts}
\usepackage{braket}

\newtheorem{remark}{Remark}

\definecolor{linkcolor}{RGB}{0,83,166}
\hypersetup{
  colorlinks = true,
  allcolors = {linkcolor}
}

\begin{document}

\title{Snapshot-QAOA: Extending QAOA to Quantum Hamiltonian Simulation}

\author{Reuben Tate}
\email[]{rtate@lanl.gov}
\affiliation{Information Sciences, Los Alamos National Laboratory, Los Alamos, NM, U.S.A.}

\author{Quinn Langfitt}
\affiliation{Computer Science Department, Northwestern University, Evanston, IL, U.S.A.}

\author{Elijah Pelofske}
\email[]{epelofske@lanl.gov}
\affiliation{Information Systems \& Modeling, Los Alamos National Laboratory, Los Alamos, NM, U.S.A.}

\author{Ammar Kirmani}
\affiliation{Physics of Condensed Matter and Complex Systems, Los Alamos National Laboratory,
Los Alamos, NM, U.S.A.}

\author{Andreas Bärtschi}
\email[]{baertschi@lanl.gov}
\affiliation{Information Sciences, Los Alamos National Laboratory, Los Alamos, NM, U.S.A.}

\author{John Golden}
\email[]{golden@lanl.gov}
\affiliation{Information Sciences, Los Alamos National Laboratory, Los Alamos, NM, U.S.A.}

\author{Stephan Eidenbenz}
\email[]{eidenben@lanl.gov}
\affiliation{Information Sciences, Los Alamos National Laboratory, Los Alamos, NM, U.S.A.}

\begin{abstract}

We present Snapshot-QAOA, a variation of the Quantum Approximate Optimization Algorithm (QAOA) that finds approximate minimum energy eigenstates of a large set of quantum Hamiltonians. 
Traditionally, QAOA targets the task of approximately solving combinatorial optimization problems. Snapshot-QAOA enables a significant expansion of the use case space for QAOA to more general quantum Hamiltonians, where the goal is to approximate the ground-state. 
Snapshot-QAOA retains the desirable variational-algorithm qualities of QAOA such as a small parameter count and relatively shallow circuit depth. 
Snapshot-QAOA is thus a better trainable alternative to the (Near Intermediate-Scale Quantum) NISQ-era Variational Quantum Eigensolver (VQE) algorithm, while retaining a significant circuit-depth advantage over the (Quantum Error Corrected) QEC-era Quantum Phase Estimation (QPE) algorithm. 
Our fundamental approach is inspired by the idea of Trotterization of a continuous-time linear adiabatic anneal schedule. Snapshot-QAOA restricts the QAOA evolution to not phasing out the mixing Hamiltonian completely at the end of the evolution, instead evolving only a partial typical linear QAOA schedule, thus creating a type of snapshot of the typical QAOA evolution. In this way, the cost of the quantum Hamiltonian is encoded by the phase separator to address the diagonal terms in the problem Hamiltonian and by the mixer to address the non-diagonal terms in the problem Hamiltonian. We focus on QAOA with the transverse field mixer, and thus simulations of quantum Hamiltonians that contain single-site Pauli-X terms. By measuring the expectation value of the system in both X and Z-basis, we can estimate the ground-state energy of transverse field quantum Hamiltonians. 
The accuracy of ground-state energy finding with snapshot-QAOA is evaluated using extensive numerical simulations on a $16$ qubit J1-J2 frustrated transverse field Ising model.

\end{abstract}

\maketitle


\section{Introduction}
\label{section:introduction}
The Quantum Approximate Optimization Algorithm (QAOA) \cite{QAOA, farhi2015quantum}, which was also generalized under the same acronym to the Quantum Alternating Operator Ansatz \cite{Hadfield_2019}, was introduced as a near-term quantum hardware-friendly algorithm that could, given sufficient parameter training and circuit depth, sample optimal or good approximate solutions of combinatorial optimization problems.

Given a classical cost function $C(z)$, the standard QAOA is comprised of the following components. First, an initial state $\ket{\psi}$, then a phase separating Hamiltonian $H_C$ that gives the cost value for all spin configurations by giving the basis states phases $e^{-i\gamma C(z)}$. Next, a mixing Hamiltonian $H_M$ facilitates state transitions; this is also known as the driving Hamiltonian. Typically, this is the transverse field mixer $\sum_i X_i$. An integer $p\geq1$ is the number of rounds to run the algorithm - this can also be thought of as the \emph{QAOA depth}, and vector of real numbers $\vec{\beta} = (\beta_1,...,\beta_p)$ of length $p$ that parameterizes $H_M$, and a vector of real numbers $\vec{\gamma} = (\gamma_1,...,\gamma_p)$, also of length $p$ that parameterizes $H_C$. We will typically call these parameters \emph{QAOA angles}.

Typically, QAOA is studied as a quantum algorithm that is intended to sample solutions, ideally optimal or nearly optimal solutions, of classical discrete combinatorial optimizations problems (i.e. the decision variables are binary, or spins). A considerable amount of study has been applied to this form of QAOA, including potential evidence for speedups over other quantum algorithms or existing classical optimization solvers \cite{shaydulin2024evidence, QAOA_exponential_speedup_over_unstructured_search, boulebnane2024solving, montanaro2024quantumspeedupssolvingnearsymmetric} as well as generalizations of QAOA to include more complex mixing Hamiltonians and different initial states \cite{Hadfield_2019, PhysRevA.101.012320, GM_QAOA, He_2023, tate2023warm, hadfield2017quantum}. 

In this work we propose Snapshot-QAOA, which allows a significant expansion of the typical capabilities of QAOA to include a broad class of quantum Hamiltonians (meaning Hamiltonians with non-diagonal terms). For small systems, the ground state (and energy) can be found using exact diagonalization of the Hamiltonian; however, this is intractable for larger systems as the memory and runtime requirements typically scale exponentially in the system size for general Hamiltonians. The fundamental computational ideas used in Snapshot-QAOA are based on adiabatic quantum computation and transverse field driven quantum annealing in particular \cite{Kadowaki_1998, Santoro_2002, farhi2000quantumcomputationadiabaticevolution, morita2008mathematical, rajak2023quantum}. 

Snapshot-QAOA is well-suited for Hamiltonians of the form $H = c_0H_0+c_1H_1$, i.e. a linear combination of two sub-Hamiltonians $H_0$ and $H_1$, where the ground state of $H_0$ is non-degenerate and easy to prepare, and where the corresponding unitaries $e^{-i\alpha H_0}$ and $e^{-i\alpha H_1}$ are easily implementable on a quantum computer for any $\alpha \in \mathbb{R}$. The key insight is that, for such Hamiltonians $H$, for any $T>0$, there exists a linear annealing schedule with an associated time-dependent Hamiltonian $\mathcal{H}_T(t)$ for which the ``snapshot" of $\mathcal{H}_T(t)$ at some time $t = \tau$ is exactly $H$, i.e., $H = \mathcal{H}_T(\tau)$. We show that this relationship allows one to easily determine suitable $T$-dependent values of the usual QAOA circuit parameters (typically denoted by $\vec{\gamma}$ and $\vec{\beta}$), thus causing Snapshot-QAOA to converge towards the ground state energy as the number of QAOA layers, $p$, increases. Thus, Snapshot-QAOA only has one parameter ($T$) that requires tuning. This last property means that one does not necessarily need to perform variational learning on the usual QAOA circuit parameters ($\vec{\gamma}$ and $\vec{\beta}$) in order to achieve good performance; however, we show that, like with standard QAOA, optimizing $\vec{\gamma}$ and $\vec{\beta}$ further can indeed yield improved performance over the Trotterized non-optimized schedule.

\begin{figure}[ht!]
\centering
\includegraphics[width=0.999\linewidth]{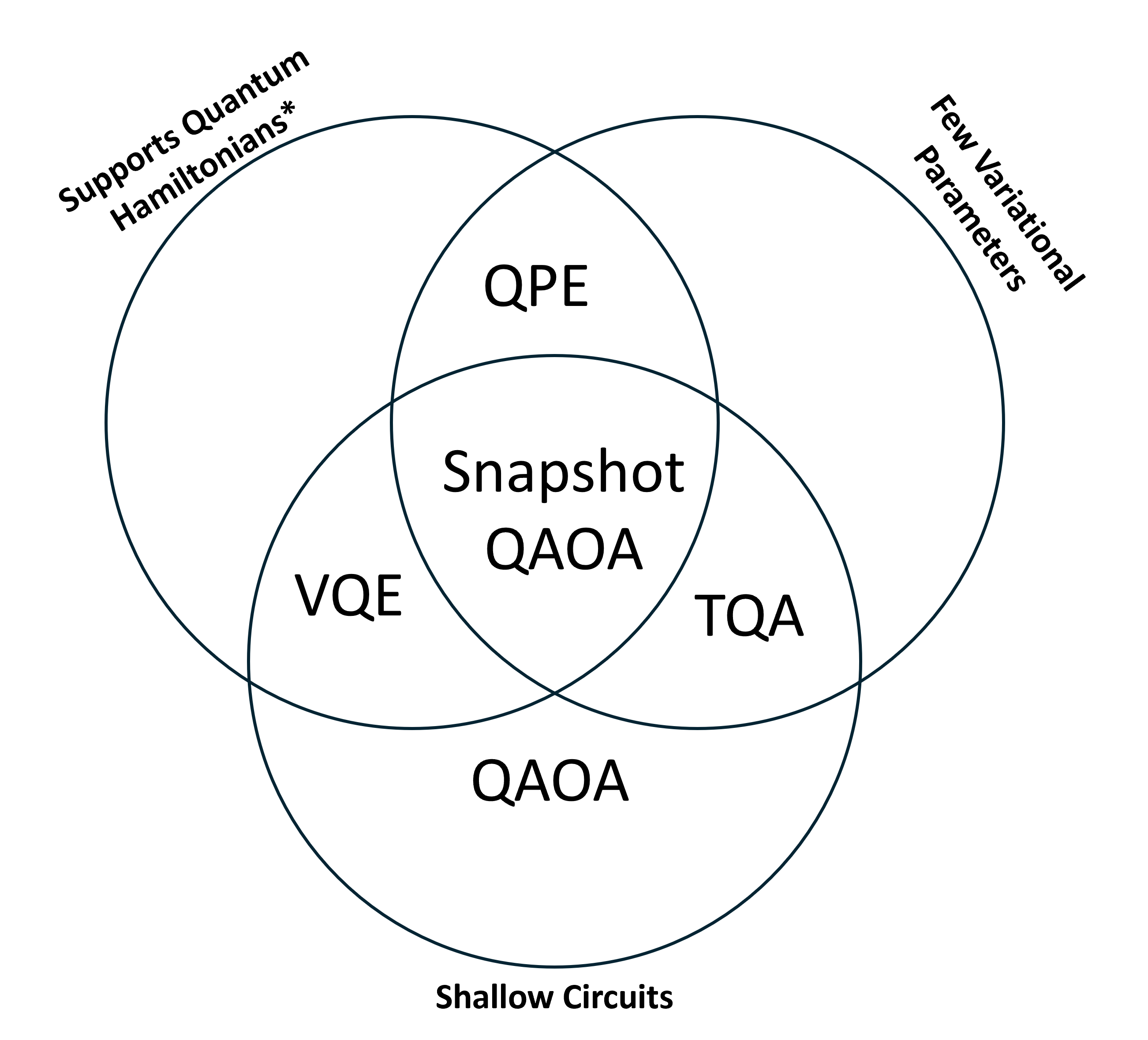}
    \caption{Venn diagram depicting the various properties of Snapshot-QAOA that other alternative algorithms do not have. Regarding the $(*)$ in the diagram above, we do acknowledge that QPE and VQE are able to support a broader class of quantum Hamiltonians (e.g. any sum of Pauli strings) compared to Snapshot-QAOA. We also remark that, per layer, QAOA typically has much fewer variational parameters compared to VQE; the former typically has $O(1)$ parameters per layer whereas the latter typically has $\Omega(n)$ parameters per layer where $n$ is the number of qubits.}
    \label{fig:venn_snapshot}
\end{figure}

The above Snapshot-QAOA properties are significant because while there do exist relatively short-depth quantum algorithm approaches to approximate the ground-states of quantum Hamiltonians, namely variational quantum algorithms such as the Variational Quantum Eigensolver (VQE) \cite{Peruzzo_2014, McClean_2016}, typically these algorithms contain a large number of trainable parameters (usually at least $\Omega(n)$ parameters per VQE layer where $n$ is the number of qubits) which makes the machine learning loop quite difficult in general, especially for standard black-box optimizers. VQE also suffers from other issues related to the difficulty of optimizing the parameters such as the existence of barren plateaus \cite{cerezo2021cost, hashimoto2026comprehensivenumericalstudiesbarren, yao2026gradientanalysisbarrenplateau, Larocca_2022} and local optima \cite{wierichs2020avoiding, Hirsbrunner_2024, kirmani2025variational}. In the other extreme, Quantum Phase Estimation (QPE) is a quantum algorithm that is used as a sub-routine for finding ground-states of quantum Hamiltonians, but requires extremely large circuits\footnote{QPE needs a state preparation circuit to prepare a state that has suitable overlap with the ground state. The cost of just the state preparation circuit alone is comparable Snapshot-QAOA, let alone the entire QPE circuit.} to implement and is therefore typically considered in the realm of only being feasible with full-scale Quantum Error Correction (QEC) \cite{kitaev1995quantummeasurementsabelianstabilizer}. Lastly, algorithms such as Trotterized Quantum Annealing (TQA) \cite{QA_initialization_of_QAOA} also has a single parameter $T$ like Snapshot-QAOA and have NISQ-friendly shallow circuits; however, the approach (as defined in \cite{QA_initialization_of_QAOA}) is limited only to classical Hamiltonians. The Venn diagram in Fig.~\ref{fig:venn_snapshot} summarizes the properties of the algorithms discussed so far.

Just like standard QAOA and VQE (and unlike QPE), we emphasize that Snapshot-QAOA is a \emph{heuristic} meaning that, despite its convergence guarantees in the infinite circuit depth limit ($p = \infty$), there are no known theoretical guarantees\footnote{We remark that there exist known lower-bounds on the number of standard QAOA layers $p$ required to achieve a certain approximation ratio \cite{benchasattabuse2023lower}; however, such bounds are frequently loose and/or trivial, and thus are not enough to elevate QAOA past its status as a heuristic. } in the general finite $p$ regime.

Snapshot-QAOA, in contrast, retains many properties of QAOA such as having a number of control parameters, $\vec{\gamma}$ and $\vec{\beta}$, that is independent of the number of sites in the system being simulated for each layer. Snapshot-QAOA also exhibits properties of, and is based on, a Trotterization of a continuous time evolution that in the limit will reach the ground-state of the quantum Hamiltonian being simulated, similar to standard QAOA which has the same properties for a linear adiabatic schedule. (Section~\ref{sec:results_further_optimize_betas_gammas}). 

The time-dependent Hamiltonian $\mathcal{H}_T$ mentioned previously is dependent on the total anneal time $T$; this parameter is also a parameter for Snapshot-QAOA that can be freely chosen. In the pure annealing case (with no noise or decoherence), the ground state energy is obtained as $T \to \infty$; however for Snapshot-QAOA with fixed number of rounds $p$, one must be careful as increased $T$ also means a larger Trotterization step $\Delta t \propto T/p$. Finding the optimal $T^*$ for any fixed $p$ is a non-trivial problem; however, we find that approximating the optimal $T^*$ with a simple heuristic allows one to obtain satisfactory empirical results.

As a case study, this work empirically explores the behavior of Snapshot-QAOA on a $4 \times 4$ frustrated Transverse-Field Ising Model (TFIM) 2D $J_1$-$J_2$ square lattice with periodic boundary conditions. We show that Snapshot-QAOA indeed converges towards the ground state of this TFIM with high enough QAOA layers $p$.

Lastly, one additional benefit to Snapshot-QAOA is its ease of implementation: it can be implemented using most already-existing QAOA software packages,  libraries, and circuit level instantiations with little to no source code modifications.

This study is an extended version of a conference study that was published in IEEE QAI~\cite{snapshot_QAOA}. Code and data are publicly available on Zenodo~\cite{pelofske_2025_15293030}.

\subsection*{Related Work}
Several prior studies have utilized some form of QAOA to simulate quantum Hamiltonians - specifically meaning Hamiltonians with more than one Pauli basis, e.g., multiple Pauli bases, and measured expectation values in those multiple bases. The authors of \cite{Pagano_2020} study depth $p=1$ and $p=2$ QAOA circuits, including on a trapped-ion quantum computer, for measuring observables of a Hamiltonian that contained both Pauli X and Y bases (although, quantum Hamiltonian ground state approximation was not the explicit goal of these experiments). This approach uses the standard variational learning loop procedure to find a set of $\vec{\beta}, \vec{\gamma}$ that minimizes the energy of the Hamiltonian. Reference \cite{pexe2024usingfeedbackbasedquantumalgorithm} numerically studies a feedback-based variant of QAOA known as FALQON \cite{Magann_2022} to simulate a 1D ANNNI (axial next-nearest-neighbor Ising) model, which contains multiple Pauli basis terms. The primary drawback of the feedback based FALQON algorithm is that it requires extremely high circuit depth. Refs.~\cite{kannan2024quantumapproximateoptimizationalgorithm, marwaha2024performancevariationalalgorithmslocal} both proposed QAOA inspired quantum algorithms that can approximate the ground state of quantum Hamiltonians. While distinct from Snapshot-QAOA, the ideas shared by these studies and ours are similar.

The work in \cite{Lotshaw_2022} examines the task of computing low-energy states of a magnetically frustrated Hamiltonian using QAOA, including the addition of a longitudinal field. However, their Hamiltonian is still strictly diagonal. Several studies have numerically simulated QAOA based on the original motivation of QAOA, which can be viewed as a Trotterization of an adiabatic schedule; these are typically based on standard quantum annealing schedules that work well at sufficiently long time evolution such as in a linear schedule \cite{QA_initialization_of_QAOA}. While reference \cite{QA_initialization_of_QAOA} only simulates diagonal Hamiltonians, the \emph{long time evolution} that is a Trotterized version of the continuous time adiabatic schedule shares a similar motivation with our proposed Snapshot-QAOA. 

Current analog quantum computers are able to perform similar simulations, with continuous time evolution, as Snapshot-QAOA. For example, D-Wave quantum annealers which are manufactured based on superconducting flux qubits \cite{johnson2011quantum, Bunyk_2014} can approximately simulate quantum Hamiltonians that contain $\sum_i X_i$ terms \cite{King_2018, PRXQuantum.2.030317, pelofske2024simulatingheavyhextransversefield} - this is possible since the driving Hamiltonian used in the hardware is the transverse field mixer, therefore performing state readouts during an intermediate point in the anneal can allow measurement of observables of the quantum Hamiltonian, not just the classical diagonal Hamiltonian. Programmable Rydberg atom arrays can also perform certain types of quantum Hamiltonian simulations \cite{Ebadi_2021, Semeghini_2021}. However, there are several limitations of these analog quantum simulators, such as limited qubit coherence times \cite{Bluvstein_2022, king2023quantum}, a finite measurement time that can contribute to part of the system evolution, and in the case of D-Wave quantum annealers, qubits can only be measured in the computational $Z$-basis. 

The $J_1-J_2$ Hamiltonian we consider in this study has also recently been tested with VQE in ref.~\cite{kirmani2025variational}; in addition to needing much fewer parameters, later in this study we definitively demonstrate that Snapshot-QAOA gives energies much closer to the ground state compared to VQE.

\begin{figure}[ht!]
    \centering
    \includegraphics[width=0.999\linewidth]{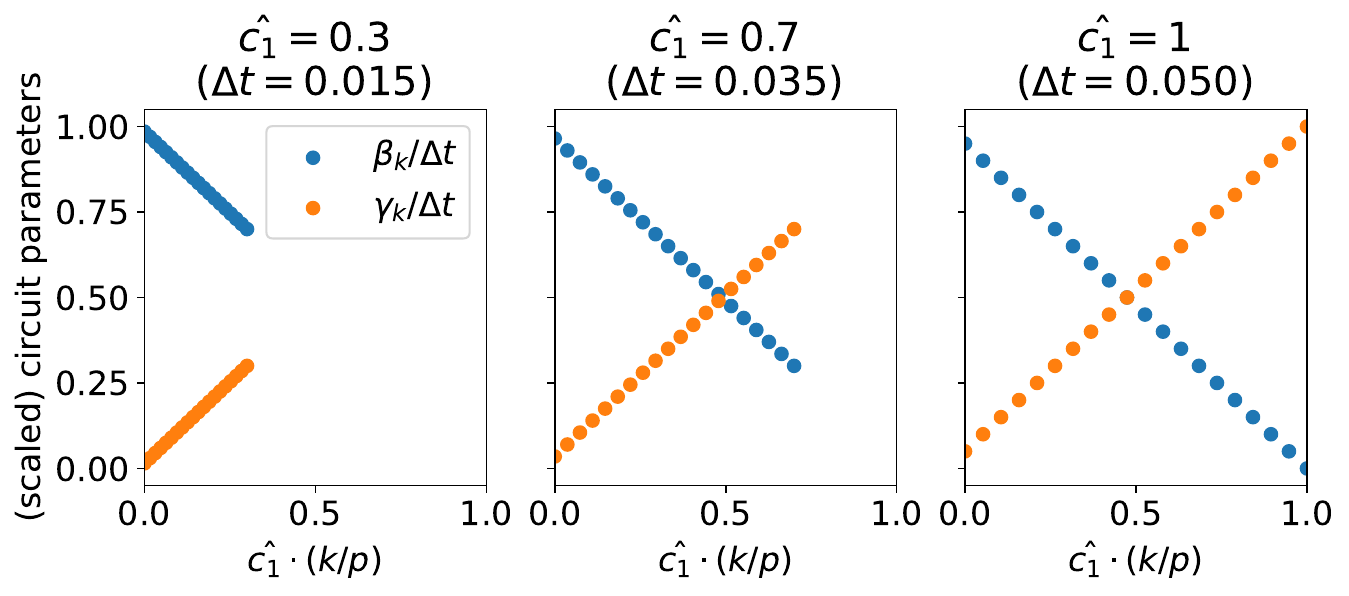}
    \caption{A visualization connecting the $\vec{\gamma}$ and $\vec{\beta}$ parameters of Snapshot-QAOA to partial quantum annealing. The $\vec{\gamma}$ and $\vec{\beta}$ parameters of Snapshot-QAOA follow a linear ramp for varying $\hat{c}_1$ values ($\hat{c}_1=0.3,0.7,1.0$). The last subplot, with $\hat{c}_1=1.0$, corresponds to TQA \cite{QA_initialization_of_QAOA}. Each subplot above uses $p=20$ layers with $T=1$. The parameters are scaled by $\Delta t = \tau / p = \hat{c_1} T / p = \hat{c_1}/20$ so that the values along the vertical axis range lie in the interval $[0,1]$. The horizontal axis represents the fraction of the total anneal schedule (from $t=0$ to $t=T$) that the parameter corresponds to. At each $k=1,2,\dots,p$, the scaled parameters $\beta_k/\Delta t$ and $\gamma_k/\Delta t$ add up to 1, so the vertical axis can be interpreted as the ``relative influence" that each of the two alternating unitaries has at each layer of Snapshot-QAOA.}
    \label{fig:schedule_of_gammas_and_betas}
\end{figure}

\section{Background}
\label{section:background}
\subsection{Quantum Annealing}
Suppose we wish to find the ground state $\ket{\psi_1}$ of some Hamiltonian $H_1$. To accomplish this, we consider a different  Hamiltonian $H_0$ (with the same dimensions as $H_1$) whose ground state $\ket{\psi_0}$ we can easily prepare. Now consider the following time-dependent Hamiltonian $\mathcal{H}_T$ given by:
\begin{equation}
\label{eqn:time_dependent_ham}
\mathcal{H}_T(t) = (1-t/T)H_0 + (t/T)H_1,
\end{equation}
where $T$ is the total annealing time.  We say that $\mathcal{H}_T$ has no level-crossings if the two smallest eigenvalues (with multiplicity) of $\mathcal{H}_T(t)$ have a non-zero gap between them for all $t \in [0,T]$. If $\mathcal{H}_T$ has no level crossings, then the quantum adiabatic theorem states that as long as the anneal time $T$ is long enough, then we will always remain in the ground state of $H_T(t)$ at each time $t$; in particular, we will evolve from the ground state of $\mathcal{H}_T(0) = H_0$ (i.e. $\ket{\psi_0}$) to the ground state of $\mathcal{H}_T(T) = H_1$ (i.e. $\ket{\psi_1}$) \cite{born1928beweis, kato1950adiabatic}.

\subsection{QAOA}
Let $c: \{0,1\}^n \to \mathbb{R}$ be a classical cost function and let $H_1$ be a (diagonal) Hamiltonian with the property that $H_1\ket{b} = c(b)\ket{b}$ for all $b \in \{0,1\}^n$. Observe that minimizing $c$ is equivalent to finding the ground state of $H_1$; the intention of QAOA is to find this ground state. Similar to quantum annealing, some $H_0$ is selected where the ground state $\ket{\psi_0}$ is easy to construct (e.g. $H_0 = -\sum_i X_i$ and $\ket{\psi_0} = \ket{+}^{\otimes n}$). To this end, the following $p$-layer QAOA is run with parameters $\vec{\gamma} = (\gamma_1, \dots, \gamma_p)$ and $\vec{\beta} = (\beta_1,\dots, \beta_p)$:
$$\ket{\psi_p(\vec{\gamma},\vec{\beta})} =   \left[\prod_{k=1}^p\exp\left( -i\beta_k H_0\right) \exp\left(-i\gamma_k H_1 \right) \right] \ket{\psi_0}.$$

The state is then measured $\ket{\psi_p(\vec{\gamma},\vec{\beta})}$ in the computational basis to obtain bitstrings $x \in \{0,1\}^n$. The parameters $\vec{\gamma}$ and $\vec{\beta}$ above are typically optimized by an outer classical loop that attempts to minimize the expected cost returned by the QAOA circuit, i.e., $\bra{\psi_p(\vec{\gamma},\vec{\beta})} H_1 \ket{\psi_p(\vec{\gamma},\vec{\beta})}$. To understand the mechanism underlying QAOA, it is important to understand the relationship between QAOA and quantum annealing, which we discuss in the next section. 

\subsection{Connection Between QAOA and Quantum Annealing}
\label{sec:QAOA_QAA_Connection}

By letting $\Delta t = T/p$ and picking
\begin{align}
\label{eqn:gamma_and_beta_full_anneal}
\beta_k &= \Delta t \cdot (1-k/p)\\
\gamma_k &= \Delta t \cdot (k/p) \notag,
\end{align}

one can show that the $p$-layer QAOA (with the above parameters) is a Trotterized version of running quantum annealing with a total anneal time of $T$. A brief derivation of these formulas for $\gamma_k$ and $\beta_k$ can be found in Sack et al.'s work \cite{QA_initialization_of_QAOA}; they refer to this variant of QAOA with $(\vec{\gamma},\vec{\beta})$ chosen by Eq.~\ref{eqn:gamma_and_beta_full_anneal} as \emph{Trotterized Quantum Annealing} (TQA). Farhi et al. first discussed the connection between QAOA and the quantum annealing in the original QAOA paper \cite{QAOA}; this connection is used to establish the fact that as $p\to \infty$, the state produced by QAOA will approach the ground state corresponding to the optimal solution of the classical optimization problem. However, the advantage (and motivation) of QAOA is that it is also able to do well in the low-$p$ regime since $\vec{\gamma}$ and $\vec{\beta}$ are variables we can freely optimize \cite{QAOA}.

Our method, which we call Snapshot-QAOA, generalizes the standard QAOA by further exploiting the connection between QAOA and adiabatic quantum annealing, which allows one to find ground states of certain classes of ``quantum" (i.e., non-diagonal) Hamiltonians. We discuss the details of our approach in the next section.

\section{Snapshot-QAOA}
In this section, we go through the general theoretical framework for Snapshot-QAOA, which implements a partial quantum annealing schedule in the Quantum Approximate Optimization Algorithm (QAOA) \cite{QAOA}. We begin by discussing the types of Hamiltonians $H$ that our approach can be applied to (Sec.~\ref{sec:general_hamiltonian_structure}), then we discuss how such $H$'s can be viewed as a ``snapshot" at some point time of an adiabatic quantum annealing schedule (Sec.~\ref{sec:annealingSnapshot}),  and lastly, we provide formulae for the $(\vec{\gamma},\vec{\beta})$ values that correspond to a partial annealing schedule (Sec.~\ref{sec:QAOA_with_Partial_Anneal_Schedule}).

\subsection{General Hamiltonian Structure}
\label{sec:general_hamiltonian_structure}
In this work, we explore the problem of finding ground states and energies of certain kinds of Hamiltonians. Namely, we consider Hamiltonians $H$ of the form:
\begin{equation}\label{eqn:HamiltonianForm}
    H = c_0H_0+c_1H_1,
\end{equation}
where $H_0,H_1$ are Hermitian and $c_0,c_1 \in \mathbb{R}$. We let $\ket{\psi_H}$ denote the target ground state of $H$ with corresponding ground state energy $\mathcal{E}_H$.

\begin{remark}Note that the coefficients $c_0$ and $c_1$ of Eq.~\ref{eqn:HamiltonianForm} are artificial in a sense, i.e., we could replace them with any other choice of coefficients $c_0'$ and $c_1'$ and instead write $H=c_0'H_0'+c_1'H_1'$ by setting $H_0' := \frac{c_0}{c_0'}H_0$ and $H_1' := \frac{c_1}{c_1'}H_1$. However, for many problems, there is often a ``natural" choice of $c_0$ and $c_1$ that we will use.
\end{remark}

In addition to having the form above, there are five other criteria that the Hamiltonians should satisfy:
\begin{enumerate}[noitemsep]
    \item The ground state, $\ket{\psi_0}$, of $H_0$ should be non-degenerate, i.e., it is the unique ground state.
    \item The preparation of the ground state, $\ket{\psi_0}$, of $H_0$ should be easily implementable on a quantum computer.
    \item For any constant $\alpha \in \mathbb{R}$, the unitary operators $e^{-i\alpha H_0}$ and $e^{-i\alpha H_1}$ should be easily implementable on a quantum device.
    \item The Hamiltonians $H_0$ and $H_1$ should not commute with one another.
    \item For all $T>0$, the corresponding time-dependent Hamiltonian with total anneal time $T$, i.e., \begin{equation}\label{eqn:time_dependent_ham}\mathcal{H}_T(t) = (1-t/T)H_0 + (t/T)H_1,\end{equation}  should not have any level-crossings, i.e., the two smallest eigenvalues (with multiplicity) of $\mathcal{H}_T(t)$ have a non-zero gap between them for all $t \in [0,T]$.
\end{enumerate}

We leave the notion of ``easily implementable" up to the reader. Whether or not our approach is suitable for a particular application will depend on the specific quantum hardware, noise level tolerances, time, computing resource limits, and potentially other factors.

We provide concrete examples of Hamiltonians that satisfy the above properties in Sec.~\ref{section:applicable_hamiltonians}.

\subsection{Annealing Snapshot}
\label{sec:annealingSnapshot}

Given a Hamiltonian of the form in Eq.~\ref{eqn:HamiltonianForm}, we can view it as the ``snapshot" of some time-dependent Hamiltonian corresponding to a linear-annealing schedule. To this end, we first construct what we call the \emph{normalized} Hamiltonian by scaling by the appropriate factor so that the coefficients add up to 1:
$$\hat{H} = \hat{c_0}H_0 + \hat{c_1}H_1,$$
where,
$$\hat{c_0} = \frac{c_0}{c_0+c_1} \quad \text{and} \quad \hat{c_1} = \frac{c_1}{c_0+c_1}.$$

For any $T>0$, consider the corresponding time-dependent Hamiltonian with a total-anneal time of $T$ given by Eq.~\ref{eqn:time_dependent_ham}.
The above Hamiltonian is the time-dependent Hamiltonian that occurs when quantum annealing with a linear schedule is applied with a starting Hamiltonian of $H_0 = \mathcal{H}_T(0)$ and an ending Hamiltonian of $H_1= \mathcal{H}_T(T)$. 
Observe that when substituting $t = \hat{c_1}T$ into the equation, then our normalized Hamiltonian $\hat{H}$ can be viewed as a snapshot of the above Hamiltonian, i.e., $\hat{H} = \mathcal{H}_T(\hat{c_1}T),$ in other words, $\hat{H}$ is the Hamiltonian that occurs at a $\hat{c_1}$-fraction of the way through the anneal process. We will use $\tau = \hat{c_1}T$ to denote the partial anneal time so that $\hat{H} = \mathcal{H}_T(\tau)$.

\begin{remark}
When $c_0,c_1 > 0$, then $0 < \hat{c_1} < 1$, and so it makes sense to refer to $\hat{H}$ as being the result of ``partial annealing". However, without the positivity restrictions, it is possible for $\tau < 0$  or $\tau > T$, which are in a sense ``non-physical". This work only investigates examples where $0 < \hat{c_1} < 1$ (or equivalently $0 < \tau < T$); however, such non-physical cases may be of interest for future work.
\end{remark}

\subsection{Description of Snapshot-QAOA}
\label{sec:QAOA_with_Partial_Anneal_Schedule}
Given a (quantum) Hamiltonian $H = c_0H_0+c_1H_1$ satisfying the properties in Sec.~\ref{sec:general_hamiltonian_structure}, we now present a variant of QAOA, which we call Snapshot-QAOA, that aims to find the ground state of $H$. In Snapshot-QAOA with $p$ steps, given a parameter $T>0$, we construct the state 
$$\ket{\psi_p(T)} =   \left[\prod_{k=1}^p\exp\left( -i\beta_k H_0\right) \exp\left(-i\gamma_k H_1 \right) \right] \ket{\psi_0},$$
where each of the $\gamma_k$'s and $\beta_k$'s are $T$-dependent values that are chosen in a specific way (which we describe in later sections), and $\ket{\psi_0}$ is the ground state of $H_0$. Observe that the above circuit is equivalent to the standard QAOA circuit with ``mixer" $H_0$ and ``phase separator" $H_1$; however, unlike standard QAOA, the phase-separating Hamiltonian $H_1$ is \emph{not} equal to the Hamiltonian $H$ whose ground state we wish to find (and thus the optimal variational parameters $\vec{\gamma}$ and $\vec{\beta}$ will be different).

In standard QAOA, by choosing $\beta_k = \frac{T}{p} \cdot (1-k/p)$ and $\gamma_k = \frac{T}{p} \cdot (k/p)$, one can show that QAOA is equivalent to a Trotterization of a linear annealing schedule with total anneal time of $T$; a brief derivation of these formulas for $\gamma_k$ and $\beta_k$ can be found in Sack et al.'s work \cite{QA_initialization_of_QAOA}. Since $\hat{H} = \mathcal{H}_T(\tau)$, the derivation can be modified to find a choice of $(\vec{\gamma},\vec{\beta})$ that causes Snapshot-QAOA to become a Trotterized version of a \emph{partial} annealing process running from $t=0$ to $t=\tau$ of the time-dependent Hamiltonian $\mathcal{H}_T(t)$:
\begin{align}
\label{eqn:partialAnnealParams}
    \beta_k &= \frac{\tau}{p} \cdot \left(1-\frac{k\hat{c}_1 }{p}\right) \\
    \gamma_k &= \frac{\tau}{p} \cdot \left( \frac{k\hat{c}_1 }{p}\right) \nonumber
\end{align}
Fig.~\ref{fig:schedule_of_gammas_and_betas} shows how the $\gamma_k$ and $\beta_k$ values change (with $k$) as a linear ramp. Note that $T$ is a parameter that can be freely chosen in Snapshot-QAOA; once $T$ (and $p,c_0,c_1,H_0,H_1$) are chosen, then all the $\beta_k$'s and $\gamma_k$'s are uniquely determined. If desired, one can then choose to further optimize $\vec{\beta}$ and $\vec{\gamma}$ to minimize the state energy further; unless otherwise stated, we will use the term Snapshot-QAOA to refer to the algorithm \emph{without} any further optimization of $\vec{\gamma}$ and $\vec{\beta}$. 

The energy $\mathcal{E}_p(T)$ of the resulting Snapshot-QAOA state $\ket{\psi_p(T)}$ is given by:
$$\mathcal{E}_p(T) = \bra{ \psi_p(T)} H \ket{\psi_p(T)}.$$

In the case that $\tau=T$ (and hence $\hat{c}_0=0$ and $\hat{c}_1=1$), one can show that our Snapshot-QAOA method is equivalent to the method by Sack et al. \cite{QA_initialization_of_QAOA} which they refer to as Trotterized Quantum Annealing (TQA). 

In the next few sections, we explore how Snapshot-QAOA simulations perform on certain quantum Hamiltonians of interest, and Sec.~\ref{sec:analytical_results} examines theoretical properties of Snapshot-QAOA (e.g. periodicity, convergence, etc.). To maintain consistency with our terminology, we will instead refer to the $\hat{c}_1=1$ setting of Snapshot-QAOA as Final-Snapshot-QAOA. Note that Final-Snapshot-QAOA has an additional freedom in mixer choice since if $\hat{c}_0=0$, changing $H_0$ does not change the Hamiltonian $H$ as $H=c_0H_0+c_1H_1 = c_1H_1$.

\section{Snapshot-QAOA Applied to Specific Quantum Hamiltonians}
\label{section:simulation_methods}

\subsection{Applicable Hamiltonians for Snapshot-QAOA}
\label{section:applicable_hamiltonians}
In Sec.~\ref{sec:general_hamiltonian_structure}, we discussed rather abstract properties that the Hamiltonian $H$ should satisfy for Snapshot-QAOA. We will take a moment to give some concrete examples and forms of Hamiltonians for which Snapshot-QAOA is applicable.

Of particular interest is the class of Transverse-Field Ising Models (TFIM) which consists of Pauli-$ZZ$ couplings determined by a weighted interaction graph and Pauli-$X$ terms that correspond to a transverse magnetic field; we define this class of Hamiltonians more formally in the next subsection. TFIM models have applications in fields such as chemistry and materials science, e.g., TFIMs can be used to model and study the magnetization of $\text{Ca}_3\text{Co}_2\text{O}_6$ \cite{nekrashevich2022reaching,PRXQuantum.2.030317}. In general, TFIM is a rather broad and heavily-studied category of Hamiltonians: many specific kinds of TFIMs arise as a result of considering variations in the weights and structure of the interaction graph, such as the $J_1-J_2-J_3$ TFIM on the ruby lattice \cite{duft2024order}, the 1D ANNNI model in the transverse field \cite{sen1989ising, suzuki2012quantum, sen1992numerical, Chandra_2007, Rieger_1996}, the 3D Ising Spin Glass in a transverse magnetic field \cite{guo1994quantum}, TFIM with a triangular lattice \cite{liu2020intrinsic}, and TFIM on a honeycomb lattice \cite{coletta2011phase}. Snapshot-QAOA can be used on any TFIM, regardless of the choice of structure and interaction strengths determined by the interaction graph, and thus, all of the example Hamiltonians listed above can be used with Snapshot-QAOA.

The Snapshot-QAOA algorithm is not just limited to TFIMs. In particular, Snapshot-QAOA can be used with any Hamiltonian of the form $H = c_0H_0 + c_1H_1$ where,
$$H_0 = \sum_i a_i P_i + \sum_{i,j} b_{ij} P_iP_j + \sum_{i,j,k} c_{ijk} P_i P_j P_k + \dots$$
$$H_1 = \sum_i d_i Q_i + \sum_{i,j} e_{ij} Q_iQ_j + \sum_{i,j,k} f_{ijk} Q_i Q_j Q_k + \dots,$$
where the coefficients ($a_i, b_{ij},\dots$) are real-valued,  $P,Q \in \{X,Y,Z\}$ are two distinct Pauli-types (i.e., $P \neq Q$),
and where $H_0$ is known to not have any degenerate ground states. The Antiferromagnetic $ZZXZ$ model \cite{schiffer2024quantum} is an example of a non-TFIM Hamiltonian that falls within the form above.

Hamiltonians outside of the above form can also be used for Snapshot-QAOA as long as the sub-Hamiltonians $H_0$ and $H_1$ are both easily implementable on a digital quantum computer and satisfy the remaining conditions in Sec.~\ref{sec:general_hamiltonian_structure}, e.g., one could take $H = c_0H_0 + c_1H_1$ with $H_0$ being described by $\sum_{(i,j) \in E} X_iX_j + Y_iY_j$ where $E$ is the edge-set of a ring or clique \cite{rieffel2020xy} and $H_1$ representing a classical thresholding function \cite{golden2021threshold}.

\subsection{Transverse Field Ising Model Ground State Simulation with Snapshot-QAOA}
\label{section:simulation_methods_TFIM_details}

We now focus on the Transverse Field Ising Model (TFIM) - this class of Hamiltonians contains arbitrary $Z_i$ spin interactions, and $\sum_i X_i$ terms. Snapshot-QAOA can address approximate ground-state finding of this general class of Hamiltonians. 

Given a graph $G = (V,E)$ with edge weights $w: E \to \mathbb{R}$, and a constant $B_x$, we can construct the following TFIM Hamiltonian:
$$H = B_x\sum_{j \in V}X_j + \sum_{(j,k) \in E} w_{jk} Z_jZ_k .$$

By identifying 

$$c_0 = B_x, H_0 = \sum_{j \in V}X_j, c_1 = 1, H_1 = \sum_{(j,k) \in E} w_{jk} Z_iZ_j,$$ we see that this class of Hamiltonians adheres to the form specified in Eq.~\ref{eqn:HamiltonianForm}. By normalizing the constants as is done in Sec.~\ref{sec:annealingSnapshot}, we obtain the normalized Hamiltonian $\hat{H} = \hat{c_0}H_0+\hat{c_1}H_1$ with $\hat{c_0} = 1 - \hat{c_1}$ where,
$$\hat{c_1} = \frac{1}{1+B_x}.$$

Observe that for $B_x > 0$, we have that $0 < \hat{c_1} < 1$, thus, as discussed in Sec.~\ref{sec:annealingSnapshot}, $\hat{H}$ can be realized as a snapshot of the annealing process of some time-dependent Hamiltonian. As a result of Sec.~\ref{sec:QAOA_with_Partial_Anneal_Schedule}, starting with the ground state of $H_0$ (which is $\ket{-}^{\otimes n}$), the ground-state of $\hat{H}$ can be approximated using a QAOA circuit with circuit parameters set to those in Eq.~\ref{eqn:partialAnnealParams} with the substitution $\hat{c_1} = \frac{1}{1+B_x}$.

\section{Analytical Results}
\label{sec:analytical_results}
In this section, we present theoretical results regarding how the energy $\mathcal{E}_p(T)$ (of the state $\ket{\psi_p(T)}$ output by Snapshot-QAOA) changes as $p$ and $T$ vary. Sec.~\ref{sec:limitingBehavior} gives us assurance that the ground energy is approached as $p$ tends to infinity. Sec.~\ref{sec:theoretical_props_of_anneal_time} shows that the parameter search space for $T$ can typically be restricted to a bounded region that can easily be calculated.

\subsection{Limiting Behavior as $T\to\infty$ and $p\to\infty$}
\label{sec:limitingBehavior}
Let $\mathcal{E}_\text{QA}(T) = \lim_{p \to \infty} \mathcal{E}_p(T)$; then $\mathcal{E}_\text{QA}(T)$ is the result of the Trotter limit which is exactly equal to the energy achieved by doing quantum annealing with time-dependent Hamiltonian $\mathcal{H}_T$ (from time $t=0$ to $t=\tau$). Since the corresponding time-dependent Hamiltonian $\mathcal{H}_T$ is assumed to not have any level-crossings, it then follows from the quantum adiabatic theorem that $\lim_{T\to\infty}\mathcal{E}_\text{QA}(T) = \lim_{T\to\infty}\lim_{p\to\infty} \mathcal{E}_p(T) = \mathcal{E}_H$, i.e., one achieves the ground state energy of $H$ in the limit. We remark that the ordering of the limits $(\lim_{T\to\infty})$ and $(\lim_{p\to\infty})$ is important; in particular  $\lim_{p\to\infty}\lim_{T\to\infty} \mathcal{E}_p(T)$ is undefined since $\lim_{T\to\infty} \mathcal{E}_p(T)$ will typically not exist (see Fig.~\ref{fig:examples_T_vs_Energy_full_period}). However, what can be said is that $\lim_{p \to \infty} \min_{T \in \mathbb{R}} \mathcal{E}_p(T) = \mathcal{E}_H$; this can easily be shown via the squeeze theorem of calculus:

$$\mathcal{E}_H \leq \min_{T'}\mathcal{E}_p(T') \leq \mathcal{E}_p(T), \quad \forall p, \  \forall T,$$
$$\implies \mathcal{E}_H \leq \lim_{p \to \infty} \min_{T'}\mathcal{E}_p(T') \leq  \lim_{p \to \infty} \mathcal{E}_p(T), \quad   \forall T$$
$$\implies  \mathcal{E}_H \leq  \lim_{p \to \infty} \min_{T'}\mathcal{E}_p(T') \leq  \lim_{T \to \infty}\lim_{p \to \infty} \mathcal{E}_p(T)= \mathcal{E}_H $$
$$\implies \lim_{p \to \infty} \min_{T}\mathcal{E}_p(T) =  \mathcal{E}_H .$$

\subsection{Periodicity of $T$}
\label{sec:theoretical_props_of_anneal_time}
Suppose that the QAOA unitaries corresponding to $H_0$ and $H_1$ are both periodic in $\beta$ and $\gamma$, i.e., $e^{-i\beta H_0}$ is periodic in $\beta$ with period $\rho_0$ and $e^{-i\gamma H_1}$ is periodic in $\gamma$ with period $\rho_1$ (up to some global phase). Then, unlike the quantum annealing case, it can be shown that the energy of Snapshot-QAOA is periodic in $T$ with a period $\rho$ that can be easily calculated, i.e., $\mathcal{E}_p(T) = \mathcal{E}_p(\rho+T)$ for all $T$. For the period $\rho$ that we calculate below, we do not make any claims regarding whether or not such a $\rho$ is a \emph{minimal} period or not.

For $k=1,2,\dots,p$, let $\hat{\beta}_k$ and $\hat{\gamma}_k$ be the values that partial-annealing QAOA would return in the case that $T=1$. Observe that if $\beta_k$ and $\gamma_k$ are the values returned by partial-annealing QAOA for arbitrary $T$, then $\beta_k = T \hat{\beta}_k$ and $\gamma_k = T\hat{\beta_k}$ (this follows from how $\gamma_k$ and $\beta_k$ are defined in Eq.~\ref{eqn:partialAnnealParams}). It then follows that the QAOA unitary $e^{-i\beta_k H_0} = e^{-i\hat{\beta}_kT H_0}$ is periodic in $T$ with period $\rho_0/\hat{\beta}_k$ and similarly, $e^{-i\gamma_k H_1} = e^{-i\hat{\gamma}_kT H_1}$ is periodic in $T$ with period $\rho_1/\hat{\gamma}_k$. Thus, as a function of $T$, the overall state (and hence the energy) of Snapshot-QAOA has a period that is the least-common multiple (LCM) of periods of each unitary, i.e., 

\begin{equation}
\label{eqn:period_formula}
\rho = \text{LCM}(\rho_0/\hat{\beta}_1, \dots, \rho_0/\hat{\beta}_p, \rho_1/\hat{\gamma}_1, \dots, \rho_1/\hat{\gamma}_p).
\end{equation}

It can also be shown that \emph{within} each period $\rho$, there exists a mirror-symmetry, i.e., $\mathcal{E}_p(T) = \mathcal{E}_p(\rho-T)$. From Eq.~\ref{eqn:partialAnnealParams}, it is clear that changing the anneal time from $T$ to $-T$ has the effect of negating all of the $\gamma_k$ and $\beta_k$ parameters, which is known to not have an effect on energies output by the QAOA circuit (see \cite{lee2023depth} for an elegant proof). Thus, $\mathcal{E}_p(T) = \mathcal{E}_p(-T) = \mathcal{E}_p(\rho-T)$ for all $T$.

We remark\footnote{The results in the associated paragraph are well-known; we refer the reader to Section 7 of the Supplementary Materials of \cite{tate2025theoretical} for a detailed proof.} that in the case where $H_0$ is the transverse field mixer, i.e., $H_0 = -\sum_j X_j$, then we can take\footnote{While the minimal period of the transverse field \emph{unitary} is $\pi$, in the case of the Hamiltonian $H^\star(J_2, B_x)$, the overall \emph{energy} is actually $\pi/2$ periodic with respect to $\beta$. This is because, up to a global phase, $\exp(-i\frac{\pi}{2} \sum_j X_j) = \prod_j X_j$ which commutes with both $H_0$ and $H_1$. In such a case, it is safe to use the value $\rho_0 = \pi/2$ in Equation \ref{eqn:period_formula} if one only cares about periodicity of the energy and not the state.} $\rho_0 = \pi$. In the case that $H_1$ is a diagonal matrix with integer entries, then one can take $\rho_1 = 2\pi$.

For the specific Hamiltonian $H^\star(J_2,B_x)$ defined in Sec.~\ref{section:simulation_methods_frustrated_Hamiltonian_test_case}, if $J_2 = \frac{a}{b} \in \mathbb{Q}$ is written as a reduced-fraction, then one can take $\rho_1 = b \cdot (\pi/2)$; this is because in the case $b=1$, the periodic square lattice is $4$-regular, and flipping any single spin changes the diagonal energy by a multiple of $4$. Since the ferromagnetic configuration has energy divisible by $4$, it follows that all eigenvalues of $H_1$ lie in $4\mathbb{Z}$ and  consequently, the phase $e^{-i\gamma H_1}$ has minimal period $\pi/2$, a factor of four smaller than the generic $2\pi$ periodicity for integer-valued diagonal Hamiltonians.

\begin{figure*}[ht!]
    \centering
    \includegraphics[width=0.6\linewidth,trim={3cm 2cm 3cm 2cm},clip]{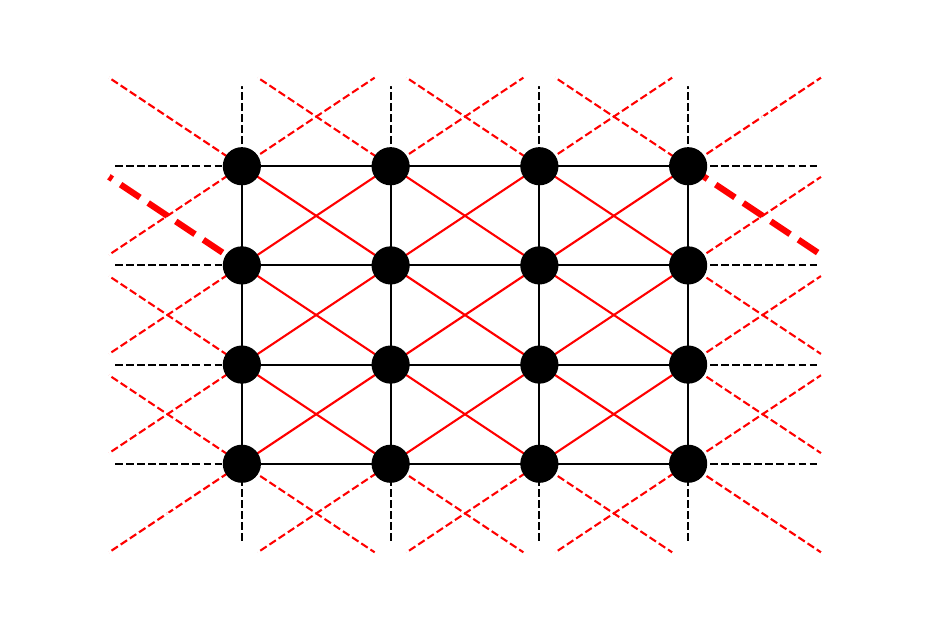}
    \caption{The underlying graph $G$ for the sub-Hamiltonian $H_1$ of the $4\times 4$ frustrated TFIM $H^\star$ that defines toroidal periodic boundary conditions of the lattice. The black lines denote nearest-neighbor connections with weight $J_1=1$ and the red lines denote next-nearest-neighbor connections with weight $J_2$. Each solid line corresponds to exactly one edge in $G$. The periodic boundary conditions are represented via dashed lines; each dashed line has a corresponding dashed line that corresponds to the same underlying edge in the graph. The bolded red lines are one such example of a pair of dashed lines that correspond to the same underlying edge. }
    \label{fig:grid_graph_4_by_4}
\end{figure*}

\section{Numerical Results}
\label{section:results}
We present numerical simulations of Snapshot-QAOA on a specific test case of a frustrated quantum Hamiltonian $H^\star$, details described in  Sec.~\ref{section:simulation_methods_frustrated_Hamiltonian_test_case}. All numerical simulations are performed using an adapted version of the Julia programming language \cite{bezanson2017julia} based QAOA simulator \texttt{JuliQAOA} \cite{JuliQAOA}. All numerical computations were performed using parallelization and HPC (High Performance Computing) resources. All numerical experiments use statevector simulations; no shot noise is present. All eigenstates of the $16$-qubit Hamiltonians are computed through LAPACK exact diagonalization routines in the Python 3 library Numpy \cite{lapackcite, harris2020array}. This method gives us the exact ground state energy for each Hamiltonian defined by $J_2, B_x$ which we notate as $\mathcal{E}_0(H^\star(J_2,B_x))$. 

Some results (Sec.~\ref{sec:results_further_optimize_betas_gammas}) will consider Snapshot-QAOA with further optimization of the $\vec{\gamma}$ and $\vec{\beta}$ parameters; in such cases further optimization is performed via a local search using the gradient-based Broyden–Fletcher–Goldfarb–Shanno (BFGS) optimizer \cite{fletcher2000practical} with $\vec{\gamma}$ and $\vec{\beta}$ initialized according to Eq.~\ref{eqn:partialAnnealParams}. 

\subsection{Test Case: The Frustrated 2D TFIM $J_1$-$J_2$ Square Lattice}
\label{section:simulation_methods_frustrated_Hamiltonian_test_case}
Here we choose a specific Hamiltonian to demonstrate the capabilities of Snapshot-QAOA; the 2D $J_1$-$J_2$ frustrated transverse field Ising model \cite{Oitmaa_2020_TF, PhysRevE.99.012134, PhysRevE.97.022124, PhysRevB.92.174419, PhysRevB.94.214419}:

\begin{align}
\label{eq:quantum_J1_J2_Hamiltonian}
H = -J_1\sum_{\langle ij \rangle}\sigma_i^z\sigma_j^z + J_2\sum_{\langle\!\langle ij\rangle\!\rangle}\sigma_i^z\sigma_j^z
+ B_x \sum_i\sigma_i^x
\end{align}

where $\langle \rangle$ denotes the set of nearest neighbors in Euclidean space and $\langle\!\langle \rangle\!\rangle$ defines the corresponding set of next-nearest-neighbors. This model has competing magnetic interactions between the ferromagnetic nearest neighbor terms and the next-nearest-neighbor terms. We set $J_1=1$ and then vary $J_2$ as the frustration parameter. $B_x$ controls the strength of quantum fluctuations. This Hamiltonian serves as a good test case for the simulation capabilities of Snapshot-QAOA because it is geometrically frustrated and has a maximally frustrated region with high degeneracy when $J_2=0.5$ for small $B_x$. The ground-state ordering of this model is ferromagnetic when $J_2<0.5$, and when $J_2>0.5$ the ground-state ordering is antiferromagnetic-like with stripe like patterns. When $B_x$ is sufficiently large the ordering is paramagnetic. The Hamiltonian used in the numerical simulations below has a graph $G$ that is a $4\times 4$ grid with periodic boundary conditions where the edge to each vertex's nearest neighbor (i.e., the 4 neighbors in each of the 4 cardinal directions) has weight $J_1$ and the edge to each vertex's next-nearest neighbor (i.e., the 4 neighbors that are immediately diagonal to the vertex in the grid) has weight $J_2$; a depiction of this graph can be found in Fig.~\ref{fig:grid_graph_4_by_4}. Periodic boundary conditions are preferred so as to remove edge effects and to approach the thermodynamic limit. We will denote this specific Hamiltonian as $H^\star$ or as $H^\star(J_2,B_x)$ when we wish to refer to the Hamiltonian with specific $J_2$ and $B_x$ values. 

Quantum Monte Carlo (QMC) and Density Matrix Renormalization Group (DMRG) techniques can also be used to obtain the ground-state energy of this quantum Hamiltonian, both methods can be readily applied to various frustrated magnetic systems~\cite{MonteCarlofrustrated, DMRG_frustrate}. However, DMRG is limited in scale for dimensions greater than $1$, and quantum monte carlo can suffer from instance specific failure modes such as slow convergence caused by low transverse fields. It is therefore of interest to develop quantum algorithms that can perform better than the existing classical methods (QMC and DMRG), and this frustrated $J_1-J_2$ model serves as a good quantum Hamiltonian model with tunable frustration to explore the algorithmic properties of Snapshot-QAOA. 

We remark that one would ideally want to run our approach on much larger grid sizes than $4 \times 4$ since it is believed that the ground state of larger grid sizes (with periodic boundary conditions) yields an increasing better approximation of the ground state behavior of the infinite-size system (i.e., a grid whose vertex set is $\mathbb{Z} \times \mathbb{Z}$), which would yield valuable insights as to the frustration effects particularly near the maximally frustrated region.

\begin{figure*}[ht!]
    \centering
    \includegraphics[width=0.32\linewidth]{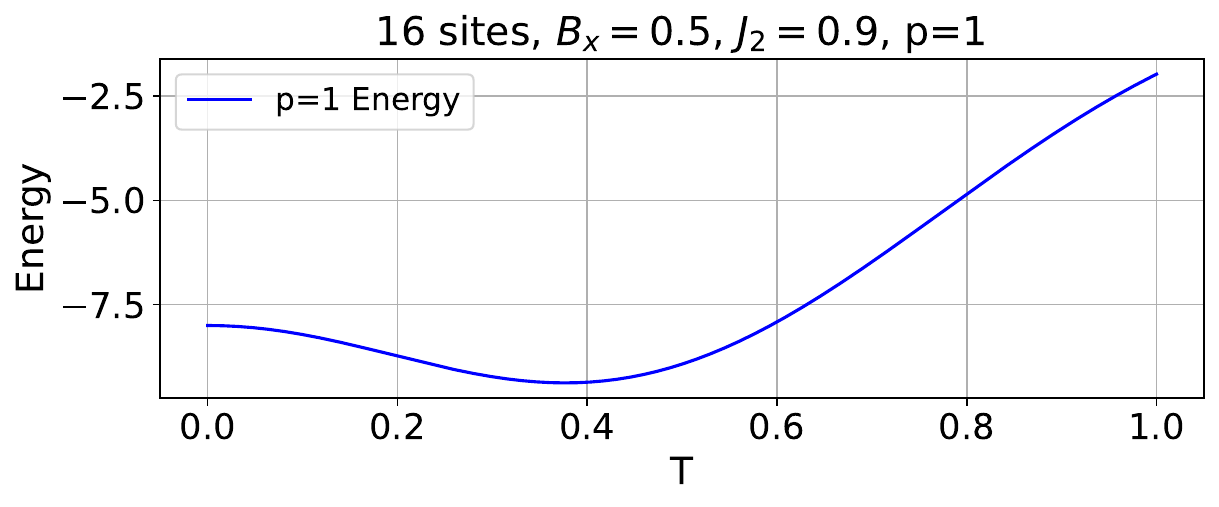}
    \includegraphics[width=0.32\linewidth]{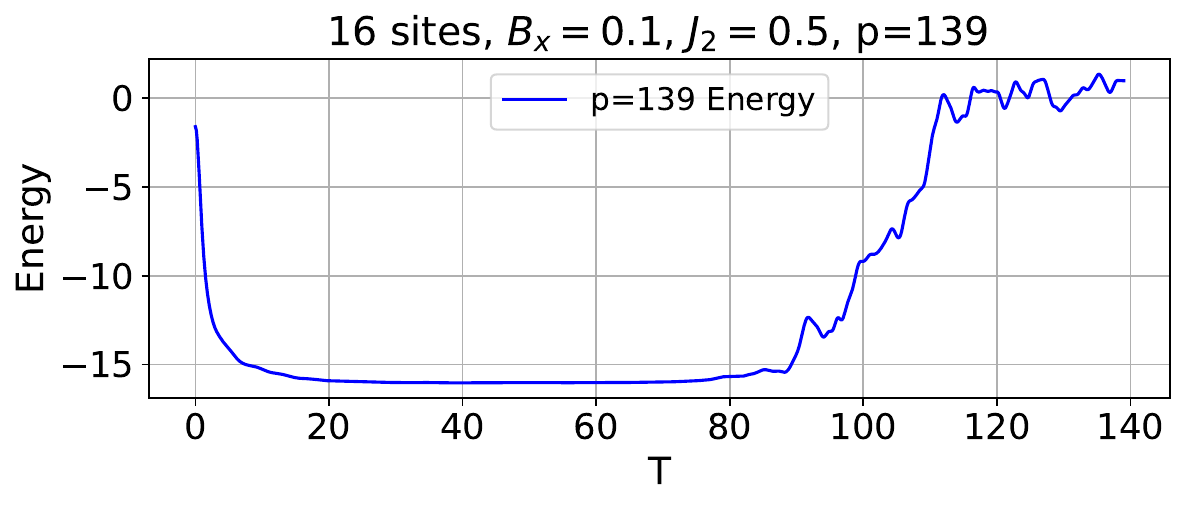}
    \includegraphics[width=0.32\linewidth]{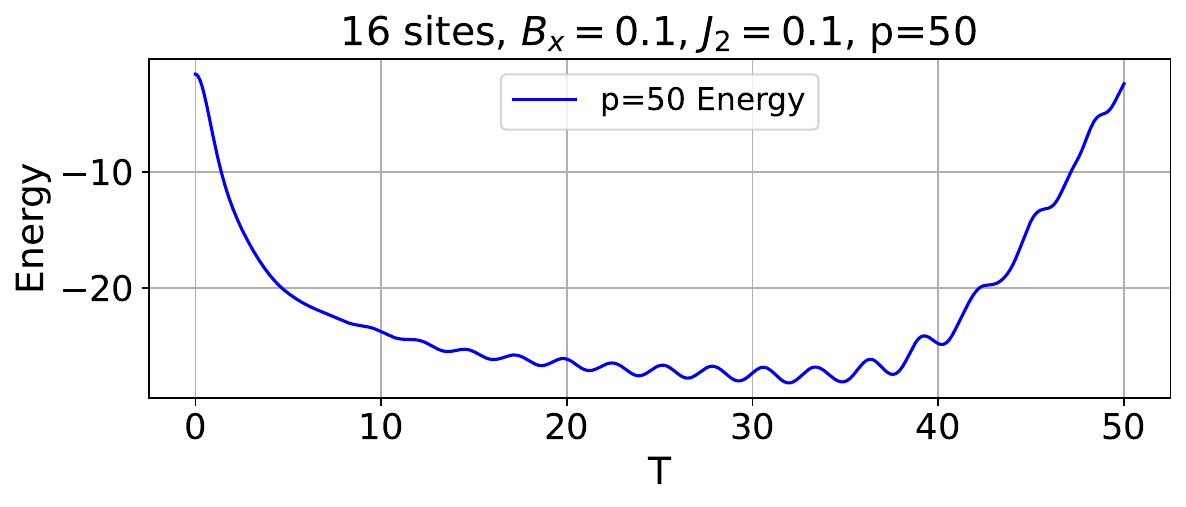}
    \caption{ \textbf{Representative $[0, p]$ T parameter search space for Snapshot-QAOA}. These are several examples of Hamiltonian energy (y-axis) vs. the total-anneal-time parameter $T$ (x-axis), for different values of $p$, $J_2$, and $B_x$. Each sub-plot is showing the Hamiltonian expectation value for $1000$ linearly spaced T values between $0$ and $p$.  }
    \label{fig:examples_T_vs_Energy_up_to_p}
\end{figure*}

\begin{figure*}[ht!]
    \centering
    \includegraphics[width=0.32\linewidth]{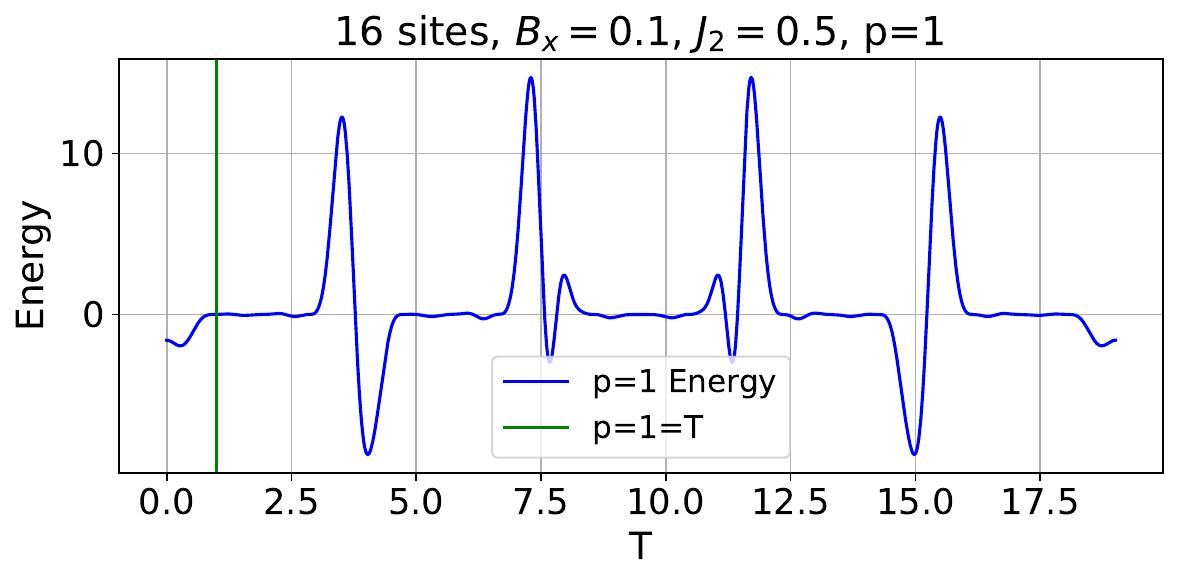}
    \includegraphics[width=0.32\linewidth]{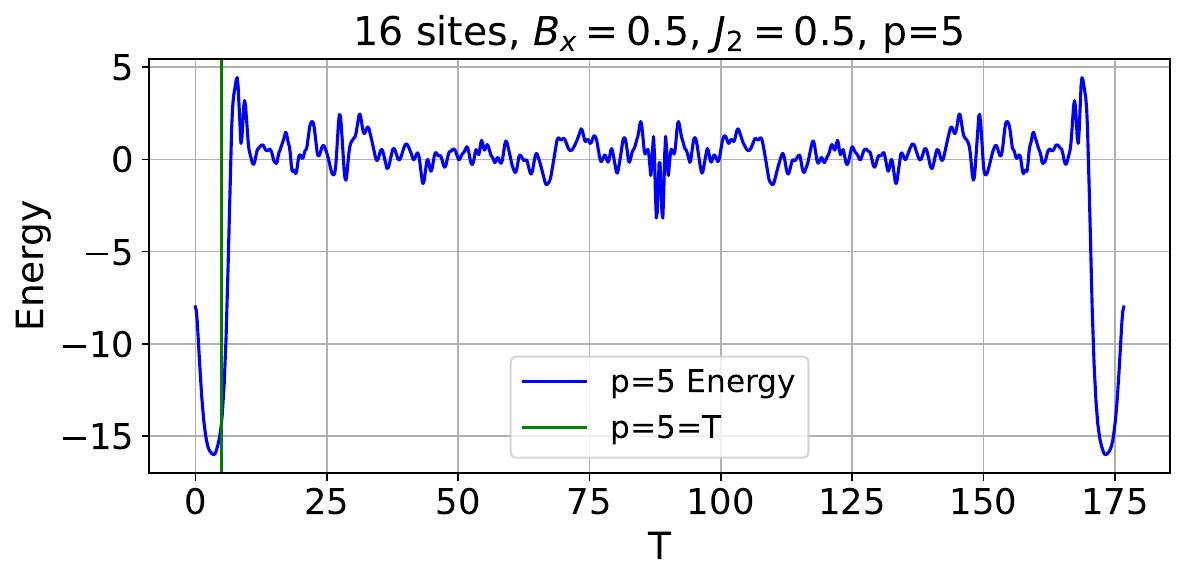}
    \includegraphics[width=0.32\linewidth]{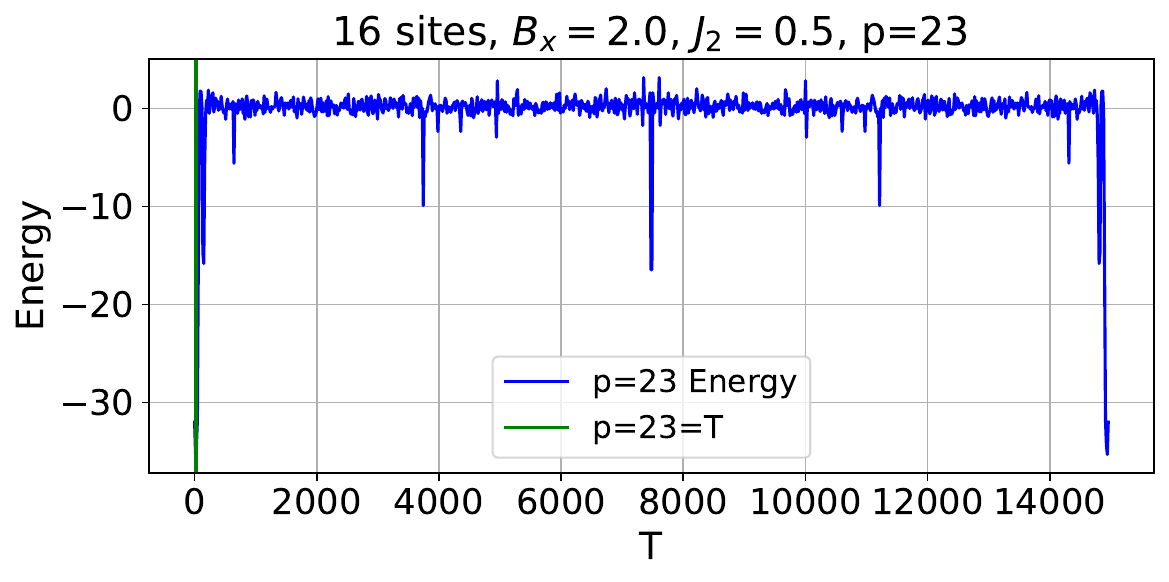}
    \caption{ \textbf{Snapshot-QAOA total-anneal-time $T$ parameter search space over the full time period in the interval from $[0, \rho]$ (defined by Eq.~\eqref{eqn:period_formula}).} Hamiltonian energy (y-axis) vs. the total-anneal-time parameter $T$ (x-axis), for different values of $p$, $J_2$, and $B_x$. Each plot is showing the Hamiltonian expectation value for $1000$ linearly spaced T values between $0$ and the minimum T-period $\rho$. The x-axis index where $T=p$ is marked by a vertical green line. Notice that there is a clear mirror symmetry in each $T$ search space. }
    \label{fig:examples_T_vs_Energy_full_period}
\end{figure*}

\begin{figure*}[ht!]
    \centering
    \includegraphics[width=0.8\linewidth]{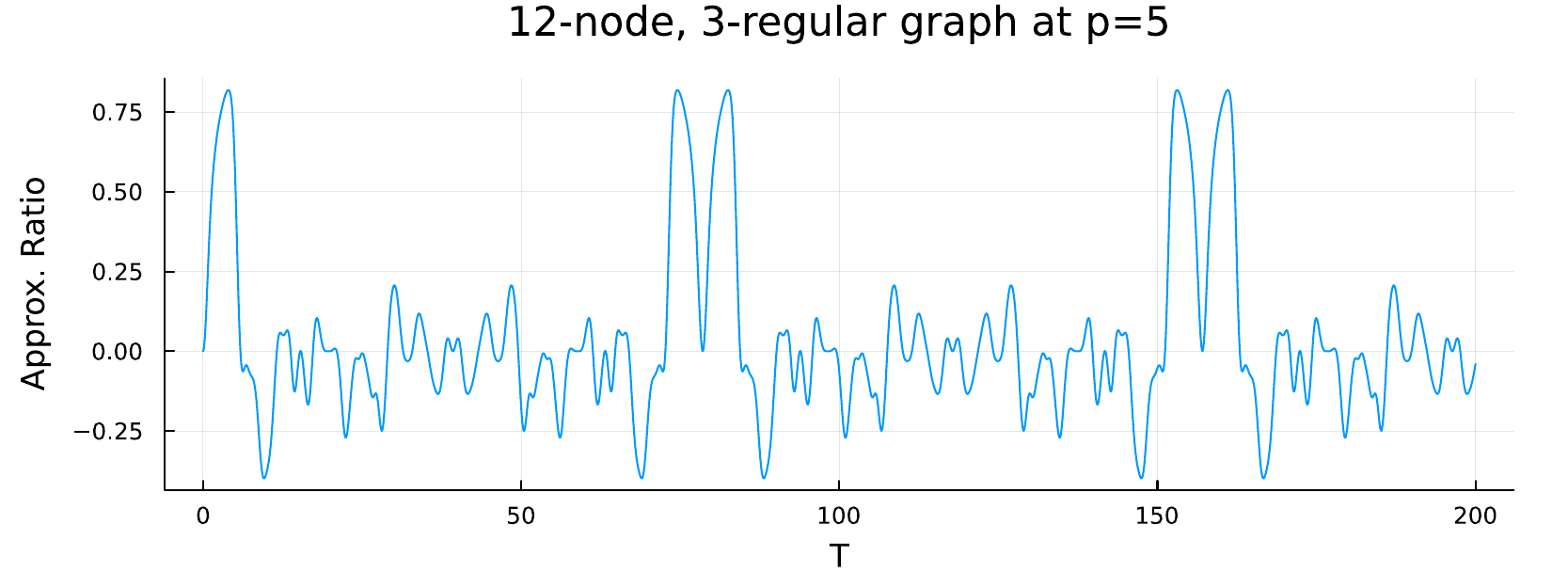}
    \caption{The approximation ratio ($y$-axis) obtained by TQA (i.e. Snapshot-QAOA with $\tau=T$) for varying values of $T$ ($x$-axis) on a 12-node 3-regular graph with $p=5$ layers. The approximation ratio is defined in the same (non-traditional) manner as in \cite{QA_initialization_of_QAOA}; where they use $H = \sum_{(i,j)\in E} Z_i Z_j$ (instead of $H = \sum_{(i,j)\in E} \frac{1}{2}(1-Z_i Z_j)$) for the Max-Cut Hamiltonian and the approximation is given by $\bra{\psi(T)}H\ket{\psi(T)}/\mathcal{E}_H$ where $\ket{\psi(T)}$ is the output of the algorithm with parameter $T$ and $\mathcal{E}_H$ is the ground state energy of the classical maximum cut Hamiltonian $H$. }
    \label{fig:replicate_sack_result_maxcut}
\end{figure*}

\subsection{Optimizing $T$ with Fixed $p$}
\label{sec:results_finding_optimal_T_for_fixed_p}
In the adiabatic quantum annealing setting, due to the quantum adiabatic theorem, one is guaranteed to obtain the ground state in the limit as $T\to\infty$. However, in our QAOA setting with a fixed choice of $p$, this is no longer the case due to the fact that as $T$ increases, so does the step-size $\Delta t = \tau/p = \hat{c}_1T/p$, leading to an increase in Trotterization errors. The fixed $p$ regime is of interest since increased $p$ leads to increased circuit depth (and hence increased runtime and quantum decoherence/noise), whereas, interestingly, the total annealing time $T$ has no (direct) effect on the circuit depth of Snapshot-QAOA. 

Sack et al. \cite{QA_initialization_of_QAOA} explored this phenomena in TQA setting (i.e. $\tau=T$) in the context of 3-regular Max-Cut. Similar to Sack et al., we found that Snapshot-QAOA on our Hamiltonian $H^\star$ of interest had an energy-minimizing $T$ in the range of $[0,p]$ as seen in Fig.~\ref{fig:examples_T_vs_Energy_up_to_p} which plots overall anneal time $T$ vs the energy $\mathcal{E}_p(T)$ for $T \in [0,p]$. In Fig.~\ref{fig:examples_T_vs_Energy_up_to_p} (left), we see the same phenomena observed by Sack et al. \cite{QA_initialization_of_QAOA} where as $T$ increases, the energy monotonically decreases to some minima before starting to increase again. However, by changing $J_2, B_x,$ and $p$, we sometimes observe slightly different behavior. For example, in Fig.~\ref{fig:examples_T_vs_Energy_up_to_p} (right), there are small, rapid oscillations in the overall ``dip" in the plot; such oscillations may make it difficult to optimize using certain optimization methods (e.g., gradient descent), in particular for larger $p$ if the landscape contains many more oscillations. Another example is Fig.~\ref{fig:examples_T_vs_Energy_up_to_p}-(center) where the minimum (over $[0,p]$) appears to lie in a sort of plateau and the plot has a more unpredictable and erratic behavior for $T \in [p/2,p]$. Algorithmically, Eq.~\ref{eqn:partialAnnealParams} directly gives us the parameters $\vec{\gamma}$ and $\vec{\beta}$ for Snapshot-QAOA. This is a crucial advantage of Snapshot-QAOA, which is that this parameter setting requires no extensive variational learning. However, Fig.~\ref{fig:examples_T_vs_Energy_up_to_p} shows us that we do still have one important parameter, $T$, which we can optimize, and doing so for a fixed $p$ depth improves our estimate of the ground-state energy considerably.

Given the results above where $T \in [0,p]$, one may think that as $T$ increases beyond $T > p$ that the energy becomes increasingly worse due to the increased proliferation of Trotter-like errors. However, Section~\ref{sec:theoretical_props_of_anneal_time} proved that this cannot possibly be the case since the energy is periodic in $T$ (with the period $\rho$ given by Equation~\ref{eqn:period_formula}). This warrants further investigation. In particular, it is not clear what the behavior of the $T$-vs-Energy curve is past $T > p$ and it is not clear a priori if the optimal $T^*$ always lies in or near the interval $[0,p]$. 

In Fig.~\ref{fig:examples_T_vs_Energy_full_period}, we show plots for several values of $p,B_x,J_2$, illustrating the diversity in behavior of the energy as $T$ changes throughout its period. It can be observed that as $p$ increases, so does the number of sudden sharp peaks and valleys in the energy curve over the period $\rho$. This shows that in principle, in order to find the globally optimal $T^*$, we would need to perform an exhaustive search over the interval $[0, \rho]$, or more precisely $\frac{\rho}{2}$ because the energy landscape has a mirror symmetry. We also replicate and extend Sack et al.'s work, extending the inset figure of Figure 3  of \cite{QA_initialization_of_QAOA} to larger $T$ values in Fig.~\ref{fig:replicate_sack_result_maxcut}; as a result we also find the same complex relationship between $T$ and the energy for TQA (i.e. the $\tau=T$ setting).

In Fig.~\ref{fig:examples_T_vs_Energy_full_period}, we highlight a specific plot (with $p=1,J_2=0.5,B_x=0.1$) where the optimal $T^*$ is not close to being in the interval $[0,p]$. This shows that looking in the interval $T \in [0,p]$ is not always sufficient if one wishes to find $T^*$. We hypothesize that the cases where the optimal $T^* > p$ could be due to transitory interference effects. Empirically we observe that typically $T^*$ is near $\frac{p}{2}$. 

\begin{figure*}[ht!]
    \centering
    \includegraphics[width=0.8\linewidth]{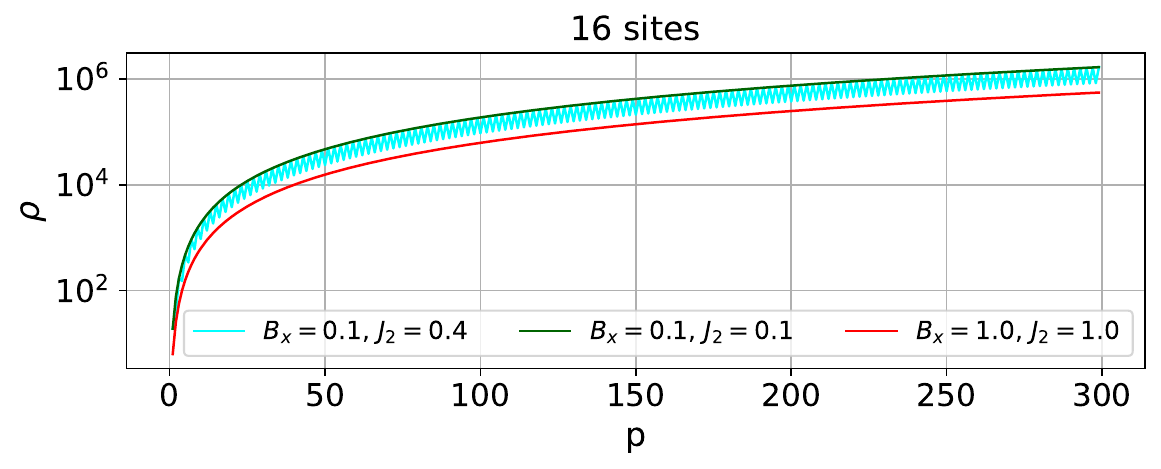}
    \caption{Scaling of the minimum T period $\rho$ (y-axis) as a function of $p$ (x-axis) for several representative Hamiltonian parameter choices of $J_2$ and $B_x$. Log scale on the y-axis. This shows that $\rho$, and therefore in principle the maximum size of the $T$ search space, grows quite quickly as a function of $p$. }
    \label{fig:T_period_scaling}
\end{figure*}

Due to the challenges described above, one can imagine various methods for optimizing $T$ with differing trade-offs between accuracy and runtime. One (expensive) option is to perform a fine grid-search over the entire period $\rho$; however, this is prohibitive since the size of the period $\rho$ grows very quickly with $p$ as illustrated in Fig.~\ref{fig:T_period_scaling}. Alternatively, one could perform a high-resolution grid-search over a small interval in the beginning of the period; this is what we do for the remaining results in this study to approximate the optimal $T^*$ value. More specifically, we perform a grid-search over the interval $T \in [0,p]$ with a step-size of $\Delta T = 0.01$; we denote the approximate optimal $T^*$ found by this procedure as $T^*_{\approx}$.

\begin{figure*}[ht!]
    \centering
    \includegraphics[width=0.495\linewidth]{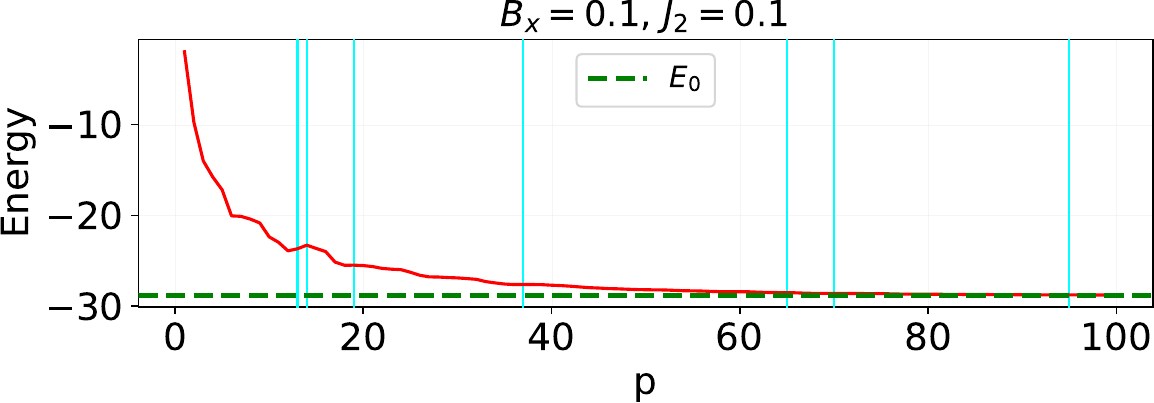}
    \includegraphics[width=0.495\linewidth]{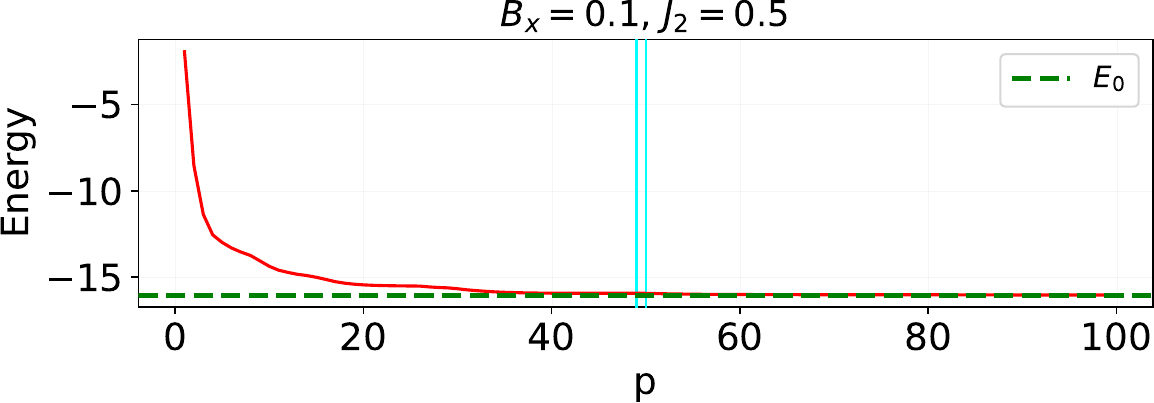}
    \includegraphics[width=0.495\linewidth]{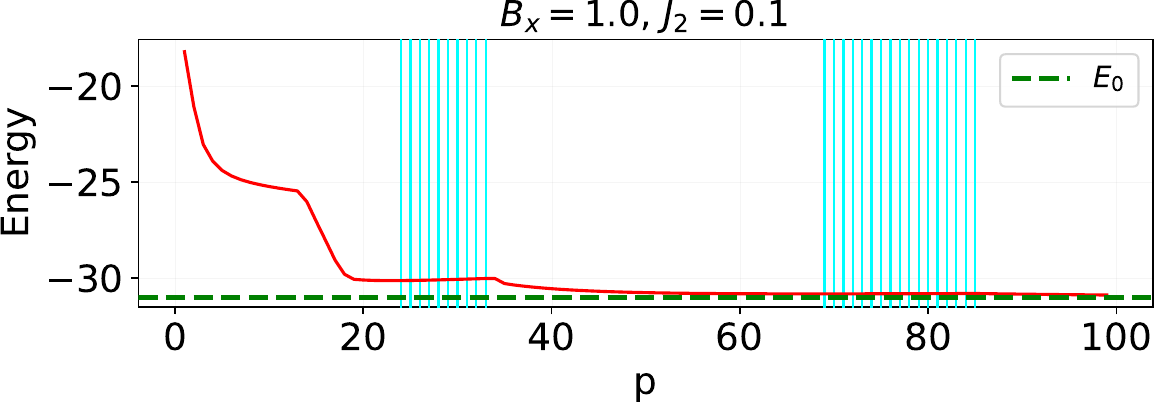}
    \includegraphics[width=0.495\linewidth]{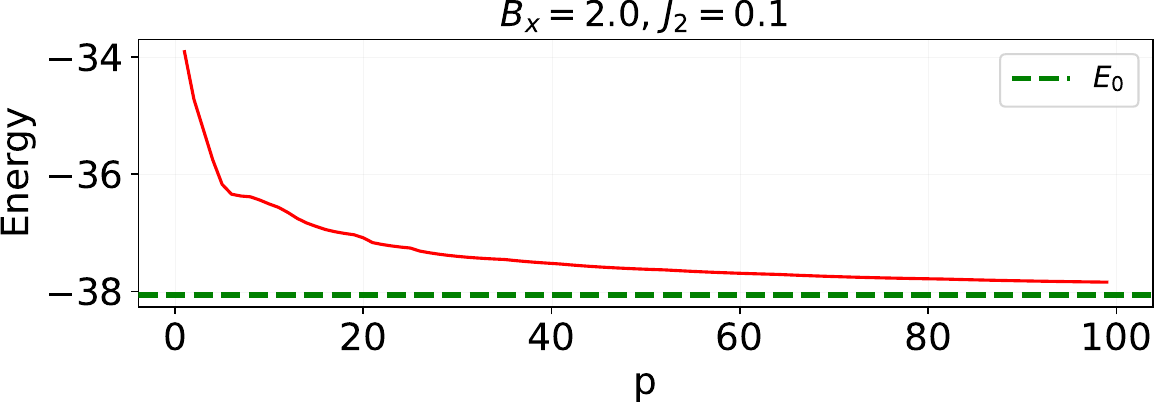}
    \caption{\textbf{Convergence of Snapshot-QAOA to the true ground-state energy}. Energy (y-axis) as a function of $p$ (x-axis), using $T^*_{\approx}$. Known ground-state energy, computed using exact diagonalization $\mathcal{E}_0(H^\star(J_2,B_x))$ is shown by the horizontal green line. If any energy value for a particular $p+1$ is higher energy than $p$, meaning the simulation decreased in quality, that $p$ index is marked by a vertical cyan line.
    }
    \label{fig:energy_vs_p_convergence_0_p_gridsearch_T}
\end{figure*}

\subsection{Convergence of Snapshot-QAOA as $p$ Increases}
\label{sec:results_convergence_of_energy_as_p_increases}

The next question we want to consider is how quickly Snapshot-QAOA, using $T^*_{\approx}$, converges to the true ground-state as $p$ is increased. Fig.~\ref{fig:energy_vs_p_convergence_0_p_gridsearch_T} reports this along with the ground-state energy for several representative Hamiltonian parameters which shows that Snapshot-QAOA converges close to the true ground-state for this $16$-qubit Hamiltonian within $p \approx 100$. This is encouraging, it shows that even relatively short depth Snapshot-QAOA, using non-variational learning on $\vec{\gamma}, \vec{\beta}$ get relatively close to the ground-state. However, we find that in some cases the expectation value does not monotonically improve with respect to $p$ (denoted by vertical cyan lines in Fig.~\ref{fig:energy_vs_p_convergence_0_p_gridsearch_T}). This is not ideal because in principle we would want to avoid rounds where the energy gets worse. Still, the trend is towards convergence despite temporary non-monotonicity. Snapshot-QAOA is not guaranteed to be monotonic, and moreover, these specific examples of non-monotonicity could be due to the sub-optimal approximation of $T^*$ by $T^*_{\approx}$. It is an open question whether monotonic improvement in $p$ is always expected for Snapshot-QAOA if the globally optimal $T^*$ parameter is found, or whether this is a consequence of the heuristic used to obtain $T^*_\approx$. Appendix~\ref{section:appendix_energy_per_site} reports the full range of ground-states energy across this quantum Hamiltonian computed with Snapshot-QAOA, using $T^*_\approx$.

\begin{figure*}[ht!]
    \centering
    \includegraphics[width=0.32\linewidth]{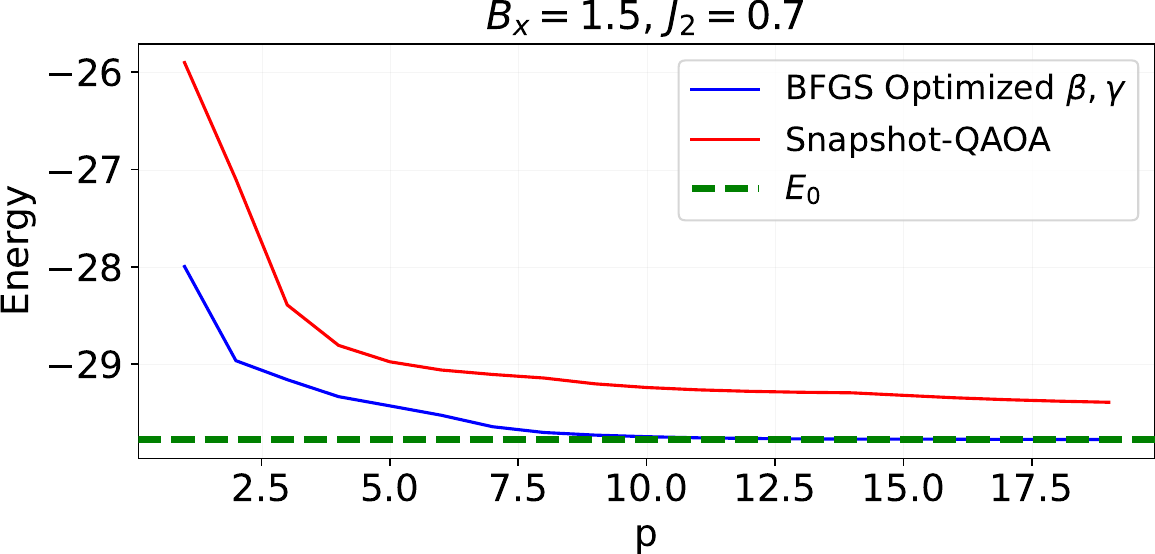}
    \includegraphics[width=0.32\linewidth]{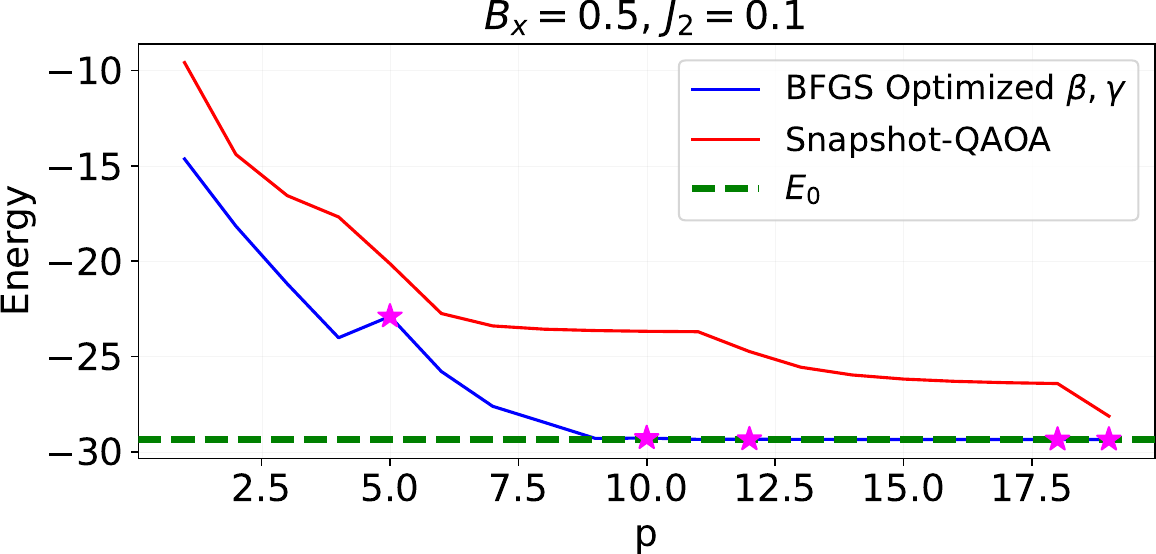}
    \includegraphics[width=0.32\linewidth]{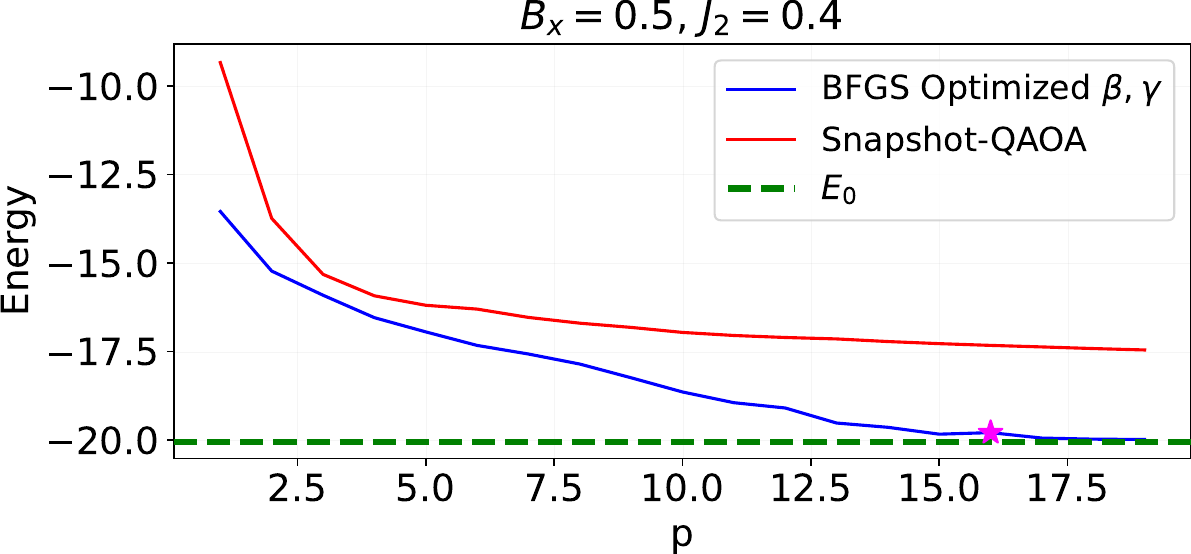}
    \caption{Full optimization of $\vec{\gamma}$ and $\vec{\beta}$ using BFGS, starting from the $T^*_\approx$ Snapshot-QAOA angles, for $p=1$ up to $p=20$. Energy (y-axis) as a function of $p$ (x-axis), where the known ground-state energy $\mathcal{E}_0(H^\star(J_2,B_x))$ is marked by a green horizontal line. Red line shows the Snapshot-QAOA energy from having only optimized $T$ and found $T^*_\approx$. Blue line shows the energy of the Hamiltonian after BFGS has optimized the vectors of QAOA angles $\vec{\gamma}$ and $\vec{\beta}$, where those angles are initialized from the $T^*_\approx$ angles. Magenta asterisks on the blue line denote any $p+1$ indices where is a decrease in performance (i.e., an increase in energy) compared to the previous $p$, specifically for the BFGS angle optimized line. These results show that further optimization of $\vec{\gamma}$ and $\vec{\beta}$ can substantially improve upon the energy obtained from Snapshot-QAOA.}
    \label{fig:BFGS_optimized_betas_gammas_converegence}
\end{figure*}

\begin{figure*}[ht!]
  \centering
  \begin{minipage}[t]{0.495\linewidth}
    \centering
    \footnotesize{Snapshot-QAOA \textbf{without} Optimization of $\vec{\gamma}$ and $\vec{\beta}$}\par
    \vspace{2pt}
    \includegraphics[width=\linewidth]{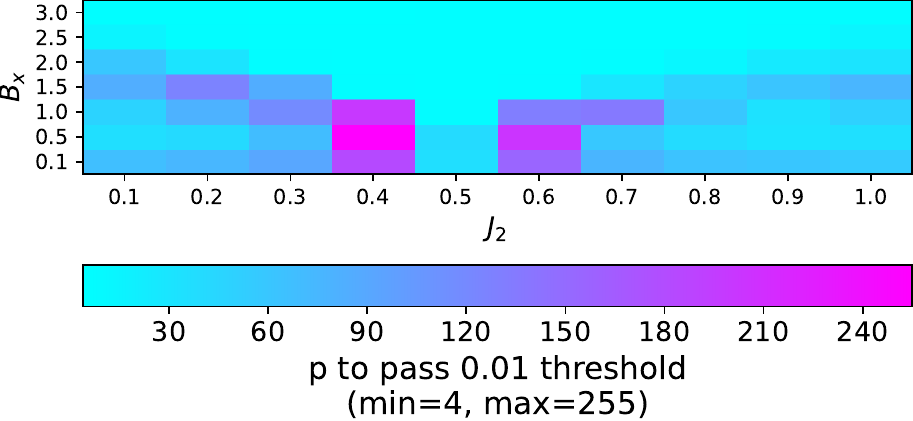}
  \end{minipage}\hfill
  \begin{minipage}[t]{0.495\linewidth}
    \centering
    \footnotesize{Snapshot-QAOA \textbf{with} Optimization of $\vec{\gamma}$ and $\vec{\beta}$}\par
    \vspace{2pt}
    \includegraphics[width=\linewidth]{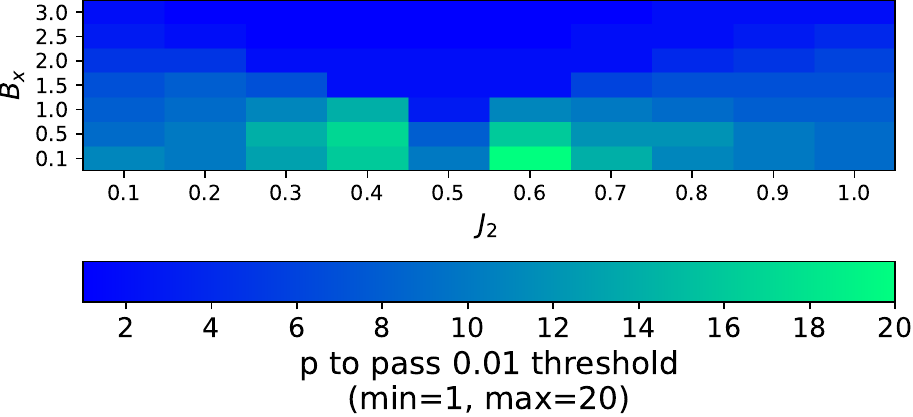}
  \end{minipage}
  \caption{\textbf{Snapshot-QAOA $p$ depth required to obtain high-quality ($0.01$ threshold) ground state energy approximation of the quantum Hamiltonian as a function of frustration ($J_2$) and transverse field ($B_x$).} Notably, the regions near small $B_x$ phase transitions are hardest for Snapshot-QAOA to approximate. This accuracy is shown using a heatmap representation showing the number of $p$ steps required for Snapshot-QAOA (using $T^*_\approx$) both with (right) and without (left) further optimization of the $\vec{\gamma}$ and $\vec{\beta}$ parameters using BFGS (starting from $T^*_\approx$), to pass a threshold of $1\%$ within the true ground-state $\mathcal{E}_0(H^\star(J_2,B_x))$. }
  \label{fig:p_steps_required_to_pass_accuracy_threshold}
\end{figure*}

\begin{figure*}[ht!]
    \centering
    \includegraphics[width=0.49\linewidth]{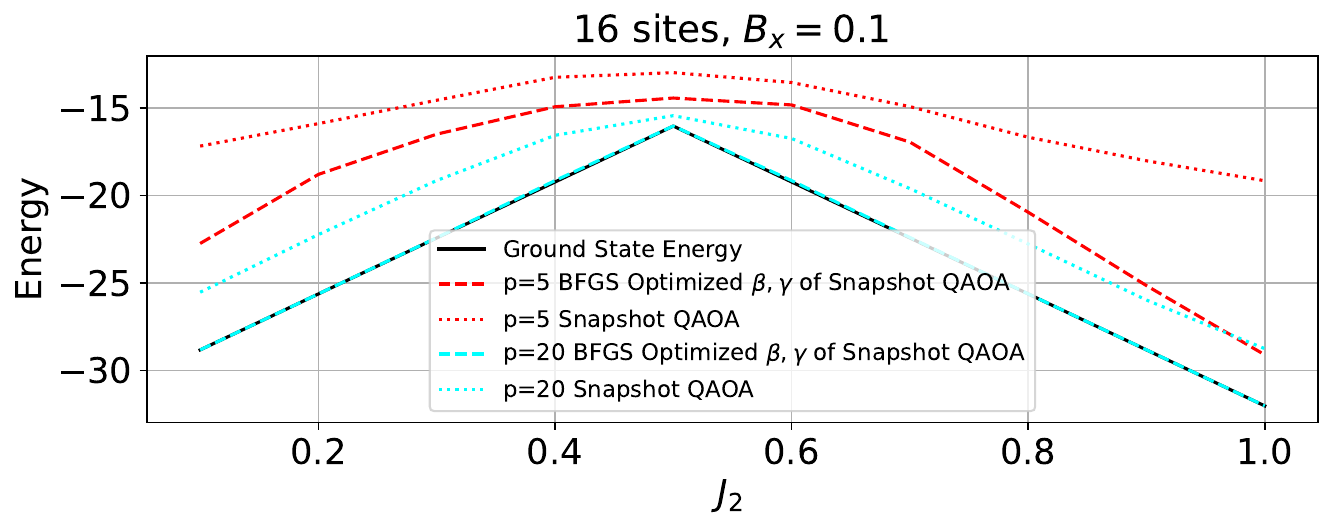}
    \includegraphics[width=0.49\linewidth]{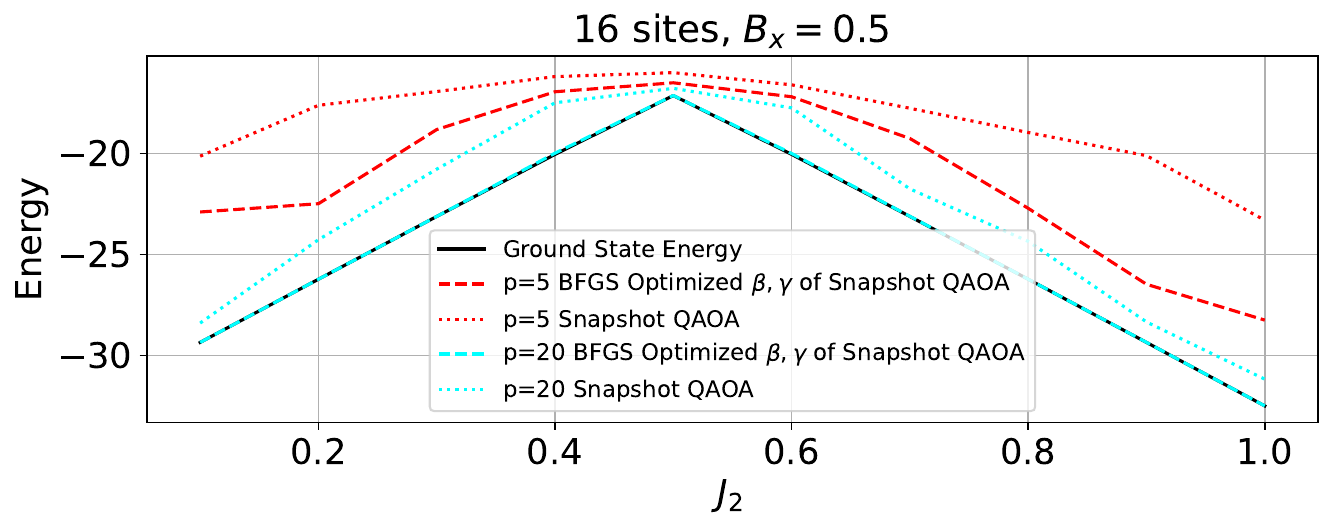}
    \includegraphics[width=0.49\linewidth]{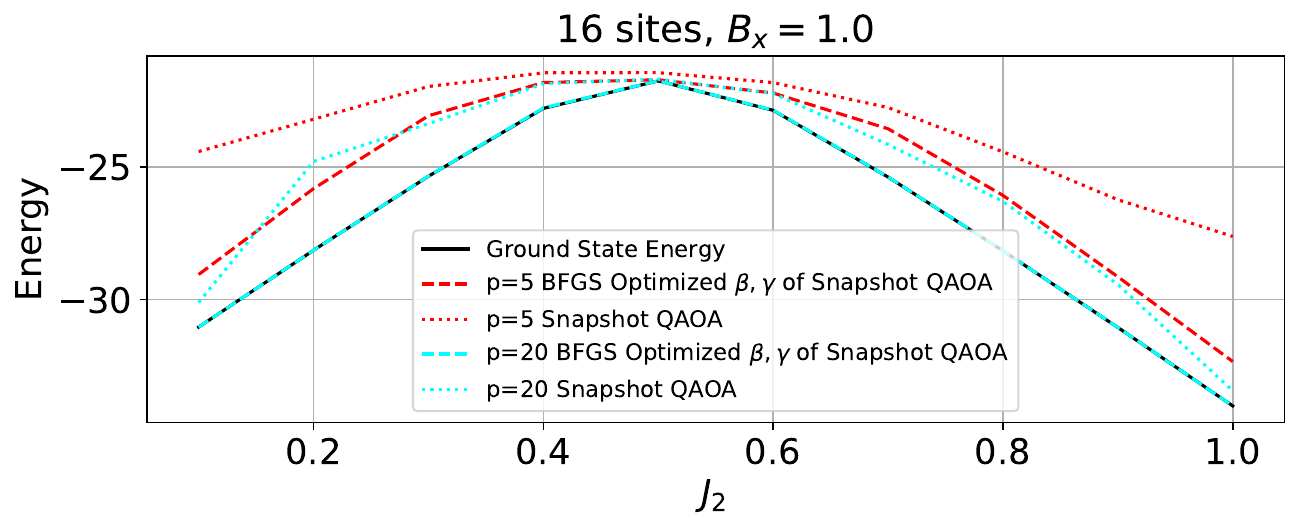}
    \includegraphics[width=0.49\linewidth]{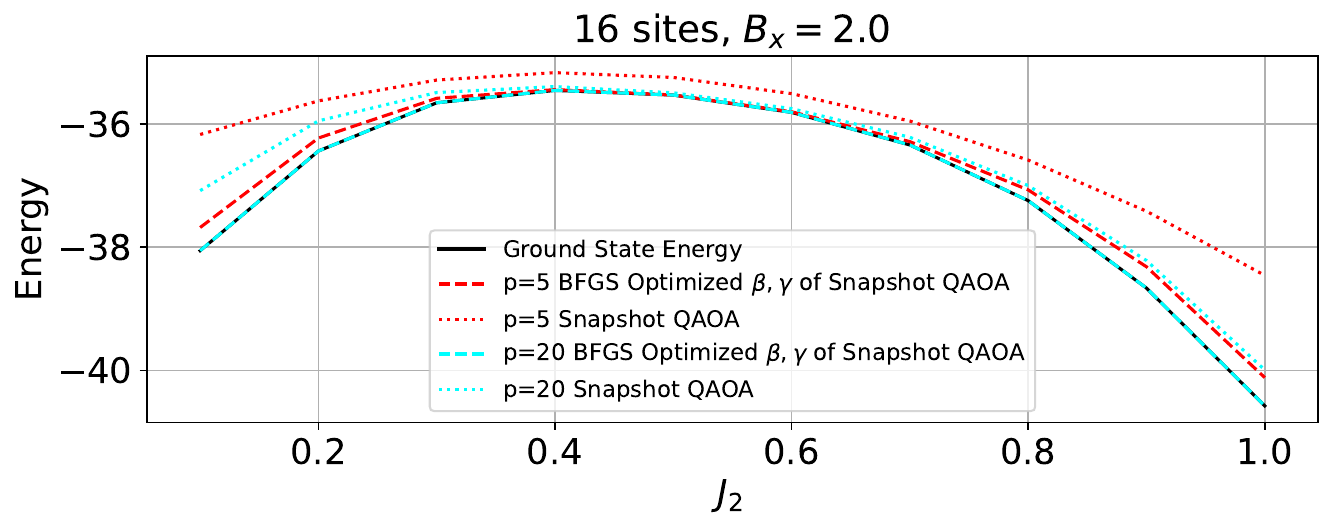}
    \caption{Energies returned by Snapshot-QAOA with (dashed lines) and without (dotted lines) further optimization of the $\gamma$ and $\beta$ parameters using BFGS, compared to the exact ground truth energy (solid black line). Each curve color (other than black) corresponds to a different $p$ value, for $p=5$ (red) and $p=20$ (cyan). Each subplot corresponds to a different $B_x$ value, with the $x$-axis denoting varying values of $J_2$ and the $y$-axis corresponding to energy values. Each Snapshot-QAOA run is performed using the approximate optimal total anneal time of $T^*_\approx$. }
    \label{fig:unoptimized_and_optimized_vs_ground_truth}
\end{figure*}

\subsection{Optimization of $\vec{\gamma}$ and $\vec{\beta}$ Parameters in Snapshot-QAOA}
\label{sec:results_further_optimize_betas_gammas}

In this section we consider allowing full optimization of $\vec{\gamma}$ and $\vec{\beta}$ at each $p$ using BFGS. We initialize $\vec{\gamma}$ and $\vec{\beta}$ according to Eq.~\ref{eqn:partialAnnealParams}, using $T^*_\approx$. Fig.~\ref{fig:BFGS_optimized_betas_gammas_converegence} plots the optimized $\vec{\gamma}, \vec{\beta}$ performance compared to the linear-ramp only $T^*_\approx$ Snapshot-QAOA. This shows that optimizing over the full set of parameters substantially improves the rate of convergence to the ground-state. However, notably there are still some instances of non-monotonicity (denoted by magenta asterisks) for the optimized Snapshot-QAOA angle performance.

A natural question is how Snapshot-QAOA (with and without further optimization of $\vec{\gamma}$ and $\vec{\beta}$) depends on $p$, as well as the frustrated Hamiltonian parameters $J_2$, $B_x$. We address this by measuring the number of $p$ steps required to reach a certain accuracy threshold within the true ground-state energy.
The accuracy threshold is defined as $\epsilon \cdot \mathcal{E}_0(H^\star(J_2,B_x))$, where $\mathcal{E}_0(H^\star(J_2,B_x))$ is the ground-state energy of $H$ obtained from exact diagonalization. For example, $\epsilon = 0.05$ means that the accuracy threshold is $5\%$ above the ground-state energy. In this way we can quantify how many $p$ steps are required to approximate the ground state of the Hamiltonian using Snapshot-QAOA. 
Fig.~\ref{fig:p_steps_required_to_pass_accuracy_threshold} plots the number of $p$ steps required to reach an energy that is within $1\%$ (specifically $1\%$ higher energy, or better) of the ground-state energy of the Hamiltonian $H^\star(J_2,B_x)$ using Snapshot-QAOA with and without further optimization of $\vec{\gamma}$ and $\vec{\beta}$. Remarkably, this representation of the complexity search space to approximate the ground state of this frustrated model uncovers a reasonably accurate representation of the boundaries of magnetic phase diagram of this model \cite{Oitmaa_2020_TF, PhysRevE.99.012134}. Namely, the paramagnetic phase, also referred to as the $\Gamma-$phase, requires the smallest number of $p$ steps to approximate the ground state, shown by the cyan region in Fig.~\ref{fig:p_steps_required_to_pass_accuracy_threshold}-(left). 
At small transverse field, at $J_2=0.5$ the maximally frustrated region defines a discontinuous ferromagnetic-antiferromagnetic transition and is highly degenerate. We may expect that ground-state approximation at this maximally frustrated point to be computationally challenging due to magnetic frustration and high degeneracy. However, Fig.~\ref{fig:p_steps_required_to_pass_accuracy_threshold} shows the exact opposite; when $J_2=0.5$, the rounds required to reach a good approximation threshold are quite low. The region that requires the most computational resources, as measured by the required $p$ depth, is the region \emph{near} the maximally frustrated point (on either side of $J_2=0.5$) at low transverse field. This hints at a potential significant computational capability of Snapshot-QAOA; providing high accuracy ground-state energy estimates at maximally frustrated regions of quantum Hamiltonians. Frustrated regions in the phase diagram of quantum Hamiltonians are one of the conditions that makes it challenging to study ground-state properties using classical methods such as Monte Carlo.

Notably, when further optimization of $\vec{\gamma}$ and $\vec{\beta}$ are allowed, Snapshot-QAOA was able to reach good approximation of the ground-state energy of the $16$-qubit frustrated TFIM within only $20$ $p$ steps for all evaluated regions of the Hamiltonian parameter space, whereas Fig.~\ref{fig:p_steps_required_to_pass_accuracy_threshold}-(top) shows that Snapshot-QAOA without such further optimization, while it does converge, required up to $250$ $p$ steps.

In Fig.~\ref{fig:unoptimized_and_optimized_vs_ground_truth}, the Snapshot-QAOA energies (both with and without further optimization of $\vec{\gamma}$ and $\vec{\beta}$) and the exact ground state energies are plotted together, with different energy curves for varying values of $p$. Similar to what was observed in Figure Fig.~\ref{fig:p_steps_required_to_pass_accuracy_threshold}, in Fig.~\ref{fig:unoptimized_and_optimized_vs_ground_truth} it is clear that the energies produced by Snapshot-QAOA converge to the true ground state energy much faster whenever further optimization of $\vec{\gamma}$ and $\vec{\beta}$ is allowed.

\begin{figure*}[ht!]
     \centering
     \includegraphics[width=0.49\linewidth]{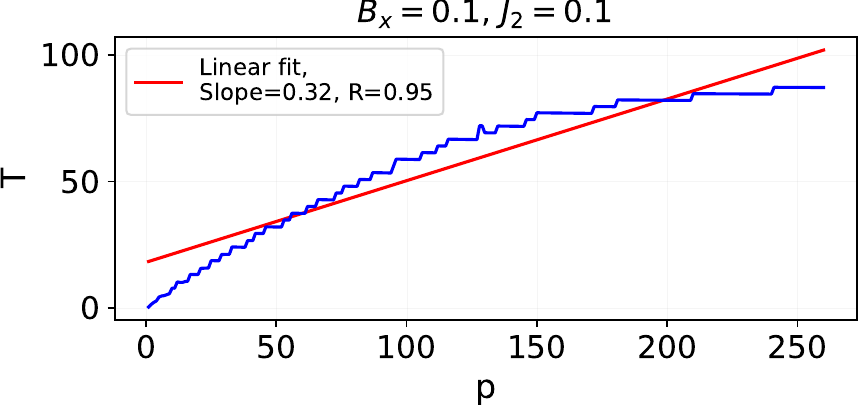}
     \includegraphics[width=0.49\linewidth]{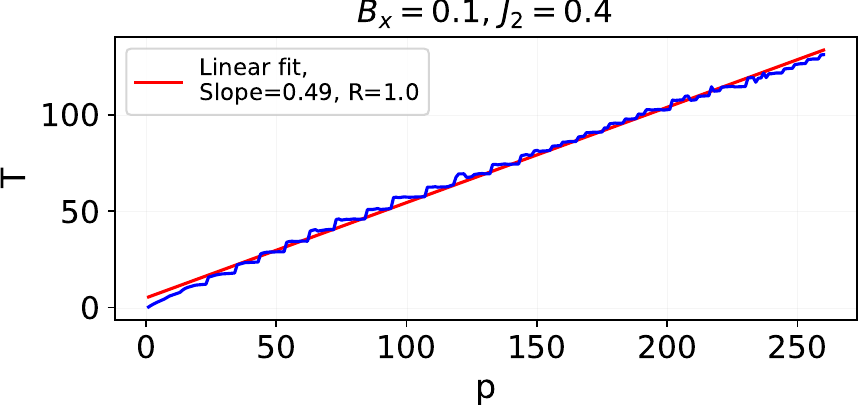}
     \includegraphics[width=0.49\linewidth]{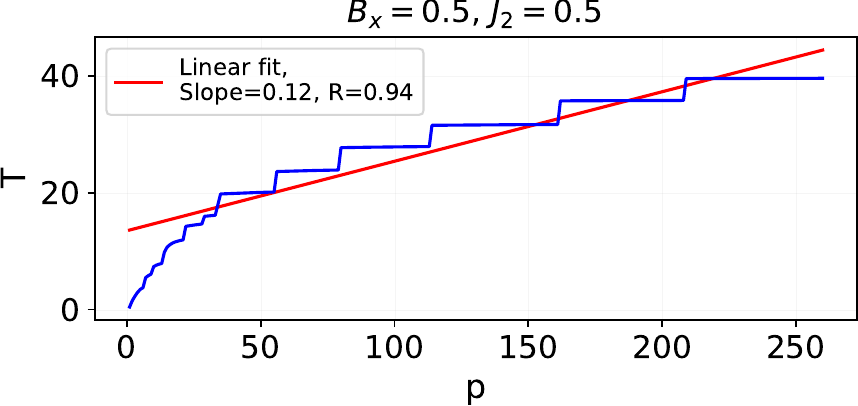}
     \includegraphics[width=0.49\linewidth]{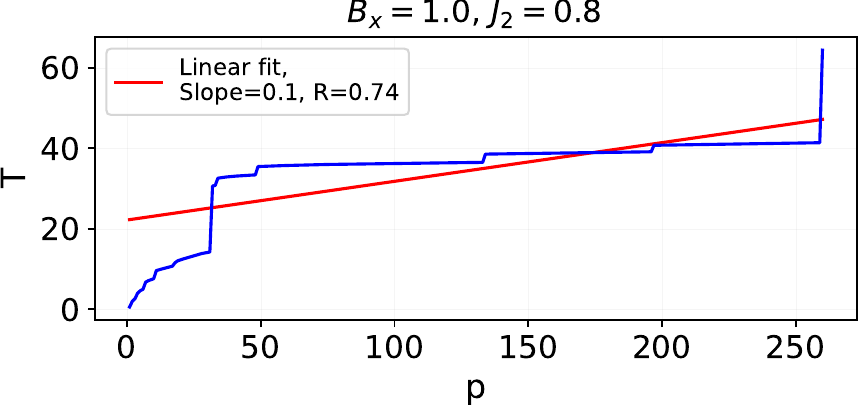}
     \caption{Linear regression between $p$ (x-axis) and $T^*_\approx$ (y-axis), for all steps of $p$ up to $p=260$. All of these relationships appear to be approximately stepwise functions, but the step size is not consistent. Some of the step sizes are small enough that the linear function approximates the relationship well, whereas for others the relationship is much weaker. }
     \label{fig:linear_regression_plots}
 \end{figure*}

\begin{figure*}[ht!]
    \centering
    \includegraphics[width=0.495\linewidth]{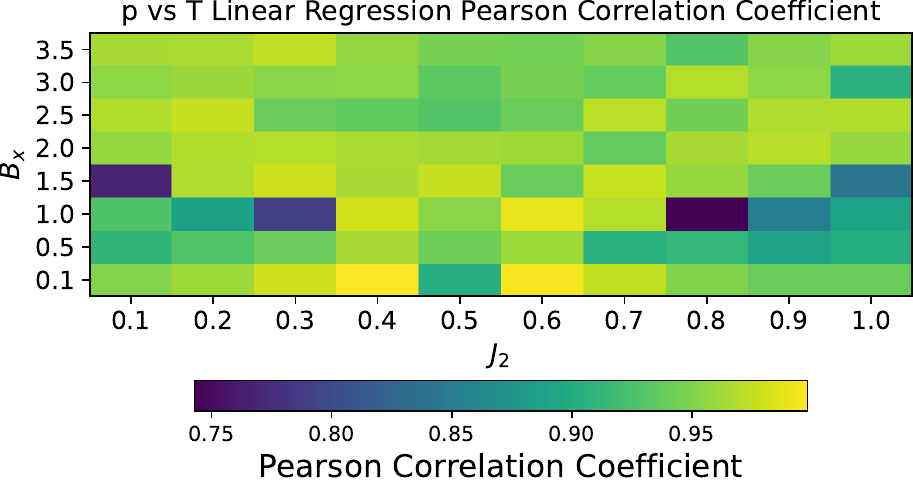}
    \includegraphics[width=0.495\linewidth]{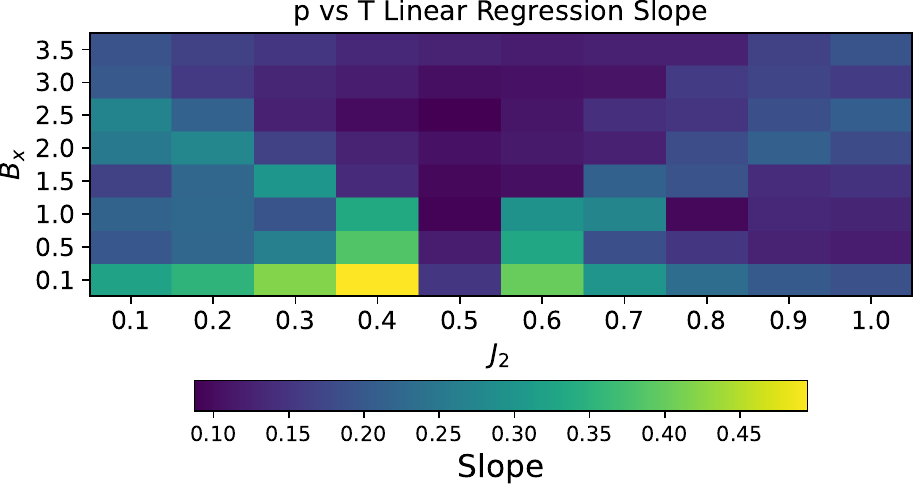}
    \caption{\textbf{Analysis of the linear relationship between $T$ and $p$ for Snapshot-QAOA approximating the ground-state of the square frustrated $16$-qubit TFIM.} We report this in the form of heatmaps of linear regression quantities between $p$ and $T^*_\approx$ from $p=1$ up to $p=260$. Pearson correlation coefficient $R$ (left) and linear regression best-fit line slope (right). }
    \label{fig:linear_regression_slope_R_heatmaps}
\end{figure*}

\subsection{Linear Regression Between $T^*$ and $p$ in the Interval from $[0, p]$}
\label{sec:results_linear_relationship_T_and_p}
In the case of TQA (i.e., Snapshot-QAOA with $\tau=T$) for the 3-regular Max-Cut problem, Sack et al. \cite{QA_initialization_of_QAOA} found that the relationship between $p$ and the optimal $T^*$ is approximately linear; in particular, they empirically show that $T^* \approx \delta t \cdot p$ where $\delta t \approx 0.75$. Since  $\Delta t = T/p$ (for Final-Snapshot-QAOA), then the linear relationship between $T^*$ and $p$ suggests running Final-Snapshot-QAOA with a $p$-independent time-step of $\Delta t = \delta t$; such a linear relationship would also imply a $p$-independent time-step for general Snapshot-QAOA of $\Delta t = \tau/p = (\hat{c}_1 T)/p = \hat{c}_1 \cdot \delta t$.

For general Snapshot-QAOA, if a linear-relationship strongly holds for all $T^*$ and $p$ and we have an accurate estimate of the ``slope" $\delta t$, then this gets around the challenges regarding finding the optimal $T^*$ as discussed in Sec.~\ref{sec:results_finding_optimal_T_for_fixed_p}. In the rest of this section, we investigate the relationship between $T^*$ and $p$ to see if any such linear relationship exists. Since the true $T^*$ is challenging to obtain and verify (as discussed in Sec.~\ref{sec:results_finding_optimal_T_for_fixed_p}), we instead work with the estimate $T^*_\approx$.

In Fig.~\ref{fig:linear_regression_plots}, we show the results of the linear regression (and the underlying data) for 4 different choices of $(J_2,B_x)$ between $p$ and $T^*_\approx$. The linear regression fit is quantified by the Pearson correlation coefficient where $1.0$ corresponds to a perfect positive linear fit to the data and $0.0$ would indicate no linear relationship between $p$ and $T^*_\approx$. In some cases, e.g., Fig.~\ref{fig:linear_regression_plots}-(top-right), we find a nearly perfect linear relationship. In other cases, e.g., Fig.~\ref{fig:linear_regression_plots} (bottom-left), the linear relationship is much weaker but with the overall trend still appearing roughly linear. Meanwhile, Fig.~\ref{fig:linear_regression_plots} (bottom-right) shows that there exist cases where the relationship between $T_\approx^*$ and $p$ exhibits irregular behavior, with large jumps and long plateaus. These results mean that for some parameters, a reasonably simple linear extrapolation could find good values of $T^*_\approx$, but in general the results indicate that more complex extrapolations would be required -- and in particular, the relationship between $p$ and $T^*_\approx$ depends on the underlying Hamiltonian being simulated. 

In Fig.~\ref{fig:linear_regression_slope_R_heatmaps}, we provide a heatmap that shows how the slope $\delta t$ (calculated via linear regression) changes with changing $J_2$ and $B_x$. Additionally, we provide a complementary heatmap showing the $R$ coefficient of the linear regression. Notably, the heatmap of the slope $\delta t$ approximately recovers the divisions between the magnetic phases of the underlying frustrated $J_1$-$J_2$ model \cite{Oitmaa_2020_TF, PhysRevE.99.012134}, just as the accuracy threshold plot of Fig.~\ref{fig:p_steps_required_to_pass_accuracy_threshold} did. The Hamiltonian parameter set that has the most reliable linear relationship between $T$ and $p$ is $B_x=0.1, J_2=0.4$, which has a $R$ coefficient of $0.998$. The Hamiltonian parameter that has the worst linear fit is $B_x=1, J_2=0.8$ with an R coefficient of $0.743$.

\begin{figure*}[ht!]
    \centering
    \includegraphics[width=0.32\linewidth]{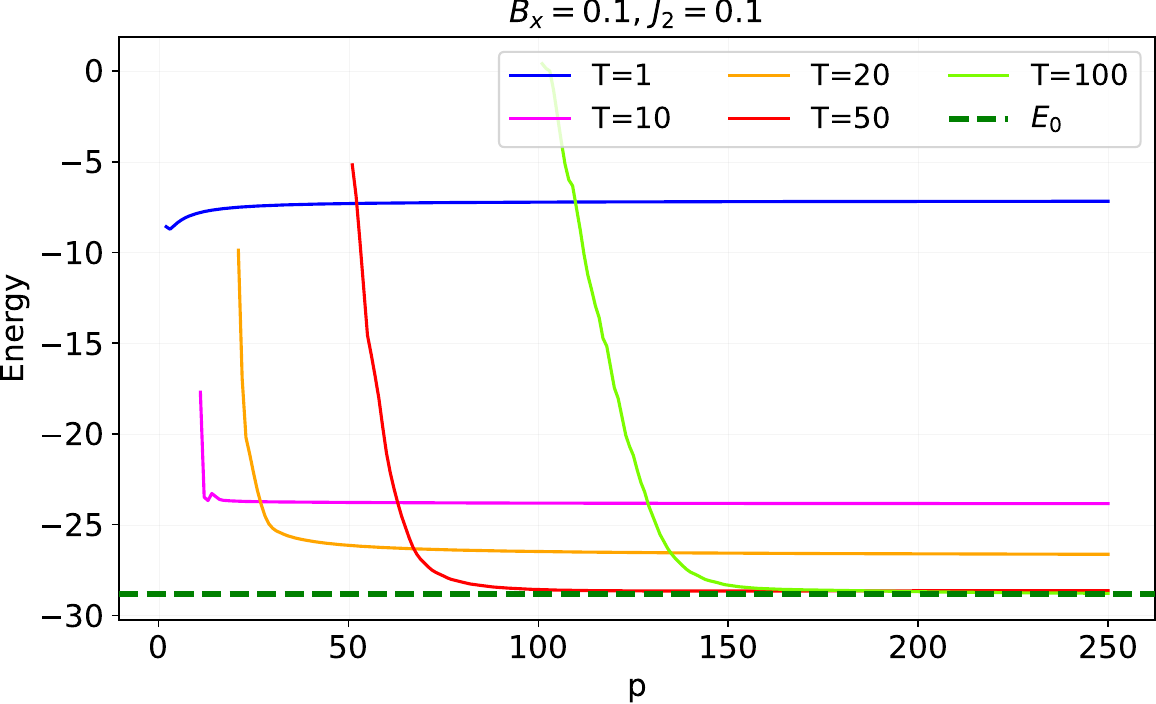}
    \includegraphics[width=0.32\linewidth]{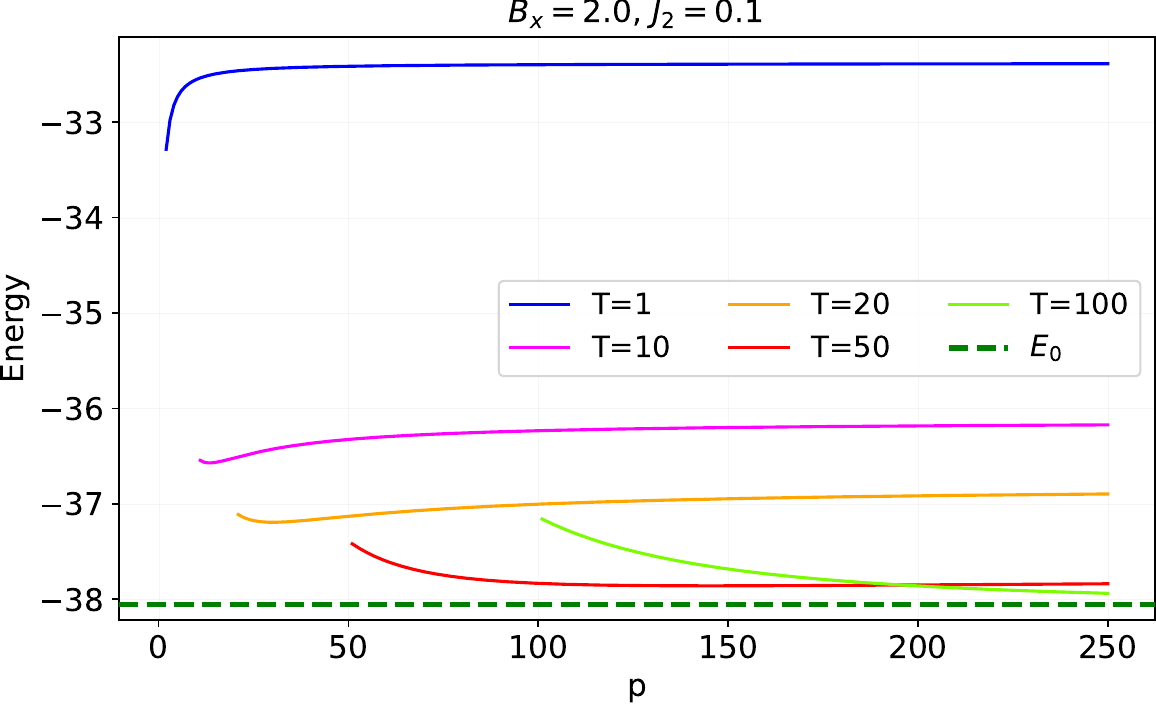}
    \includegraphics[width=0.32\linewidth]{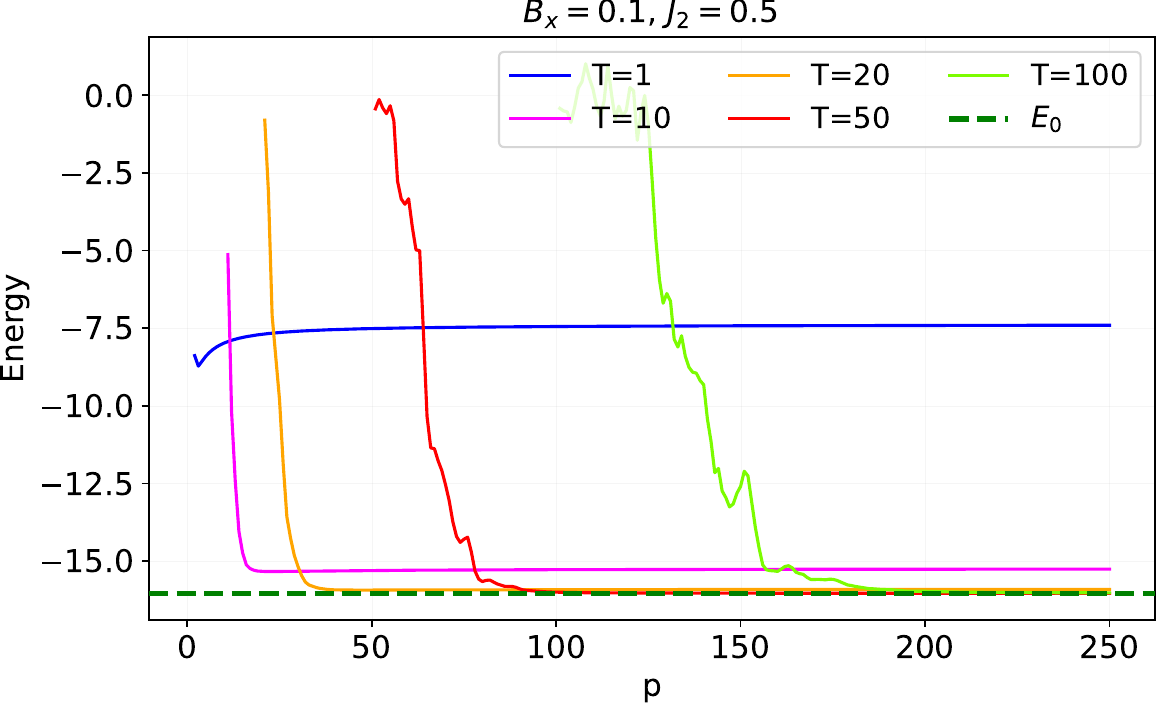}
    \caption{\textbf{Representative examples of fixed $T$ with increasing $p$.} Hamiltonian energy (y-axis) as a function of $p$ (x-axis) for three representative Hamiltonian parameters. Green horizontal line denotes the ground-state energy $\mathcal{E}_0(H^\star(J_2,B_x))$. }
    \label{fig:fixed_T_energy_vs_p}
\end{figure*}

\subsection{Behavior of $\mathcal{E}_p(T)$ with $T$ fixed}
Fig.~\ref{fig:fixed_T_energy_vs_p} depicts how the energy $\mathcal{E}_p(T)$ changes with $p$, with each curve representing a different fixed value of $T$. There are a few key observations. First, we observe that each curve appears to be converging to some fixed value; this is expected as $\lim_{p \to \infty} \mathcal{E}_p(T)$ converges to the energy $\mathcal{E}_\text{QA}(T)$ obtained by quantum annealing which was discussed in Sec.~\ref{sec:limitingBehavior}. We remark that such curves are not necessarily monotonically decreasing. For large enough $p$ (i.e., the right-most parts of the figure), we see that increasing $T$ brings the energy closer\footnote{The right side of each sub-plot of Fig.~\ref{fig:fixed_T_energy_vs_p} seems to suggest that $\mathcal{E}_\text{QA}(T)$ is monotonically decreasing in $T$; however, we do not claim that this is always the case as the adiabatic theorem only guarantees a limiting behavior and not a monotonic behavior. Prior work has shown for success probabilities (an alternative metric of success), that there is non-monotonic behavior with changing $T$ \cite{crosson2014different}.} to the ground state energy $\mathcal{E}_0(H^\star(J_2,B_x))$; again, this is as expected as $\lim_{T\to\infty}\mathcal{E}_\text{QA}(T) = \lim_{T\to\infty}\lim_{p\to\infty} \mathcal{E}_p(T) = \mathcal{E}_H$ as discussed in Sec.~\ref{sec:limitingBehavior}.

\section{Discussion}
\label{section:discussion}

This study has proposed an extension of QAOA that allows approximate ground-state finding of Hamiltonians beyond those corresponding to problems in (classical) combinatorial optimization. We have shown that Snapshot-QAOA can approximate the ground-state of quantum (non-diagonal) Hamiltonians very well, as evidenced by numerical simulations of a frustrated 2D $J_1$-$J_2$ TFIM model in various regions of its magnetic phase diagram. Importantly, Snapshot-QAOA is quite NISQ-hardware friendly (Noisy Intermediate Scale Quantum \cite{Preskill_2018}) - and can be used without any variational learning procedure, or, optionally the full $\vec{\beta}, \vec{\gamma}$ schedule can be further optimized using variational learning. 

In theory, other mixing Hamiltonians could satisfy the required properties of Snapshot-QAOA - and therefore the ground states of more complex quantum Hamiltonians could also be approximated, not just transverse field driven systems. Additionally, in theory the condition that ``the unitaries $e^{-\alpha H_0}$ and $e^{-\alpha H_1}$ should be easily implementable" can be relaxed to further extend the applicability of Snapshot-QAOA to more complex Hamiltonians; however with such Hamiltonians, more circuit complexity is required to approximate such unitaries (e.g., via Trotterization). One advantage of Snapshot-QAOA with the TFIM model is that the transverse field mixer consists of a single layer of single-qubit gates per $p$. In future work, one can also relax the condition that $H_0$ has no degenerate ground states; Snapshot-QAOA can still be executed on such Hamiltonians (such as the Quantum Compass model \cite{dorier2005quantum}), but it is unclear if the convergence property of Snapshot-QAOA still holds under such conditions.  It may also be of interest to study Hamiltonians for which the partial annealing time $\tau$ is ``nonphysical", i.e., $\tau \notin [0,T]$.

One of the most important future topics of study for Snapshot-QAOA is determining more sophisticated methods of approximating optimal $T$, and moreover determining under what conditions the optimal $T$ is within the range $[0, p]$ and under what conditions that is not the case. As a rule-of-thumb, we currently recommend that practitioners set $T$ to be slightly smaller than $p$ and if time permits, to try to optimize $T$ in the interval $[0,p]$. Another open question concerns understanding the full complexity of the search landscapes of $T$, both for Snapshot-QAOA but also for Trotterized quantum annealing \cite{QA_initialization_of_QAOA} for standard QAOA for sampling combinatorial optimization problems. 
What is clear is that optimizing $T$ is easier than directly optimizing $\vec{\gamma}$ and $\vec{\beta}$ in the sense that $T$ contains fewer tunable parameters, namely being the sole parameter, whereas the full set of (standard) QAOA angles contains $2p$ parameters. Another open question is how much performance can be gained by heavily optimizing the $\vec{\beta}$, $\vec{\gamma}$ from Snapshot-QAOA compared to only using optimized $T^*$ and if these optimized values of $\vec{\beta}$ and $\vec{\gamma}$ are amenable to parameter transfer; it is known for standard QAOA that optimized parameters are indeed transferable \cite{galda2021transferability,shaydulin2023parameter, Claes2021instance, wurtz2021fixedangleconjectureqaoa, chernyavskiy2023method, boulebnane2021predictingparametersquantumapproximate}, (this is also known as parameter concentration \cite{Akshay_2021}).

For the particular Snapshot-QAOA simulations we performed on the frustrated $J_1$-$J_2$ TFIM model, there are several open questions. The clearest open question is why Snapshot-QAOA is able to efficiently, with a small number of p steps, approximate the system at the phase transitions for small $B_x$ and $J_2=0.5$ -- but also requires a large number of $p$ steps for the states very near (but not \emph{at}) these phase transitions. Another open question is why the linear regression slopes and $p$ steps to pass certain accuracy thresholds both approximately replicate the regions of the underlying magnetic phase diagram of the model (Figures \ref{fig:p_steps_required_to_pass_accuracy_threshold}, \ref{fig:linear_regression_slope_R_heatmaps}). 

Lastly, it would be an interesting question to consider how Snapshot-QAOA could be used as an initial state for QPE \cite{kitaev1995quantummeasurementsabelianstabilizer}.

\section*{Acknowledgments}
\label{sec:acknowledgments}
This work was supported by the U.S. Department of Energy through the Los Alamos National Laboratory. Los Alamos National Laboratory is operated by Triad National Security, LLC, for the National Nuclear Security Administration of U.S. Department of Energy (Contract No. 89233218CNA000001). Research presented in this article was supported by the NNSA's Advanced Simulation and Computing Beyond Moore's Law Program at Los Alamos National Laboratory and by the Laboratory Directed Research and Development program of Los Alamos National Laboratory under project number 20230049DR. This research used resources provided by the Darwin testbed at Los Alamos National Laboratory (LANL) which is funded by the Computational Systems and Software Environments subprogram of LANL's Advanced Simulation and Computing program (NNSA/DOE). LANL report LA-UR-26-20991.

\appendix

\begin{figure}[ht!]
    \centering
    \includegraphics[width=0.999\linewidth]{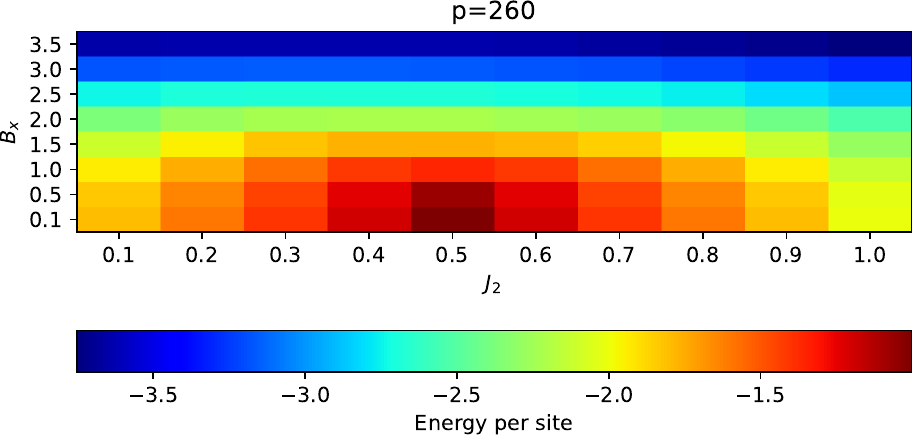}
    \caption{ Heatmap showing the ground-state energy (normalized to be energy per qubit), approximated with Snapshot-QAOA, across a large region of the magnetic phase diagram of the 2D $J_1-J_2$ quantum Hamiltonian.  }
    \label{fig:energy_per_site}
\end{figure}

\section{Energy per Qubit Computed with Snapshot-QAOA}
\label{section:appendix_energy_per_site}

Fig.~\ref{fig:energy_per_site} reports the normalized ground-state energy across the different phases of the $J_1-J_2$ model. Here, $p=260$ was used, with the linear-ramp Snapshot-QAOA schedule given by $T^*_\approx$.


\bibliographystyle{apsrev4-2-titles}
\bibliography{references}

\begin{thebibliography}{85}%
\makeatletter
\providecommand \@ifxundefined [1]{%
 \@ifx{#1\undefined}
}%
\providecommand \@ifnum [1]{%
 \ifnum #1\expandafter \@firstoftwo
 \else \expandafter \@secondoftwo
 \fi
}%
\providecommand \@ifx [1]{%
 \ifx #1\expandafter \@firstoftwo
 \else \expandafter \@secondoftwo
 \fi
}%
\providecommand \natexlab [1]{#1}%
\providecommand \enquote  [1]{``#1''}%
\providecommand \bibnamefont  [1]{#1}%
\providecommand \bibfnamefont [1]{#1}%
\providecommand \citenamefont [1]{#1}%
\providecommand \href@noop [0]{\@secondoftwo}%
\providecommand \href [0]{\begingroup \@sanitize@url \@href}%
\providecommand \@href[1]{\@@startlink{#1}\@@href}%
\providecommand \@@href[1]{\endgroup#1\@@endlink}%
\providecommand \@sanitize@url [0]{\catcode `\\12\catcode `\$12\catcode `\&12\catcode `\#12\catcode `\^12\catcode `\_12\catcode `\%12\relax}%
\providecommand \@@startlink[1]{}%
\providecommand \@@endlink[0]{}%
\providecommand \url  [0]{\begingroup\@sanitize@url \@url }%
\providecommand \@url [1]{\endgroup\@href {#1}{\urlprefix }}%
\providecommand \urlprefix  [0]{URL }%
\providecommand \Eprint [0]{\href }%
\providecommand \doibase [0]{https://doi.org/}%
\providecommand \selectlanguage [0]{\@gobble}%
\providecommand \bibinfo  [0]{\@secondoftwo}%
\providecommand \bibfield  [0]{\@secondoftwo}%
\providecommand \translation [1]{[#1]}%
\providecommand \BibitemOpen [0]{}%
\providecommand \bibitemStop [0]{}%
\providecommand \bibitemNoStop [0]{.\EOS\space}%
\providecommand \EOS [0]{\spacefactor3000\relax}%
\providecommand \BibitemShut  [1]{\csname bibitem#1\endcsname}%
\let\auto@bib@innerbib\@empty
\bibitem [{\citenamefont {Farhi}\ \emph {et~al.}(2014)\citenamefont {Farhi}, \citenamefont {Goldstone},\ and\ \citenamefont {Gutmann}}]{QAOA}%
  \BibitemOpen
  \bibfield  {author} {\bibinfo {author} {\bibfnamefont {E.}~\bibnamefont {Farhi}}, \bibinfo {author} {\bibfnamefont {J.}~\bibnamefont {Goldstone}},\ and\ \bibinfo {author} {\bibfnamefont {S.}~\bibnamefont {Gutmann}},\ }\bibfield  {title} {\emph {\bibinfo {title} {{A Quantum Approximate Optimization Algorithm}}},\ }\bibfield  {journal} {\bibinfo  {journal} {arXiv preprint}\ }\href {https://doi.org/10.48550/ARXIV.1411.4028} {10.48550/ARXIV.1411.4028} (\bibinfo {year} {2014}),\ \Eprint {https://arxiv.org/abs/1411.4028} {arXiv:1411.4028} \BibitemShut {NoStop}%
\bibitem [{\citenamefont {Farhi}\ \emph {et~al.}(2015)\citenamefont {Farhi}, \citenamefont {Goldstone},\ and\ \citenamefont {Gutmann}}]{farhi2015quantum}%
  \BibitemOpen
  \bibfield  {author} {\bibinfo {author} {\bibfnamefont {E.}~\bibnamefont {Farhi}}, \bibinfo {author} {\bibfnamefont {J.}~\bibnamefont {Goldstone}},\ and\ \bibinfo {author} {\bibfnamefont {S.}~\bibnamefont {Gutmann}},\ }\bibfield  {title} {\emph {\bibinfo {title} {{A Quantum Approximate Optimization Algorithm Applied to a Bounded Occurrence Constraint Problem}}},\ }\bibfield  {journal} {\bibinfo  {journal} {arXiv preprint}\ }\href {https://doi.org/10.48550/arXiv.1412.6062} {10.48550/arXiv.1412.6062} (\bibinfo {year} {2015}),\ \Eprint {https://arxiv.org/abs/1412.6062} {arXiv:1412.6062} \BibitemShut {NoStop}%
\bibitem [{\citenamefont {Hadfield}\ \emph {et~al.}(2019)\citenamefont {Hadfield}, \citenamefont {Wang}, \citenamefont {O{\textquotesingle}Gorman}, \citenamefont {Rieffel}, \citenamefont {Venturelli},\ and\ \citenamefont {Biswas}}]{Hadfield_2019}%
  \BibitemOpen
  \bibfield  {author} {\bibinfo {author} {\bibfnamefont {S.}~\bibnamefont {Hadfield}}, \bibinfo {author} {\bibfnamefont {Z.}~\bibnamefont {Wang}}, \bibinfo {author} {\bibfnamefont {B.}~\bibnamefont {O{\textquotesingle}Gorman}}, \bibinfo {author} {\bibfnamefont {E.}~\bibnamefont {Rieffel}}, \bibinfo {author} {\bibfnamefont {D.}~\bibnamefont {Venturelli}},\ and\ \bibinfo {author} {\bibfnamefont {R.}~\bibnamefont {Biswas}},\ }\bibfield  {title} {\emph {\bibinfo {title} {{From the Quantum Approximate Optimization Algorithm to a Quantum Alternating Operator Ansatz}}},\ }\href {https://doi.org/10.3390/a12020034} {\bibfield  {journal} {\bibinfo  {journal} {Algorithms}\ }\textbf {\bibinfo {volume} {12}},\ \bibinfo {pages} {34} (\bibinfo {year} {2019})},\ \Eprint {https://arxiv.org/abs/1709.03489} {arXiv:1709.03489} \BibitemShut {NoStop}%
\bibitem [{\citenamefont {Shaydulin}\ \emph {et~al.}(2024)\citenamefont {Shaydulin}, \citenamefont {Li}, \citenamefont {Chakrabarti}, \citenamefont {DeCross}, \citenamefont {Herman}, \citenamefont {Kumar}, \citenamefont {Larson}, \citenamefont {Lykov}, \citenamefont {Minssen}, \citenamefont {Sun}, \citenamefont {Alexeev}, \citenamefont {Dreiling}, \citenamefont {Gaebler}, \citenamefont {Gatterman}, \citenamefont {Gerber}, \citenamefont {Gilmore}, \citenamefont {Gresh}, \citenamefont {Hewitt}, \citenamefont {Horst}, \citenamefont {Hu}, \citenamefont {Johansen}, \citenamefont {Matheny}, \citenamefont {Mengle}, \citenamefont {Mills}, \citenamefont {Moses}, \citenamefont {Neyenhuis}, \citenamefont {Siegfried}, \citenamefont {Yalovetzky},\ and\ \citenamefont {Pistoia}}]{shaydulin2024evidence}%
  \BibitemOpen
  \bibfield  {author} {\bibinfo {author} {\bibfnamefont {R.}~\bibnamefont {Shaydulin}}, \bibinfo {author} {\bibfnamefont {C.}~\bibnamefont {Li}}, \bibinfo {author} {\bibfnamefont {S.}~\bibnamefont {Chakrabarti}}, \bibinfo {author} {\bibfnamefont {M.}~\bibnamefont {DeCross}}, \bibinfo {author} {\bibfnamefont {D.}~\bibnamefont {Herman}}, \bibinfo {author} {\bibfnamefont {N.}~\bibnamefont {Kumar}}, \bibinfo {author} {\bibfnamefont {J.}~\bibnamefont {Larson}}, \bibinfo {author} {\bibfnamefont {D.}~\bibnamefont {Lykov}}, \bibinfo {author} {\bibfnamefont {P.}~\bibnamefont {Minssen}}, \bibinfo {author} {\bibfnamefont {Y.}~\bibnamefont {Sun}}, \bibinfo {author} {\bibfnamefont {Y.}~\bibnamefont {Alexeev}}, \bibinfo {author} {\bibfnamefont {J.~M.}\ \bibnamefont {Dreiling}}, \bibinfo {author} {\bibfnamefont {J.~P.}\ \bibnamefont {Gaebler}}, \bibinfo {author} {\bibfnamefont {T.~M.}\ \bibnamefont {Gatterman}}, \bibinfo {author} {\bibfnamefont {J.~A.}\ \bibnamefont {Gerber}}, \bibinfo {author} {\bibfnamefont
  {K.}~\bibnamefont {Gilmore}}, \bibinfo {author} {\bibfnamefont {D.}~\bibnamefont {Gresh}}, \bibinfo {author} {\bibfnamefont {N.}~\bibnamefont {Hewitt}}, \bibinfo {author} {\bibfnamefont {C.~V.}\ \bibnamefont {Horst}}, \bibinfo {author} {\bibfnamefont {S.}~\bibnamefont {Hu}}, \bibinfo {author} {\bibfnamefont {J.}~\bibnamefont {Johansen}}, \bibinfo {author} {\bibfnamefont {M.}~\bibnamefont {Matheny}}, \bibinfo {author} {\bibfnamefont {T.}~\bibnamefont {Mengle}}, \bibinfo {author} {\bibfnamefont {M.}~\bibnamefont {Mills}}, \bibinfo {author} {\bibfnamefont {S.~A.}\ \bibnamefont {Moses}}, \bibinfo {author} {\bibfnamefont {B.}~\bibnamefont {Neyenhuis}}, \bibinfo {author} {\bibfnamefont {P.}~\bibnamefont {Siegfried}}, \bibinfo {author} {\bibfnamefont {R.}~\bibnamefont {Yalovetzky}},\ and\ \bibinfo {author} {\bibfnamefont {M.}~\bibnamefont {Pistoia}},\ }\bibfield  {title} {\emph {\bibinfo {title} {Evidence of scaling advantage for the quantum approximate optimization algorithm on a classically intractable
  problem}},\ }\href {https://doi.org/10.1126/sciadv.adm6761} {\bibfield  {journal} {\bibinfo  {journal} {Science Advances}\ }\textbf {\bibinfo {volume} {10}},\ \bibinfo {pages} {eadm6761} (\bibinfo {year} {2024})},\ \Eprint {https://arxiv.org/abs/2308.02342} {arXiv:2308.02342} \BibitemShut {NoStop}%
\bibitem [{\citenamefont {Golden}\ \emph {et~al.}(2023{\natexlab{a}})\citenamefont {Golden}, \citenamefont {B{\"{a}}rtschi}, \citenamefont {Eidenbenz},\ and\ \citenamefont {O'Malley}}]{QAOA_exponential_speedup_over_unstructured_search}%
  \BibitemOpen
  \bibfield  {author} {\bibinfo {author} {\bibfnamefont {J.}~\bibnamefont {Golden}}, \bibinfo {author} {\bibfnamefont {A.}~\bibnamefont {B{\"{a}}rtschi}}, \bibinfo {author} {\bibfnamefont {S.}~\bibnamefont {Eidenbenz}},\ and\ \bibinfo {author} {\bibfnamefont {D.}~\bibnamefont {O'Malley}},\ }in\ \href {https://doi.org/10.1109/QCE57702.2023.00063} {\emph {\bibinfo {booktitle} {IEEE International Conference on Quantum Computing and Engineering QCE'23}}}\ (\bibinfo {year} {2023})\ pp.\ \bibinfo {pages} {496--505},\ \Eprint {https://arxiv.org/abs/2202.00648} {arXiv:2202.00648} \BibitemShut {NoStop}%
\bibitem [{\citenamefont {Boulebnane}\ and\ \citenamefont {Montanaro}(2024)}]{boulebnane2024solving}%
  \BibitemOpen
  \bibfield  {author} {\bibinfo {author} {\bibfnamefont {S.}~\bibnamefont {Boulebnane}}\ and\ \bibinfo {author} {\bibfnamefont {A.}~\bibnamefont {Montanaro}},\ }\bibfield  {title} {\emph {\bibinfo {title} {{Solving Boolean Satisfiability Problems With The Quantum Approximate Optimization Algorithm}}},\ }\href {https://doi.org/10.1103/PRXQuantum.5.030348} {\bibfield  {journal} {\bibinfo  {journal} {PRX Quantum}\ }\textbf {\bibinfo {volume} {5}},\ \bibinfo {pages} {030348} (\bibinfo {year} {2024})},\ \Eprint {https://arxiv.org/abs/2208.06909} {arxiv:2208.06909} \BibitemShut {NoStop}%
\bibitem [{\citenamefont {Montanaro}\ and\ \citenamefont {Zhou}(2024)}]{montanaro2024quantumspeedupssolvingnearsymmetric}%
  \BibitemOpen
  \bibfield  {author} {\bibinfo {author} {\bibfnamefont {A.}~\bibnamefont {Montanaro}}\ and\ \bibinfo {author} {\bibfnamefont {L.}~\bibnamefont {Zhou}},\ }\href {https://arxiv.org/abs/2411.04979} {\bibinfo {title} {{Quantum speedups in solving near-symmetric optimization problems by low-depth QAOA}}} (\bibinfo {year} {2024}),\ \Eprint {https://arxiv.org/abs/2411.04979} {arXiv:2411.04979 [quant-ph]} \BibitemShut {NoStop}%
\bibitem [{\citenamefont {Wang}\ \emph {et~al.}(2020)\citenamefont {Wang}, \citenamefont {Rubin}, \citenamefont {Dominy},\ and\ \citenamefont {Rieffel}}]{PhysRevA.101.012320}%
  \BibitemOpen
  \bibfield  {author} {\bibinfo {author} {\bibfnamefont {Z.}~\bibnamefont {Wang}}, \bibinfo {author} {\bibfnamefont {N.~C.}\ \bibnamefont {Rubin}}, \bibinfo {author} {\bibfnamefont {J.~M.}\ \bibnamefont {Dominy}},\ and\ \bibinfo {author} {\bibfnamefont {E.~G.}\ \bibnamefont {Rieffel}},\ }\bibfield  {title} {\emph {\bibinfo {title} {{$XY$ mixers: Analytical and numerical results for the quantum alternating operator ansatz}}},\ }\href {https://doi.org/10.1103/PhysRevA.101.012320} {\bibfield  {journal} {\bibinfo  {journal} {Phys. Rev. A}\ }\textbf {\bibinfo {volume} {101}},\ \bibinfo {pages} {012320} (\bibinfo {year} {2020})}\BibitemShut {NoStop}%
\bibitem [{\citenamefont {Bartschi}\ and\ \citenamefont {Eidenbenz}(2020)}]{GM_QAOA}%
  \BibitemOpen
  \bibfield  {author} {\bibinfo {author} {\bibfnamefont {A.}~\bibnamefont {Bartschi}}\ and\ \bibinfo {author} {\bibfnamefont {S.}~\bibnamefont {Eidenbenz}},\ }in\ \href {https://doi.org/10.1109/qce49297.2020.00020} {\emph {\bibinfo {booktitle} {2020 IEEE International Conference on Quantum Computing and Engineering (QCE)}}}\ (\bibinfo  {publisher} {IEEE},\ \bibinfo {year} {2020})\ p.\ \bibinfo {pages} {72–82}\BibitemShut {NoStop}%
\bibitem [{\citenamefont {He}\ \emph {et~al.}(2023)\citenamefont {He}, \citenamefont {Shaydulin}, \citenamefont {Chakrabarti}, \citenamefont {Herman}, \citenamefont {Li}, \citenamefont {Sun},\ and\ \citenamefont {Pistoia}}]{He_2023}%
  \BibitemOpen
  \bibfield  {author} {\bibinfo {author} {\bibfnamefont {Z.}~\bibnamefont {He}}, \bibinfo {author} {\bibfnamefont {R.}~\bibnamefont {Shaydulin}}, \bibinfo {author} {\bibfnamefont {S.}~\bibnamefont {Chakrabarti}}, \bibinfo {author} {\bibfnamefont {D.}~\bibnamefont {Herman}}, \bibinfo {author} {\bibfnamefont {C.}~\bibnamefont {Li}}, \bibinfo {author} {\bibfnamefont {Y.}~\bibnamefont {Sun}},\ and\ \bibinfo {author} {\bibfnamefont {M.}~\bibnamefont {Pistoia}},\ }\bibfield  {title} {\emph {\bibinfo {title} {{Alignment between initial state and mixer improves QAOA performance for constrained optimization}}},\ }\bibfield  {journal} {\bibinfo  {journal} {npj Quantum Information}\ }\textbf {\bibinfo {volume} {9}},\ \href {https://doi.org/10.1038/s41534-023-00787-5} {10.1038/s41534-023-00787-5} (\bibinfo {year} {2023})\BibitemShut {NoStop}%
\bibitem [{\citenamefont {Tate}\ \emph {et~al.}(2023)\citenamefont {Tate}, \citenamefont {Moondra}, \citenamefont {Gard}, \citenamefont {Mohler},\ and\ \citenamefont {Gupta}}]{tate2023warm}%
  \BibitemOpen
  \bibfield  {author} {\bibinfo {author} {\bibfnamefont {R.}~\bibnamefont {Tate}}, \bibinfo {author} {\bibfnamefont {J.}~\bibnamefont {Moondra}}, \bibinfo {author} {\bibfnamefont {B.}~\bibnamefont {Gard}}, \bibinfo {author} {\bibfnamefont {G.}~\bibnamefont {Mohler}},\ and\ \bibinfo {author} {\bibfnamefont {S.}~\bibnamefont {Gupta}},\ }\bibfield  {title} {\emph {\bibinfo {title} {Warm-started qaoa with custom mixers provably converges and computationally beats goemans-williamson's max-cut at low circuit depths}},\ }\href@noop {} {\bibfield  {journal} {\bibinfo  {journal} {Quantum}\ }\textbf {\bibinfo {volume} {7}},\ \bibinfo {pages} {1121} (\bibinfo {year} {2023})}\BibitemShut {NoStop}%
\bibitem [{\citenamefont {Hadfield}\ \emph {et~al.}(2017)\citenamefont {Hadfield}, \citenamefont {Wang}, \citenamefont {Rieffel}, \citenamefont {O'Gorman}, \citenamefont {Venturelli},\ and\ \citenamefont {Biswas}}]{hadfield2017quantum}%
  \BibitemOpen
  \bibfield  {author} {\bibinfo {author} {\bibfnamefont {S.}~\bibnamefont {Hadfield}}, \bibinfo {author} {\bibfnamefont {Z.}~\bibnamefont {Wang}}, \bibinfo {author} {\bibfnamefont {E.~G.}\ \bibnamefont {Rieffel}}, \bibinfo {author} {\bibfnamefont {B.}~\bibnamefont {O'Gorman}}, \bibinfo {author} {\bibfnamefont {D.}~\bibnamefont {Venturelli}},\ and\ \bibinfo {author} {\bibfnamefont {R.}~\bibnamefont {Biswas}},\ }in\ \href@noop {} {\emph {\bibinfo {booktitle} {Proceedings of the Second International Workshop on Post Moores Era Supercomputing}}}\ (\bibinfo {year} {2017})\ pp.\ \bibinfo {pages} {15--21}\BibitemShut {NoStop}%
\bibitem [{\citenamefont {Kadowaki}\ and\ \citenamefont {Nishimori}(1998)}]{Kadowaki_1998}%
  \BibitemOpen
  \bibfield  {author} {\bibinfo {author} {\bibfnamefont {T.}~\bibnamefont {Kadowaki}}\ and\ \bibinfo {author} {\bibfnamefont {H.}~\bibnamefont {Nishimori}},\ }\bibfield  {title} {\emph {\bibinfo {title} {{Quantum annealing in the transverse Ising model}}},\ }\href {https://doi.org/10.1103/physreve.58.5355} {\bibfield  {journal} {\bibinfo  {journal} {Physical Review E}\ }\textbf {\bibinfo {volume} {58}},\ \bibinfo {pages} {5355–5363} (\bibinfo {year} {1998})}\BibitemShut {NoStop}%
\bibitem [{\citenamefont {Santoro}\ \emph {et~al.}(2002)\citenamefont {Santoro}, \citenamefont {Marto{\v{n}}\'{a}k}, \citenamefont {Tosatti},\ and\ \citenamefont {Car}}]{Santoro_2002}%
  \BibitemOpen
  \bibfield  {author} {\bibinfo {author} {\bibfnamefont {G.~E.}\ \bibnamefont {Santoro}}, \bibinfo {author} {\bibfnamefont {R.}~\bibnamefont {Marto{\v{n}}\'{a}k}}, \bibinfo {author} {\bibfnamefont {E.}~\bibnamefont {Tosatti}},\ and\ \bibinfo {author} {\bibfnamefont {R.}~\bibnamefont {Car}},\ }\bibfield  {title} {\emph {\bibinfo {title} {{Theory of Quantum Annealing of an Ising Spin Glass}}},\ }\href {https://doi.org/10.1126/science.1068774} {\bibfield  {journal} {\bibinfo  {journal} {Science}\ }\textbf {\bibinfo {volume} {295}},\ \bibinfo {pages} {2427–2430} (\bibinfo {year} {2002})}\BibitemShut {NoStop}%
\bibitem [{\citenamefont {Farhi}\ \emph {et~al.}(2000)\citenamefont {Farhi}, \citenamefont {Goldstone}, \citenamefont {Gutmann},\ and\ \citenamefont {Sipser}}]{farhi2000quantumcomputationadiabaticevolution}%
  \BibitemOpen
  \bibfield  {author} {\bibinfo {author} {\bibfnamefont {E.}~\bibnamefont {Farhi}}, \bibinfo {author} {\bibfnamefont {J.}~\bibnamefont {Goldstone}}, \bibinfo {author} {\bibfnamefont {S.}~\bibnamefont {Gutmann}},\ and\ \bibinfo {author} {\bibfnamefont {M.}~\bibnamefont {Sipser}},\ }\href {https://arxiv.org/abs/quant-ph/0001106} {\bibinfo {title} {{Quantum Computation by Adiabatic Evolution}}} (\bibinfo {year} {2000}),\ \Eprint {https://arxiv.org/abs/quant-ph/0001106} {arXiv:quant-ph/0001106 [quant-ph]} \BibitemShut {NoStop}%
\bibitem [{\citenamefont {Morita}\ and\ \citenamefont {Nishimori}(2008)}]{morita2008mathematical}%
  \BibitemOpen
  \bibfield  {author} {\bibinfo {author} {\bibfnamefont {S.}~\bibnamefont {Morita}}\ and\ \bibinfo {author} {\bibfnamefont {H.}~\bibnamefont {Nishimori}},\ }\bibfield  {title} {\emph {\bibinfo {title} {Mathematical foundation of quantum annealing}},\ }\href@noop {} {\bibfield  {journal} {\bibinfo  {journal} {Journal of Mathematical Physics}\ }\textbf {\bibinfo {volume} {49}} (\bibinfo {year} {2008})}\BibitemShut {NoStop}%
\bibitem [{\citenamefont {Rajak}\ \emph {et~al.}(2023)\citenamefont {Rajak}, \citenamefont {Suzuki}, \citenamefont {Dutta},\ and\ \citenamefont {Chakrabarti}}]{rajak2023quantum}%
  \BibitemOpen
  \bibfield  {author} {\bibinfo {author} {\bibfnamefont {A.}~\bibnamefont {Rajak}}, \bibinfo {author} {\bibfnamefont {S.}~\bibnamefont {Suzuki}}, \bibinfo {author} {\bibfnamefont {A.}~\bibnamefont {Dutta}},\ and\ \bibinfo {author} {\bibfnamefont {B.~K.}\ \bibnamefont {Chakrabarti}},\ }\bibfield  {title} {\emph {\bibinfo {title} {Quantum annealing: An overview}},\ }\href@noop {} {\bibfield  {journal} {\bibinfo  {journal} {Philosophical Transactions of the Royal Society A}\ }\textbf {\bibinfo {volume} {381}},\ \bibinfo {pages} {20210417} (\bibinfo {year} {2023})}\BibitemShut {NoStop}%
\bibitem [{\citenamefont {Peruzzo}\ \emph {et~al.}(2014)\citenamefont {Peruzzo}, \citenamefont {McClean}, \citenamefont {Shadbolt}, \citenamefont {Yung}, \citenamefont {Zhou}, \citenamefont {Love}, \citenamefont {Aspuru-Guzik},\ and\ \citenamefont {O’Brien}}]{Peruzzo_2014}%
  \BibitemOpen
  \bibfield  {author} {\bibinfo {author} {\bibfnamefont {A.}~\bibnamefont {Peruzzo}}, \bibinfo {author} {\bibfnamefont {J.}~\bibnamefont {McClean}}, \bibinfo {author} {\bibfnamefont {P.}~\bibnamefont {Shadbolt}}, \bibinfo {author} {\bibfnamefont {M.-H.}\ \bibnamefont {Yung}}, \bibinfo {author} {\bibfnamefont {X.-Q.}\ \bibnamefont {Zhou}}, \bibinfo {author} {\bibfnamefont {P.~J.}\ \bibnamefont {Love}}, \bibinfo {author} {\bibfnamefont {A.}~\bibnamefont {Aspuru-Guzik}},\ and\ \bibinfo {author} {\bibfnamefont {J.~L.}\ \bibnamefont {O’Brien}},\ }\bibfield  {title} {\emph {\bibinfo {title} {A variational eigenvalue solver on a photonic quantum processor}},\ }\bibfield  {journal} {\bibinfo  {journal} {Nature Communications}\ }\textbf {\bibinfo {volume} {5}},\ \href {https://doi.org/10.1038/ncomms5213} {10.1038/ncomms5213} (\bibinfo {year} {2014})\BibitemShut {NoStop}%
\bibitem [{\citenamefont {McClean}\ \emph {et~al.}(2016)\citenamefont {McClean}, \citenamefont {Romero}, \citenamefont {Babbush},\ and\ \citenamefont {Aspuru-Guzik}}]{McClean_2016}%
  \BibitemOpen
  \bibfield  {author} {\bibinfo {author} {\bibfnamefont {J.~R.}\ \bibnamefont {McClean}}, \bibinfo {author} {\bibfnamefont {J.}~\bibnamefont {Romero}}, \bibinfo {author} {\bibfnamefont {R.}~\bibnamefont {Babbush}},\ and\ \bibinfo {author} {\bibfnamefont {A.}~\bibnamefont {Aspuru-Guzik}},\ }\bibfield  {title} {\emph {\bibinfo {title} {The theory of variational hybrid quantum-classical algorithms}},\ }\href {https://doi.org/10.1088/1367-2630/18/2/023023} {\bibfield  {journal} {\bibinfo  {journal} {New Journal of Physics}\ }\textbf {\bibinfo {volume} {18}},\ \bibinfo {pages} {023023} (\bibinfo {year} {2016})}\BibitemShut {NoStop}%
\bibitem [{\citenamefont {Cerezo}\ \emph {et~al.}(2021)\citenamefont {Cerezo}, \citenamefont {Sone}, \citenamefont {Volkoff}, \citenamefont {Cincio},\ and\ \citenamefont {Coles}}]{cerezo2021cost}%
  \BibitemOpen
  \bibfield  {author} {\bibinfo {author} {\bibfnamefont {M.}~\bibnamefont {Cerezo}}, \bibinfo {author} {\bibfnamefont {A.}~\bibnamefont {Sone}}, \bibinfo {author} {\bibfnamefont {T.}~\bibnamefont {Volkoff}}, \bibinfo {author} {\bibfnamefont {L.}~\bibnamefont {Cincio}},\ and\ \bibinfo {author} {\bibfnamefont {P.~J.}\ \bibnamefont {Coles}},\ }\bibfield  {title} {\emph {\bibinfo {title} {Cost function dependent barren plateaus in shallow parametrized quantum circuits}},\ }\href@noop {} {\bibfield  {journal} {\bibinfo  {journal} {Nature communications}\ }\textbf {\bibinfo {volume} {12}},\ \bibinfo {pages} {1791} (\bibinfo {year} {2021})}\BibitemShut {NoStop}%
\bibitem [{\citenamefont {Hashimoto}\ \emph {et~al.}(2026)\citenamefont {Hashimoto}, \citenamefont {Nakabayashi}, \citenamefont {Nagano}, \citenamefont {Iiyama}, \citenamefont {Sawada}, \citenamefont {Tanaka},\ and\ \citenamefont {Terashi}}]{hashimoto2026comprehensivenumericalstudiesbarren}%
  \BibitemOpen
  \bibfield  {author} {\bibinfo {author} {\bibfnamefont {H.}~\bibnamefont {Hashimoto}}, \bibinfo {author} {\bibfnamefont {A.}~\bibnamefont {Nakabayashi}}, \bibinfo {author} {\bibfnamefont {L.}~\bibnamefont {Nagano}}, \bibinfo {author} {\bibfnamefont {Y.}~\bibnamefont {Iiyama}}, \bibinfo {author} {\bibfnamefont {R.}~\bibnamefont {Sawada}}, \bibinfo {author} {\bibfnamefont {J.}~\bibnamefont {Tanaka}},\ and\ \bibinfo {author} {\bibfnamefont {K.}~\bibnamefont {Terashi}},\ }\href {https://arxiv.org/abs/2602.03291} {\bibinfo {title} {{Comprehensive Numerical Studies of Barren Plateau and Overparametrization in Variational Quantum Algorithm}}} (\bibinfo {year} {2026}),\ \Eprint {https://arxiv.org/abs/2602.03291} {arXiv:2602.03291 [quant-ph]} \BibitemShut {NoStop}%
\bibitem [{\citenamefont {Yao}\ and\ \citenamefont {Hasegawa}(2026)}]{yao2026gradientanalysisbarrenplateau}%
  \BibitemOpen
  \bibfield  {author} {\bibinfo {author} {\bibfnamefont {Y.}~\bibnamefont {Yao}}\ and\ \bibinfo {author} {\bibfnamefont {Y.}~\bibnamefont {Hasegawa}},\ }\href {https://arxiv.org/abs/2602.05288} {\bibinfo {title} {{Gradient Analysis of Barren Plateau in Parameterized Quantum Circuits with multi-qubit gates}}} (\bibinfo {year} {2026}),\ \Eprint {https://arxiv.org/abs/2602.05288} {arXiv:2602.05288 [quant-ph]} \BibitemShut {NoStop}%
\bibitem [{\citenamefont {Larocca}\ \emph {et~al.}(2022)\citenamefont {Larocca}, \citenamefont {Czarnik}, \citenamefont {Sharma}, \citenamefont {Muraleedharan}, \citenamefont {Coles},\ and\ \citenamefont {Cerezo}}]{Larocca_2022}%
  \BibitemOpen
  \bibfield  {author} {\bibinfo {author} {\bibfnamefont {M.}~\bibnamefont {Larocca}}, \bibinfo {author} {\bibfnamefont {P.}~\bibnamefont {Czarnik}}, \bibinfo {author} {\bibfnamefont {K.}~\bibnamefont {Sharma}}, \bibinfo {author} {\bibfnamefont {G.}~\bibnamefont {Muraleedharan}}, \bibinfo {author} {\bibfnamefont {P.~J.}\ \bibnamefont {Coles}},\ and\ \bibinfo {author} {\bibfnamefont {M.}~\bibnamefont {Cerezo}},\ }\bibfield  {title} {\emph {\bibinfo {title} {{Diagnosing Barren Plateaus with Tools from Quantum Optimal Control}}},\ }\href {https://doi.org/10.22331/q-2022-09-29-824} {\bibfield  {journal} {\bibinfo  {journal} {Quantum}\ }\textbf {\bibinfo {volume} {6}},\ \bibinfo {pages} {824} (\bibinfo {year} {2022})}\BibitemShut {NoStop}%
\bibitem [{\citenamefont {Wierichs}\ \emph {et~al.}(2020)\citenamefont {Wierichs}, \citenamefont {Gogolin},\ and\ \citenamefont {Kastoryano}}]{wierichs2020avoiding}%
  \BibitemOpen
  \bibfield  {author} {\bibinfo {author} {\bibfnamefont {D.}~\bibnamefont {Wierichs}}, \bibinfo {author} {\bibfnamefont {C.}~\bibnamefont {Gogolin}},\ and\ \bibinfo {author} {\bibfnamefont {M.}~\bibnamefont {Kastoryano}},\ }\bibfield  {title} {\emph {\bibinfo {title} {Avoiding local minima in variational quantum eigensolvers with the natural gradient optimizer}},\ }\href@noop {} {\bibfield  {journal} {\bibinfo  {journal} {Physical Review Research}\ }\textbf {\bibinfo {volume} {2}},\ \bibinfo {pages} {043246} (\bibinfo {year} {2020})}\BibitemShut {NoStop}%
\bibitem [{\citenamefont {Hirsbrunner}\ \emph {et~al.}(2024)\citenamefont {Hirsbrunner}, \citenamefont {Mullinax}, \citenamefont {Shen}, \citenamefont {Williams-Young}, \citenamefont {Klymko}, \citenamefont {Van~Beeumen},\ and\ \citenamefont {Tubman}}]{Hirsbrunner_2024}%
  \BibitemOpen
  \bibfield  {author} {\bibinfo {author} {\bibfnamefont {M.~R.}\ \bibnamefont {Hirsbrunner}}, \bibinfo {author} {\bibfnamefont {J.~W.}\ \bibnamefont {Mullinax}}, \bibinfo {author} {\bibfnamefont {Y.}~\bibnamefont {Shen}}, \bibinfo {author} {\bibfnamefont {D.~B.}\ \bibnamefont {Williams-Young}}, \bibinfo {author} {\bibfnamefont {K.}~\bibnamefont {Klymko}}, \bibinfo {author} {\bibfnamefont {R.}~\bibnamefont {Van~Beeumen}},\ and\ \bibinfo {author} {\bibfnamefont {N.~M.}\ \bibnamefont {Tubman}},\ }\bibfield  {title} {\emph {\bibinfo {title} {A circuit-generated quantum subspace algorithm for the variational quantum eigensolver}},\ }\bibfield  {journal} {\bibinfo  {journal} {The Journal of Chemical Physics}\ }\textbf {\bibinfo {volume} {161}},\ \href {https://doi.org/10.1063/5.0224883} {10.1063/5.0224883} (\bibinfo {year} {2024})\BibitemShut {NoStop}%
\bibitem [{\citenamefont {Kirmani}\ \emph {et~al.}(2025)\citenamefont {Kirmani}, \citenamefont {Pelofske}, \citenamefont {B{\"a}rtschi}, \citenamefont {Eidenbenz},\ and\ \citenamefont {Zhu}}]{kirmani2025variational}%
  \BibitemOpen
  \bibfield  {author} {\bibinfo {author} {\bibfnamefont {A.}~\bibnamefont {Kirmani}}, \bibinfo {author} {\bibfnamefont {E.}~\bibnamefont {Pelofske}}, \bibinfo {author} {\bibfnamefont {A.}~\bibnamefont {B{\"a}rtschi}}, \bibinfo {author} {\bibfnamefont {S.}~\bibnamefont {Eidenbenz}},\ and\ \bibinfo {author} {\bibfnamefont {J.-X.}\ \bibnamefont {Zhu}},\ }\bibfield  {title} {\emph {\bibinfo {title} {{Variational Quantum Simulations of a Two-Dimensional Frustrated Transverse-Field Ising Model on a Trapped-Ion Quantum Computer}}},\ }\href@noop {} {\bibfield  {journal} {\bibinfo  {journal} {arXiv preprint arXiv:2505.22932}\ } (\bibinfo {year} {2025})}\BibitemShut {NoStop}%
\bibitem [{\citenamefont {Kitaev}(1995)}]{kitaev1995quantummeasurementsabelianstabilizer}%
  \BibitemOpen
  \bibfield  {author} {\bibinfo {author} {\bibfnamefont {A.~Y.}\ \bibnamefont {Kitaev}},\ }\href {https://arxiv.org/abs/quant-ph/9511026} {\bibinfo {title} {{Quantum measurements and the Abelian Stabilizer Problem}}} (\bibinfo {year} {1995}),\ \Eprint {https://arxiv.org/abs/quant-ph/9511026} {arXiv:quant-ph/9511026 [quant-ph]} \BibitemShut {NoStop}%
\bibitem [{\citenamefont {Sack}\ and\ \citenamefont {Serbyn}(2021)}]{QA_initialization_of_QAOA}%
  \BibitemOpen
  \bibfield  {author} {\bibinfo {author} {\bibfnamefont {S.~H.}\ \bibnamefont {Sack}}\ and\ \bibinfo {author} {\bibfnamefont {M.}~\bibnamefont {Serbyn}},\ }\bibfield  {title} {\emph {\bibinfo {title} {Quantum annealing initialization of the quantum approximate optimization algorithm}},\ }\href {https://doi.org/10.22331/q-2021-07-01-491} {\bibfield  {journal} {\bibinfo  {journal} {Quantum}\ }\textbf {\bibinfo {volume} {5}},\ \bibinfo {pages} {491} (\bibinfo {year} {2021})}\BibitemShut {NoStop}%
\bibitem [{\citenamefont {Benchasattabuse}\ \emph {et~al.}(2023)\citenamefont {Benchasattabuse}, \citenamefont {B{\"a}rtschi}, \citenamefont {Garc{\'\i}a-Pintos}, \citenamefont {Golden}, \citenamefont {Lemons},\ and\ \citenamefont {Eidenbenz}}]{benchasattabuse2023lower}%
  \BibitemOpen
  \bibfield  {author} {\bibinfo {author} {\bibfnamefont {N.}~\bibnamefont {Benchasattabuse}}, \bibinfo {author} {\bibfnamefont {A.}~\bibnamefont {B{\"a}rtschi}}, \bibinfo {author} {\bibfnamefont {L.~P.}\ \bibnamefont {Garc{\'\i}a-Pintos}}, \bibinfo {author} {\bibfnamefont {J.}~\bibnamefont {Golden}}, \bibinfo {author} {\bibfnamefont {N.}~\bibnamefont {Lemons}},\ and\ \bibinfo {author} {\bibfnamefont {S.}~\bibnamefont {Eidenbenz}},\ }\bibfield  {title} {\emph {\bibinfo {title} {Lower bounds on number of qaoa rounds required for guaranteed approximation ratios}},\ }\href@noop {} {\bibfield  {journal} {\bibinfo  {journal} {arXiv preprint arXiv:2308.15442}\ } (\bibinfo {year} {2023})}\BibitemShut {NoStop}%
\bibitem [{\citenamefont {Tate}\ \emph {et~al.}(2025)\citenamefont {Tate}, \citenamefont {Langfitt}, \citenamefont {Pelofske}, \citenamefont {Kirmani}, \citenamefont {Bärtschi}, \citenamefont {Golden},\ and\ \citenamefont {Eidenbenz}}]{snapshot_QAOA}%
  \BibitemOpen
  \bibfield  {author} {\bibinfo {author} {\bibfnamefont {R.}~\bibnamefont {Tate}}, \bibinfo {author} {\bibfnamefont {Q.}~\bibnamefont {Langfitt}}, \bibinfo {author} {\bibfnamefont {E.}~\bibnamefont {Pelofske}}, \bibinfo {author} {\bibfnamefont {A.}~\bibnamefont {Kirmani}}, \bibinfo {author} {\bibfnamefont {A.}~\bibnamefont {Bärtschi}}, \bibinfo {author} {\bibfnamefont {J.}~\bibnamefont {Golden}},\ and\ \bibinfo {author} {\bibfnamefont {S.}~\bibnamefont {Eidenbenz}},\ }in\ \href {https://doi.org/10.1109/QAI63978.2025.00035} {\emph {\bibinfo {booktitle} {2025 IEEE International Conference on Quantum Artificial Intelligence (QAI)}}}\ (\bibinfo {year} {2025})\ pp.\ \bibinfo {pages} {176--183}\BibitemShut {NoStop}%
\bibitem [{\citenamefont {Pelofske}\ \emph {et~al.}(2025)\citenamefont {Pelofske}, \citenamefont {Tate}, \citenamefont {Langfitt}, \citenamefont {Kirmani}, \citenamefont {Bärtschi}, \citenamefont {Golden},\ and\ \citenamefont {Eidenbenz}}]{pelofske_2025_15293030}%
  \BibitemOpen
  \bibfield  {author} {\bibinfo {author} {\bibfnamefont {E.}~\bibnamefont {Pelofske}}, \bibinfo {author} {\bibfnamefont {R.}~\bibnamefont {Tate}}, \bibinfo {author} {\bibfnamefont {Q.}~\bibnamefont {Langfitt}}, \bibinfo {author} {\bibfnamefont {A.}~\bibnamefont {Kirmani}}, \bibinfo {author} {\bibfnamefont {A.}~\bibnamefont {Bärtschi}}, \bibinfo {author} {\bibfnamefont {J.}~\bibnamefont {Golden}},\ and\ \bibinfo {author} {\bibfnamefont {S.}~\bibnamefont {Eidenbenz}},\ }\href {https://doi.org/10.5281/zenodo.15293030} {10.5281/zenodo.15293030} (\bibinfo {year} {2025})\BibitemShut {NoStop}%
\bibitem [{\citenamefont {Pagano}\ \emph {et~al.}(2020)\citenamefont {Pagano}, \citenamefont {Bapat}, \citenamefont {Becker}, \citenamefont {Collins}, \citenamefont {De}, \citenamefont {Hess}, \citenamefont {Kaplan}, \citenamefont {Kyprianidis}, \citenamefont {Tan}, \citenamefont {Baldwin}, \citenamefont {Brady}, \citenamefont {Deshpande}, \citenamefont {Liu}, \citenamefont {Jordan}, \citenamefont {Gorshkov},\ and\ \citenamefont {Monroe}}]{Pagano_2020}%
  \BibitemOpen
  \bibfield  {author} {\bibinfo {author} {\bibfnamefont {G.}~\bibnamefont {Pagano}}, \bibinfo {author} {\bibfnamefont {A.}~\bibnamefont {Bapat}}, \bibinfo {author} {\bibfnamefont {P.}~\bibnamefont {Becker}}, \bibinfo {author} {\bibfnamefont {K.~S.}\ \bibnamefont {Collins}}, \bibinfo {author} {\bibfnamefont {A.}~\bibnamefont {De}}, \bibinfo {author} {\bibfnamefont {P.~W.}\ \bibnamefont {Hess}}, \bibinfo {author} {\bibfnamefont {H.~B.}\ \bibnamefont {Kaplan}}, \bibinfo {author} {\bibfnamefont {A.}~\bibnamefont {Kyprianidis}}, \bibinfo {author} {\bibfnamefont {W.~L.}\ \bibnamefont {Tan}}, \bibinfo {author} {\bibfnamefont {C.}~\bibnamefont {Baldwin}}, \bibinfo {author} {\bibfnamefont {L.~T.}\ \bibnamefont {Brady}}, \bibinfo {author} {\bibfnamefont {A.}~\bibnamefont {Deshpande}}, \bibinfo {author} {\bibfnamefont {F.}~\bibnamefont {Liu}}, \bibinfo {author} {\bibfnamefont {S.}~\bibnamefont {Jordan}}, \bibinfo {author} {\bibfnamefont {A.~V.}\ \bibnamefont {Gorshkov}},\ and\ \bibinfo {author} {\bibfnamefont
  {C.}~\bibnamefont {Monroe}},\ }\bibfield  {title} {\emph {\bibinfo {title} {{Quantum approximate optimization of the long-range Ising model with a trapped-ion quantum simulator}}},\ }\href {https://doi.org/10.1073/pnas.2006373117} {\bibfield  {journal} {\bibinfo  {journal} {Proceedings of the National Academy of Sciences}\ }\textbf {\bibinfo {volume} {117}},\ \bibinfo {pages} {25396–25401} (\bibinfo {year} {2020})}\BibitemShut {NoStop}%
\bibitem [{\citenamefont {Pexe}\ \emph {et~al.}(2024)\citenamefont {Pexe}, \citenamefont {Rattighieri}, \citenamefont {Malvezzi},\ and\ \citenamefont {Fanchini}}]{pexe2024usingfeedbackbasedquantumalgorithm}%
  \BibitemOpen
  \bibfield  {author} {\bibinfo {author} {\bibfnamefont {G.~E.~L.}\ \bibnamefont {Pexe}}, \bibinfo {author} {\bibfnamefont {L.~A.~M.}\ \bibnamefont {Rattighieri}}, \bibinfo {author} {\bibfnamefont {A.~L.}\ \bibnamefont {Malvezzi}},\ and\ \bibinfo {author} {\bibfnamefont {F.~F.}\ \bibnamefont {Fanchini}},\ }\href {https://arxiv.org/abs/2406.17937} {\bibinfo {title} {{Using a Feedback-Based Quantum Algorithm to Analyze the Critical Properties of the ANNNI Model Without Classical Optimization}}} (\bibinfo {year} {2024}),\ \Eprint {https://arxiv.org/abs/2406.17937} {arXiv:2406.17937 [quant-ph]} \BibitemShut {NoStop}%
\bibitem [{\citenamefont {Magann}\ \emph {et~al.}(2022)\citenamefont {Magann}, \citenamefont {Rudinger}, \citenamefont {Grace},\ and\ \citenamefont {Sarovar}}]{Magann_2022}%
  \BibitemOpen
  \bibfield  {author} {\bibinfo {author} {\bibfnamefont {A.~B.}\ \bibnamefont {Magann}}, \bibinfo {author} {\bibfnamefont {K.~M.}\ \bibnamefont {Rudinger}}, \bibinfo {author} {\bibfnamefont {M.~D.}\ \bibnamefont {Grace}},\ and\ \bibinfo {author} {\bibfnamefont {M.}~\bibnamefont {Sarovar}},\ }\bibfield  {title} {\emph {\bibinfo {title} {{Feedback-Based Quantum Optimization}}},\ }\bibfield  {journal} {\bibinfo  {journal} {Physical Review Letters}\ }\textbf {\bibinfo {volume} {129}},\ \href {https://doi.org/10.1103/physrevlett.129.250502} {10.1103/physrevlett.129.250502} (\bibinfo {year} {2022})\BibitemShut {NoStop}%
\bibitem [{\citenamefont {Kannan}\ \emph {et~al.}(2024)\citenamefont {Kannan}, \citenamefont {King},\ and\ \citenamefont {Zhou}}]{kannan2024quantumapproximateoptimizationalgorithm}%
  \BibitemOpen
  \bibfield  {author} {\bibinfo {author} {\bibfnamefont {I.}~\bibnamefont {Kannan}}, \bibinfo {author} {\bibfnamefont {R.}~\bibnamefont {King}},\ and\ \bibinfo {author} {\bibfnamefont {L.}~\bibnamefont {Zhou}},\ }\href {https://arxiv.org/abs/2412.09221} {\bibinfo {title} {{A Quantum Approximate Optimization Algorithm for Local Hamiltonian Problems}}} (\bibinfo {year} {2024}),\ \Eprint {https://arxiv.org/abs/2412.09221} {arXiv:2412.09221 [quant-ph]} \BibitemShut {NoStop}%
\bibitem [{\citenamefont {Marwaha}\ \emph {et~al.}(2024)\citenamefont {Marwaha}, \citenamefont {She},\ and\ \citenamefont {Sud}}]{marwaha2024performancevariationalalgorithmslocal}%
  \BibitemOpen
  \bibfield  {author} {\bibinfo {author} {\bibfnamefont {K.}~\bibnamefont {Marwaha}}, \bibinfo {author} {\bibfnamefont {A.}~\bibnamefont {She}},\ and\ \bibinfo {author} {\bibfnamefont {J.}~\bibnamefont {Sud}},\ }\href {https://arxiv.org/abs/2412.15147} {\bibinfo {title} {{Performance of Variational Algorithms for Local Hamiltonian Problems on Random Regular Graphs}}} (\bibinfo {year} {2024}),\ \Eprint {https://arxiv.org/abs/2412.15147} {arXiv:2412.15147 [quant-ph]} \BibitemShut {NoStop}%
\bibitem [{\citenamefont {Lotshaw}\ \emph {et~al.}(2022)\citenamefont {Lotshaw}, \citenamefont {Xu}, \citenamefont {Khalid}, \citenamefont {Buchs}, \citenamefont {Humble},\ and\ \citenamefont {Banerjee}}]{Lotshaw_2022}%
  \BibitemOpen
  \bibfield  {author} {\bibinfo {author} {\bibfnamefont {P.~C.}\ \bibnamefont {Lotshaw}}, \bibinfo {author} {\bibfnamefont {H.}~\bibnamefont {Xu}}, \bibinfo {author} {\bibfnamefont {B.}~\bibnamefont {Khalid}}, \bibinfo {author} {\bibfnamefont {G.}~\bibnamefont {Buchs}}, \bibinfo {author} {\bibfnamefont {T.~S.}\ \bibnamefont {Humble}},\ and\ \bibinfo {author} {\bibfnamefont {A.}~\bibnamefont {Banerjee}},\ }\bibfield  {title} {\emph {\bibinfo {title} {{Simulations of frustrated Ising Hamiltonians using quantum approximate optimization}}},\ }\bibfield  {journal} {\bibinfo  {journal} {Philosophical Transactions of the Royal Society A: Mathematical, Physical and Engineering Sciences}\ }\textbf {\bibinfo {volume} {381}},\ \href {https://doi.org/10.1098/rsta.2021.0414} {10.1098/rsta.2021.0414} (\bibinfo {year} {2022})\BibitemShut {NoStop}%
\bibitem [{\citenamefont {Johnson}\ \emph {et~al.}(2011)\citenamefont {Johnson}, \citenamefont {Amin}, \citenamefont {Gildert}, \citenamefont {Lanting}, \citenamefont {Hamze}, \citenamefont {Dickson}, \citenamefont {Harris}, \citenamefont {Berkley}, \citenamefont {Johansson}, \citenamefont {Bunyk} \emph {et~al.}}]{johnson2011quantum}%
  \BibitemOpen
  \bibfield  {author} {\bibinfo {author} {\bibfnamefont {M.~W.}\ \bibnamefont {Johnson}}, \bibinfo {author} {\bibfnamefont {M.~H.}\ \bibnamefont {Amin}}, \bibinfo {author} {\bibfnamefont {S.}~\bibnamefont {Gildert}}, \bibinfo {author} {\bibfnamefont {T.}~\bibnamefont {Lanting}}, \bibinfo {author} {\bibfnamefont {F.}~\bibnamefont {Hamze}}, \bibinfo {author} {\bibfnamefont {N.}~\bibnamefont {Dickson}}, \bibinfo {author} {\bibfnamefont {R.}~\bibnamefont {Harris}}, \bibinfo {author} {\bibfnamefont {A.~J.}\ \bibnamefont {Berkley}}, \bibinfo {author} {\bibfnamefont {J.}~\bibnamefont {Johansson}}, \bibinfo {author} {\bibfnamefont {P.}~\bibnamefont {Bunyk}}, \emph {et~al.},\ }\bibfield  {title} {\emph {\bibinfo {title} {Quantum annealing with manufactured spins}},\ }\href {https://doi.org/10.1038/nature10012} {\bibfield  {journal} {\bibinfo  {journal} {Nature}\ }\textbf {\bibinfo {volume} {473}},\ \bibinfo {pages} {194--198} (\bibinfo {year} {2011})}\BibitemShut {NoStop}%
\bibitem [{\citenamefont {Bunyk}\ \emph {et~al.}(2014)\citenamefont {Bunyk}, \citenamefont {Hoskinson}, \citenamefont {Johnson}, \citenamefont {Tolkacheva}, \citenamefont {Altomare}, \citenamefont {Berkley}, \citenamefont {Harris}, \citenamefont {Hilton}, \citenamefont {Lanting}, \citenamefont {Przybysz},\ and\ \citenamefont {Whittaker}}]{Bunyk_2014}%
  \BibitemOpen
  \bibfield  {author} {\bibinfo {author} {\bibfnamefont {P.~I.}\ \bibnamefont {Bunyk}}, \bibinfo {author} {\bibfnamefont {E.~M.}\ \bibnamefont {Hoskinson}}, \bibinfo {author} {\bibfnamefont {M.~W.}\ \bibnamefont {Johnson}}, \bibinfo {author} {\bibfnamefont {E.}~\bibnamefont {Tolkacheva}}, \bibinfo {author} {\bibfnamefont {F.}~\bibnamefont {Altomare}}, \bibinfo {author} {\bibfnamefont {A.~J.}\ \bibnamefont {Berkley}}, \bibinfo {author} {\bibfnamefont {R.}~\bibnamefont {Harris}}, \bibinfo {author} {\bibfnamefont {J.~P.}\ \bibnamefont {Hilton}}, \bibinfo {author} {\bibfnamefont {T.}~\bibnamefont {Lanting}}, \bibinfo {author} {\bibfnamefont {A.~J.}\ \bibnamefont {Przybysz}},\ and\ \bibinfo {author} {\bibfnamefont {J.}~\bibnamefont {Whittaker}},\ }\bibfield  {title} {\emph {\bibinfo {title} {{Architectural Considerations in the Design of a Superconducting Quantum Annealing Processor}}},\ }\href {https://doi.org/10.1109/tasc.2014.2318294} {\bibfield  {journal} {\bibinfo  {journal} {IEEE Transactions on Applied
  Superconductivity}\ }\textbf {\bibinfo {volume} {24}},\ \bibinfo {pages} {1–10} (\bibinfo {year} {2014})}\BibitemShut {NoStop}%
\bibitem [{\citenamefont {King}\ \emph {et~al.}(2018)\citenamefont {King}, \citenamefont {Carrasquilla}, \citenamefont {Raymond}, \citenamefont {Ozfidan}, \citenamefont {Andriyash}, \citenamefont {Berkley}, \citenamefont {Reis}, \citenamefont {Lanting}, \citenamefont {Harris}, \citenamefont {Altomare}, \citenamefont {Boothby}, \citenamefont {Bunyk}, \citenamefont {Enderud}, \citenamefont {Fréchette}, \citenamefont {Hoskinson}, \citenamefont {Ladizinsky}, \citenamefont {Oh}, \citenamefont {Poulin-Lamarre}, \citenamefont {Rich}, \citenamefont {Sato}, \citenamefont {Smirnov}, \citenamefont {Swenson}, \citenamefont {Volkmann}, \citenamefont {Whittaker}, \citenamefont {Yao}, \citenamefont {Ladizinsky}, \citenamefont {Johnson}, \citenamefont {Hilton},\ and\ \citenamefont {Amin}}]{King_2018}%
  \BibitemOpen
  \bibfield  {author} {\bibinfo {author} {\bibfnamefont {A.~D.}\ \bibnamefont {King}}, \bibinfo {author} {\bibfnamefont {J.}~\bibnamefont {Carrasquilla}}, \bibinfo {author} {\bibfnamefont {J.}~\bibnamefont {Raymond}}, \bibinfo {author} {\bibfnamefont {I.}~\bibnamefont {Ozfidan}}, \bibinfo {author} {\bibfnamefont {E.}~\bibnamefont {Andriyash}}, \bibinfo {author} {\bibfnamefont {A.}~\bibnamefont {Berkley}}, \bibinfo {author} {\bibfnamefont {M.}~\bibnamefont {Reis}}, \bibinfo {author} {\bibfnamefont {T.}~\bibnamefont {Lanting}}, \bibinfo {author} {\bibfnamefont {R.}~\bibnamefont {Harris}}, \bibinfo {author} {\bibfnamefont {F.}~\bibnamefont {Altomare}}, \bibinfo {author} {\bibfnamefont {K.}~\bibnamefont {Boothby}}, \bibinfo {author} {\bibfnamefont {P.~I.}\ \bibnamefont {Bunyk}}, \bibinfo {author} {\bibfnamefont {C.}~\bibnamefont {Enderud}}, \bibinfo {author} {\bibfnamefont {A.}~\bibnamefont {Fréchette}}, \bibinfo {author} {\bibfnamefont {E.}~\bibnamefont {Hoskinson}}, \bibinfo {author} {\bibfnamefont
  {N.}~\bibnamefont {Ladizinsky}}, \bibinfo {author} {\bibfnamefont {T.}~\bibnamefont {Oh}}, \bibinfo {author} {\bibfnamefont {G.}~\bibnamefont {Poulin-Lamarre}}, \bibinfo {author} {\bibfnamefont {C.}~\bibnamefont {Rich}}, \bibinfo {author} {\bibfnamefont {Y.}~\bibnamefont {Sato}}, \bibinfo {author} {\bibfnamefont {A.~Y.}\ \bibnamefont {Smirnov}}, \bibinfo {author} {\bibfnamefont {L.~J.}\ \bibnamefont {Swenson}}, \bibinfo {author} {\bibfnamefont {M.~H.}\ \bibnamefont {Volkmann}}, \bibinfo {author} {\bibfnamefont {J.}~\bibnamefont {Whittaker}}, \bibinfo {author} {\bibfnamefont {J.}~\bibnamefont {Yao}}, \bibinfo {author} {\bibfnamefont {E.}~\bibnamefont {Ladizinsky}}, \bibinfo {author} {\bibfnamefont {M.~W.}\ \bibnamefont {Johnson}}, \bibinfo {author} {\bibfnamefont {J.}~\bibnamefont {Hilton}},\ and\ \bibinfo {author} {\bibfnamefont {M.~H.}\ \bibnamefont {Amin}},\ }\bibfield  {title} {\emph {\bibinfo {title} {Observation of topological phenomena in a programmable lattice of 1,800 qubits}},\ }\href
  {https://doi.org/10.1038/s41586-018-0410-x} {\bibfield  {journal} {\bibinfo  {journal} {Nature}\ }\textbf {\bibinfo {volume} {560}},\ \bibinfo {pages} {456–460} (\bibinfo {year} {2018})}\BibitemShut {NoStop}%
\bibitem [{\citenamefont {King}\ \emph {et~al.}(2021)\citenamefont {King}, \citenamefont {Batista}, \citenamefont {Raymond}, \citenamefont {Lanting}, \citenamefont {Ozfidan}, \citenamefont {Poulin-Lamarre}, \citenamefont {Zhang},\ and\ \citenamefont {Amin}}]{PRXQuantum.2.030317}%
  \BibitemOpen
  \bibfield  {author} {\bibinfo {author} {\bibfnamefont {A.~D.}\ \bibnamefont {King}}, \bibinfo {author} {\bibfnamefont {C.~D.}\ \bibnamefont {Batista}}, \bibinfo {author} {\bibfnamefont {J.}~\bibnamefont {Raymond}}, \bibinfo {author} {\bibfnamefont {T.}~\bibnamefont {Lanting}}, \bibinfo {author} {\bibfnamefont {I.}~\bibnamefont {Ozfidan}}, \bibinfo {author} {\bibfnamefont {G.}~\bibnamefont {Poulin-Lamarre}}, \bibinfo {author} {\bibfnamefont {H.}~\bibnamefont {Zhang}},\ and\ \bibinfo {author} {\bibfnamefont {M.~H.}\ \bibnamefont {Amin}},\ }\bibfield  {title} {\emph {\bibinfo {title} {{Quantum Annealing Simulation of Out-of-Equilibrium Magnetization in a Spin-Chain Compound}}},\ }\href {https://doi.org/10.1103/PRXQuantum.2.030317} {\bibfield  {journal} {\bibinfo  {journal} {PRX Quantum}\ }\textbf {\bibinfo {volume} {2}},\ \bibinfo {pages} {030317} (\bibinfo {year} {2021})}\BibitemShut {NoStop}%
\bibitem [{\citenamefont {Pelofske}\ \emph {et~al.}(2024)\citenamefont {Pelofske}, \citenamefont {Bärtschi},\ and\ \citenamefont {Eidenbenz}}]{pelofske2024simulatingheavyhextransversefield}%
  \BibitemOpen
  \bibfield  {author} {\bibinfo {author} {\bibfnamefont {E.}~\bibnamefont {Pelofske}}, \bibinfo {author} {\bibfnamefont {A.}~\bibnamefont {Bärtschi}},\ and\ \bibinfo {author} {\bibfnamefont {S.}~\bibnamefont {Eidenbenz}},\ }\href {https://arxiv.org/abs/2311.01657} {\bibinfo {title} {{Simulating Heavy-Hex Transverse Field Ising Model Magnetization Dynamics Using Programmable Quantum Annealers}}} (\bibinfo {year} {2024}),\ \Eprint {https://arxiv.org/abs/2311.01657} {arXiv:2311.01657 [quant-ph]} \BibitemShut {NoStop}%
\bibitem [{\citenamefont {Ebadi}\ \emph {et~al.}(2021)\citenamefont {Ebadi}, \citenamefont {Wang}, \citenamefont {Levine}, \citenamefont {Keesling}, \citenamefont {Semeghini}, \citenamefont {Omran}, \citenamefont {Bluvstein}, \citenamefont {Samajdar}, \citenamefont {Pichler}, \citenamefont {Ho}, \citenamefont {Choi}, \citenamefont {Sachdev}, \citenamefont {Greiner}, \citenamefont {Vuletić},\ and\ \citenamefont {Lukin}}]{Ebadi_2021}%
  \BibitemOpen
  \bibfield  {author} {\bibinfo {author} {\bibfnamefont {S.}~\bibnamefont {Ebadi}}, \bibinfo {author} {\bibfnamefont {T.~T.}\ \bibnamefont {Wang}}, \bibinfo {author} {\bibfnamefont {H.}~\bibnamefont {Levine}}, \bibinfo {author} {\bibfnamefont {A.}~\bibnamefont {Keesling}}, \bibinfo {author} {\bibfnamefont {G.}~\bibnamefont {Semeghini}}, \bibinfo {author} {\bibfnamefont {A.}~\bibnamefont {Omran}}, \bibinfo {author} {\bibfnamefont {D.}~\bibnamefont {Bluvstein}}, \bibinfo {author} {\bibfnamefont {R.}~\bibnamefont {Samajdar}}, \bibinfo {author} {\bibfnamefont {H.}~\bibnamefont {Pichler}}, \bibinfo {author} {\bibfnamefont {W.~W.}\ \bibnamefont {Ho}}, \bibinfo {author} {\bibfnamefont {S.}~\bibnamefont {Choi}}, \bibinfo {author} {\bibfnamefont {S.}~\bibnamefont {Sachdev}}, \bibinfo {author} {\bibfnamefont {M.}~\bibnamefont {Greiner}}, \bibinfo {author} {\bibfnamefont {V.}~\bibnamefont {Vuletić}},\ and\ \bibinfo {author} {\bibfnamefont {M.~D.}\ \bibnamefont {Lukin}},\ }\bibfield  {title} {\emph {\bibinfo {title}
  {Quantum phases of matter on a 256-atom programmable quantum simulator}},\ }\href {https://doi.org/10.1038/s41586-021-03582-4} {\bibfield  {journal} {\bibinfo  {journal} {Nature}\ }\textbf {\bibinfo {volume} {595}},\ \bibinfo {pages} {227–232} (\bibinfo {year} {2021})}\BibitemShut {NoStop}%
\bibitem [{\citenamefont {Semeghini}\ \emph {et~al.}(2021)\citenamefont {Semeghini}, \citenamefont {Levine}, \citenamefont {Keesling}, \citenamefont {Ebadi}, \citenamefont {Wang}, \citenamefont {Bluvstein}, \citenamefont {Verresen}, \citenamefont {Pichler}, \citenamefont {Kalinowski}, \citenamefont {Samajdar}, \citenamefont {Omran}, \citenamefont {Sachdev}, \citenamefont {Vishwanath}, \citenamefont {Greiner}, \citenamefont {Vuletić},\ and\ \citenamefont {Lukin}}]{Semeghini_2021}%
  \BibitemOpen
  \bibfield  {author} {\bibinfo {author} {\bibfnamefont {G.}~\bibnamefont {Semeghini}}, \bibinfo {author} {\bibfnamefont {H.}~\bibnamefont {Levine}}, \bibinfo {author} {\bibfnamefont {A.}~\bibnamefont {Keesling}}, \bibinfo {author} {\bibfnamefont {S.}~\bibnamefont {Ebadi}}, \bibinfo {author} {\bibfnamefont {T.~T.}\ \bibnamefont {Wang}}, \bibinfo {author} {\bibfnamefont {D.}~\bibnamefont {Bluvstein}}, \bibinfo {author} {\bibfnamefont {R.}~\bibnamefont {Verresen}}, \bibinfo {author} {\bibfnamefont {H.}~\bibnamefont {Pichler}}, \bibinfo {author} {\bibfnamefont {M.}~\bibnamefont {Kalinowski}}, \bibinfo {author} {\bibfnamefont {R.}~\bibnamefont {Samajdar}}, \bibinfo {author} {\bibfnamefont {A.}~\bibnamefont {Omran}}, \bibinfo {author} {\bibfnamefont {S.}~\bibnamefont {Sachdev}}, \bibinfo {author} {\bibfnamefont {A.}~\bibnamefont {Vishwanath}}, \bibinfo {author} {\bibfnamefont {M.}~\bibnamefont {Greiner}}, \bibinfo {author} {\bibfnamefont {V.}~\bibnamefont {Vuletić}},\ and\ \bibinfo {author} {\bibfnamefont
  {M.~D.}\ \bibnamefont {Lukin}},\ }\bibfield  {title} {\emph {\bibinfo {title} {Probing topological spin liquids on a programmable quantum simulator}},\ }\href {https://doi.org/10.1126/science.abi8794} {\bibfield  {journal} {\bibinfo  {journal} {Science}\ }\textbf {\bibinfo {volume} {374}},\ \bibinfo {pages} {1242–1247} (\bibinfo {year} {2021})}\BibitemShut {NoStop}%
\bibitem [{\citenamefont {Bluvstein}\ \emph {et~al.}(2022)\citenamefont {Bluvstein}, \citenamefont {Levine}, \citenamefont {Semeghini}, \citenamefont {Wang}, \citenamefont {Ebadi}, \citenamefont {Kalinowski}, \citenamefont {Keesling}, \citenamefont {Maskara}, \citenamefont {Pichler}, \citenamefont {Greiner}, \citenamefont {Vuletić},\ and\ \citenamefont {Lukin}}]{Bluvstein_2022}%
  \BibitemOpen
  \bibfield  {author} {\bibinfo {author} {\bibfnamefont {D.}~\bibnamefont {Bluvstein}}, \bibinfo {author} {\bibfnamefont {H.}~\bibnamefont {Levine}}, \bibinfo {author} {\bibfnamefont {G.}~\bibnamefont {Semeghini}}, \bibinfo {author} {\bibfnamefont {T.~T.}\ \bibnamefont {Wang}}, \bibinfo {author} {\bibfnamefont {S.}~\bibnamefont {Ebadi}}, \bibinfo {author} {\bibfnamefont {M.}~\bibnamefont {Kalinowski}}, \bibinfo {author} {\bibfnamefont {A.}~\bibnamefont {Keesling}}, \bibinfo {author} {\bibfnamefont {N.}~\bibnamefont {Maskara}}, \bibinfo {author} {\bibfnamefont {H.}~\bibnamefont {Pichler}}, \bibinfo {author} {\bibfnamefont {M.}~\bibnamefont {Greiner}}, \bibinfo {author} {\bibfnamefont {V.}~\bibnamefont {Vuletić}},\ and\ \bibinfo {author} {\bibfnamefont {M.~D.}\ \bibnamefont {Lukin}},\ }\bibfield  {title} {\emph {\bibinfo {title} {A quantum processor based on coherent transport of entangled atom arrays}},\ }\href {https://doi.org/10.1038/s41586-022-04592-6} {\bibfield  {journal} {\bibinfo  {journal} {Nature}\
  }\textbf {\bibinfo {volume} {604}},\ \bibinfo {pages} {451–456} (\bibinfo {year} {2022})}\BibitemShut {NoStop}%
\bibitem [{\citenamefont {King}\ \emph {et~al.}(2023)\citenamefont {King}, \citenamefont {Raymond}, \citenamefont {Lanting}, \citenamefont {Harris}, \citenamefont {Zucca}, \citenamefont {Altomare}, \citenamefont {Berkley}, \citenamefont {Boothby}, \citenamefont {Ejtemaee}, \citenamefont {Enderud} \emph {et~al.}}]{king2023quantum}%
  \BibitemOpen
  \bibfield  {author} {\bibinfo {author} {\bibfnamefont {A.~D.}\ \bibnamefont {King}}, \bibinfo {author} {\bibfnamefont {J.}~\bibnamefont {Raymond}}, \bibinfo {author} {\bibfnamefont {T.}~\bibnamefont {Lanting}}, \bibinfo {author} {\bibfnamefont {R.}~\bibnamefont {Harris}}, \bibinfo {author} {\bibfnamefont {A.}~\bibnamefont {Zucca}}, \bibinfo {author} {\bibfnamefont {F.}~\bibnamefont {Altomare}}, \bibinfo {author} {\bibfnamefont {A.~J.}\ \bibnamefont {Berkley}}, \bibinfo {author} {\bibfnamefont {K.}~\bibnamefont {Boothby}}, \bibinfo {author} {\bibfnamefont {S.}~\bibnamefont {Ejtemaee}}, \bibinfo {author} {\bibfnamefont {C.}~\bibnamefont {Enderud}}, \emph {et~al.},\ }\bibfield  {title} {\emph {\bibinfo {title} {Quantum critical dynamics in a 5,000-qubit programmable spin glass}},\ }\href@noop {} {\bibfield  {journal} {\bibinfo  {journal} {Nature}\ }\textbf {\bibinfo {volume} {617}},\ \bibinfo {pages} {61--66} (\bibinfo {year} {2023})}\BibitemShut {NoStop}%
\bibitem [{\citenamefont {Born}\ and\ \citenamefont {Fock}(1928)}]{born1928beweis}%
  \BibitemOpen
  \bibfield  {author} {\bibinfo {author} {\bibfnamefont {M.}~\bibnamefont {Born}}\ and\ \bibinfo {author} {\bibfnamefont {V.}~\bibnamefont {Fock}},\ }\bibfield  {title} {\emph {\bibinfo {title} {Beweis des adiabatensatzes}},\ }\href@noop {} {\bibfield  {journal} {\bibinfo  {journal} {Zeitschrift f{\"u}r Physik}\ }\textbf {\bibinfo {volume} {51}},\ \bibinfo {pages} {165--180} (\bibinfo {year} {1928})}\BibitemShut {NoStop}%
\bibitem [{\citenamefont {Kato}(1950)}]{kato1950adiabatic}%
  \BibitemOpen
  \bibfield  {author} {\bibinfo {author} {\bibfnamefont {T.}~\bibnamefont {Kato}},\ }\bibfield  {title} {\emph {\bibinfo {title} {On the adiabatic theorem of quantum mechanics}},\ }\href@noop {} {\bibfield  {journal} {\bibinfo  {journal} {Journal of the Physical Society of Japan}\ }\textbf {\bibinfo {volume} {5}},\ \bibinfo {pages} {435--439} (\bibinfo {year} {1950})}\BibitemShut {NoStop}%
\bibitem [{\citenamefont {Nekrashevich}\ \emph {et~al.}(2022)\citenamefont {Nekrashevich}, \citenamefont {Ding}, \citenamefont {Balakirev}, \citenamefont {Yi}, \citenamefont {Cheong}, \citenamefont {Civale}, \citenamefont {Kamiya},\ and\ \citenamefont {Zapf}}]{nekrashevich2022reaching}%
  \BibitemOpen
  \bibfield  {author} {\bibinfo {author} {\bibfnamefont {I.}~\bibnamefont {Nekrashevich}}, \bibinfo {author} {\bibfnamefont {X.}~\bibnamefont {Ding}}, \bibinfo {author} {\bibfnamefont {F.}~\bibnamefont {Balakirev}}, \bibinfo {author} {\bibfnamefont {H.~T.}\ \bibnamefont {Yi}}, \bibinfo {author} {\bibfnamefont {S.-W.}\ \bibnamefont {Cheong}}, \bibinfo {author} {\bibfnamefont {L.}~\bibnamefont {Civale}}, \bibinfo {author} {\bibfnamefont {Y.}~\bibnamefont {Kamiya}},\ and\ \bibinfo {author} {\bibfnamefont {V.~S.}\ \bibnamefont {Zapf}},\ }\bibfield  {title} {\emph {\bibinfo {title} {Reaching the equilibrium state of the frustrated triangular ising magnet ca 3 co 2 o 6}},\ }\href@noop {} {\bibfield  {journal} {\bibinfo  {journal} {Physical Review B}\ }\textbf {\bibinfo {volume} {105}},\ \bibinfo {pages} {024426} (\bibinfo {year} {2022})}\BibitemShut {NoStop}%
\bibitem [{\citenamefont {Duft}\ \emph {et~al.}(2024)\citenamefont {Duft}, \citenamefont {Koziol}, \citenamefont {Adelhardt}, \citenamefont {M{\"u}hlhauser},\ and\ \citenamefont {Schmidt}}]{duft2024order}%
  \BibitemOpen
  \bibfield  {author} {\bibinfo {author} {\bibfnamefont {A.}~\bibnamefont {Duft}}, \bibinfo {author} {\bibfnamefont {J.~A.}\ \bibnamefont {Koziol}}, \bibinfo {author} {\bibfnamefont {P.}~\bibnamefont {Adelhardt}}, \bibinfo {author} {\bibfnamefont {M.}~\bibnamefont {M{\"u}hlhauser}},\ and\ \bibinfo {author} {\bibfnamefont {K.~P.}\ \bibnamefont {Schmidt}},\ }\bibfield  {title} {\emph {\bibinfo {title} {Order-by-disorder in the antiferromagnetic j 1-j 2-j 3 transverse-field ising model on the ruby lattice}},\ }\href@noop {} {\bibfield  {journal} {\bibinfo  {journal} {Physical Review Research}\ }\textbf {\bibinfo {volume} {6}},\ \bibinfo {pages} {033339} (\bibinfo {year} {2024})}\BibitemShut {NoStop}%
\bibitem [{\citenamefont {Sen}\ and\ \citenamefont {Chakrabarti}(1989)}]{sen1989ising}%
  \BibitemOpen
  \bibfield  {author} {\bibinfo {author} {\bibfnamefont {P.}~\bibnamefont {Sen}}\ and\ \bibinfo {author} {\bibfnamefont {B.}~\bibnamefont {Chakrabarti}},\ }\bibfield  {title} {\emph {\bibinfo {title} {{Ising models with competing axial interactions in transverse fields}}},\ }\href@noop {} {\bibfield  {journal} {\bibinfo  {journal} {Physical Review B}\ }\textbf {\bibinfo {volume} {40}},\ \bibinfo {pages} {760} (\bibinfo {year} {1989})}\BibitemShut {NoStop}%
\bibitem [{\citenamefont {Suzuki}\ \emph {et~al.}(2012)\citenamefont {Suzuki}, \citenamefont {Inoue},\ and\ \citenamefont {Chakrabarti}}]{suzuki2012quantum}%
  \BibitemOpen
  \bibfield  {author} {\bibinfo {author} {\bibfnamefont {S.}~\bibnamefont {Suzuki}}, \bibinfo {author} {\bibfnamefont {J.-i.}\ \bibnamefont {Inoue}},\ and\ \bibinfo {author} {\bibfnamefont {B.~K.}\ \bibnamefont {Chakrabarti}},\ }\href@noop {} {\emph {\bibinfo {title} {{Quantum Ising phases and transitions in transverse Ising models}}}},\ Vol.\ \bibinfo {volume} {862}\ (\bibinfo  {publisher} {Springer},\ \bibinfo {year} {2012})\BibitemShut {NoStop}%
\bibitem [{\citenamefont {Sen}\ \emph {et~al.}(1992)\citenamefont {Sen}, \citenamefont {Chakraborty}, \citenamefont {Dasgupta},\ and\ \citenamefont {Chakrabarti}}]{sen1992numerical}%
  \BibitemOpen
  \bibfield  {author} {\bibinfo {author} {\bibfnamefont {P.}~\bibnamefont {Sen}}, \bibinfo {author} {\bibfnamefont {S.}~\bibnamefont {Chakraborty}}, \bibinfo {author} {\bibfnamefont {S.}~\bibnamefont {Dasgupta}},\ and\ \bibinfo {author} {\bibfnamefont {B.}~\bibnamefont {Chakrabarti}},\ }\bibfield  {title} {\emph {\bibinfo {title} {{Numerical estimate of the phase diagram of finite ANNNI chains in transverse field}}},\ }\href@noop {} {\bibfield  {journal} {\bibinfo  {journal} {Zeitschrift f{\"u}r Physik B Condensed Matter}\ }\textbf {\bibinfo {volume} {88}},\ \bibinfo {pages} {333--338} (\bibinfo {year} {1992})}\BibitemShut {NoStop}%
\bibitem [{\citenamefont {Chandra}\ and\ \citenamefont {Dasgupta}(2007)}]{Chandra_2007}%
  \BibitemOpen
  \bibfield  {author} {\bibinfo {author} {\bibfnamefont {A.~K.}\ \bibnamefont {Chandra}}\ and\ \bibinfo {author} {\bibfnamefont {S.}~\bibnamefont {Dasgupta}},\ }\bibfield  {title} {\emph {\bibinfo {title} {{Floating phase in the one-dimensional transverse axial next-nearest-neighbor Ising model}}},\ }\bibfield  {journal} {\bibinfo  {journal} {Physical Review E}\ }\textbf {\bibinfo {volume} {75}},\ \href {https://doi.org/10.1103/physreve.75.021105} {10.1103/physreve.75.021105} (\bibinfo {year} {2007})\BibitemShut {NoStop}%
\bibitem [{\citenamefont {Rieger}\ and\ \citenamefont {Uimin}(1996)}]{Rieger_1996}%
  \BibitemOpen
  \bibfield  {author} {\bibinfo {author} {\bibfnamefont {H.}~\bibnamefont {Rieger}}\ and\ \bibinfo {author} {\bibfnamefont {G.}~\bibnamefont {Uimin}},\ }\bibfield  {title} {\emph {\bibinfo {title} {{The one-dimensional ANNNI model in a transverse field: analytic and numerical study of effective Hamiltonians}}},\ }\href {https://doi.org/10.1007/s002570050252} {\bibfield  {journal} {\bibinfo  {journal} {Zeitschrift für Physik B Condensed Matter}\ }\textbf {\bibinfo {volume} {101}},\ \bibinfo {pages} {597–611} (\bibinfo {year} {1996})}\BibitemShut {NoStop}%
\bibitem [{\citenamefont {Guo}\ \emph {et~al.}(1994)\citenamefont {Guo}, \citenamefont {Bhatt},\ and\ \citenamefont {Huse}}]{guo1994quantum}%
  \BibitemOpen
  \bibfield  {author} {\bibinfo {author} {\bibfnamefont {M.}~\bibnamefont {Guo}}, \bibinfo {author} {\bibfnamefont {R.~N.}\ \bibnamefont {Bhatt}},\ and\ \bibinfo {author} {\bibfnamefont {D.~A.}\ \bibnamefont {Huse}},\ }\bibfield  {title} {\emph {\bibinfo {title} {Quantum critical behavior of a three-dimensional ising spin glass in a transverse magnetic field}},\ }\href@noop {} {\bibfield  {journal} {\bibinfo  {journal} {Physical review letters}\ }\textbf {\bibinfo {volume} {72}},\ \bibinfo {pages} {4137} (\bibinfo {year} {1994})}\BibitemShut {NoStop}%
\bibitem [{\citenamefont {Liu}\ \emph {et~al.}(2020)\citenamefont {Liu}, \citenamefont {Huang},\ and\ \citenamefont {Chen}}]{liu2020intrinsic}%
  \BibitemOpen
  \bibfield  {author} {\bibinfo {author} {\bibfnamefont {C.}~\bibnamefont {Liu}}, \bibinfo {author} {\bibfnamefont {C.-J.}\ \bibnamefont {Huang}},\ and\ \bibinfo {author} {\bibfnamefont {G.}~\bibnamefont {Chen}},\ }\bibfield  {title} {\emph {\bibinfo {title} {Intrinsic quantum ising model on a triangular lattice magnet tm mg ga o 4}},\ }\href@noop {} {\bibfield  {journal} {\bibinfo  {journal} {Physical Review Research}\ }\textbf {\bibinfo {volume} {2}},\ \bibinfo {pages} {043013} (\bibinfo {year} {2020})}\BibitemShut {NoStop}%
\bibitem [{\citenamefont {Coletta}\ \emph {et~al.}(2011)\citenamefont {Coletta}, \citenamefont {Picon}, \citenamefont {Korshunov},\ and\ \citenamefont {Mila}}]{coletta2011phase}%
  \BibitemOpen
  \bibfield  {author} {\bibinfo {author} {\bibfnamefont {T.}~\bibnamefont {Coletta}}, \bibinfo {author} {\bibfnamefont {J.-D.}\ \bibnamefont {Picon}}, \bibinfo {author} {\bibfnamefont {S.}~\bibnamefont {Korshunov}},\ and\ \bibinfo {author} {\bibfnamefont {F.}~\bibnamefont {Mila}},\ }\bibfield  {title} {\emph {\bibinfo {title} {Phase diagram of the fully frustrated transverse-field ising model on the honeycomb lattice}},\ }\href@noop {} {\bibfield  {journal} {\bibinfo  {journal} {Physical Review B—Condensed Matter and Materials Physics}\ }\textbf {\bibinfo {volume} {83}},\ \bibinfo {pages} {054402} (\bibinfo {year} {2011})}\BibitemShut {NoStop}%
\bibitem [{\citenamefont {Schiffer}\ \emph {et~al.}(2024)\citenamefont {Schiffer}, \citenamefont {Rubio}, \citenamefont {Trivedi},\ and\ \citenamefont {Cirac}}]{schiffer2024quantum}%
  \BibitemOpen
  \bibfield  {author} {\bibinfo {author} {\bibfnamefont {B.~F.}\ \bibnamefont {Schiffer}}, \bibinfo {author} {\bibfnamefont {A.~F.}\ \bibnamefont {Rubio}}, \bibinfo {author} {\bibfnamefont {R.}~\bibnamefont {Trivedi}},\ and\ \bibinfo {author} {\bibfnamefont {J.~I.}\ \bibnamefont {Cirac}},\ }\bibfield  {title} {\emph {\bibinfo {title} {The quantum adiabatic algorithm suppresses the proliferation of errors}},\ }\href@noop {} {\bibfield  {journal} {\bibinfo  {journal} {arXiv preprint arXiv:2404.15397}\ } (\bibinfo {year} {2024})}\BibitemShut {NoStop}%
\bibitem [{\citenamefont {Rieffel}\ \emph {et~al.}(2020)\citenamefont {Rieffel}, \citenamefont {Dominy}, \citenamefont {Rubin},\ and\ \citenamefont {Wang}}]{rieffel2020xy}%
  \BibitemOpen
  \bibfield  {author} {\bibinfo {author} {\bibfnamefont {E.}~\bibnamefont {Rieffel}}, \bibinfo {author} {\bibfnamefont {J.~M.}\ \bibnamefont {Dominy}}, \bibinfo {author} {\bibfnamefont {N.}~\bibnamefont {Rubin}},\ and\ \bibinfo {author} {\bibfnamefont {Z.}~\bibnamefont {Wang}},\ }\bibfield  {title} {\emph {\bibinfo {title} {Xy-mixers: analytical and numerical results for qaoa}},\ }\href@noop {} {\bibfield  {journal} {\bibinfo  {journal} {Phys. Rev. A}\ }\textbf {\bibinfo {volume} {101}},\ \bibinfo {pages} {012320} (\bibinfo {year} {2020})}\BibitemShut {NoStop}%
\bibitem [{\citenamefont {Golden}\ \emph {et~al.}(2021)\citenamefont {Golden}, \citenamefont {B{\"a}rtschi}, \citenamefont {O’Malley},\ and\ \citenamefont {Eidenbenz}}]{golden2021threshold}%
  \BibitemOpen
  \bibfield  {author} {\bibinfo {author} {\bibfnamefont {J.}~\bibnamefont {Golden}}, \bibinfo {author} {\bibfnamefont {A.}~\bibnamefont {B{\"a}rtschi}}, \bibinfo {author} {\bibfnamefont {D.}~\bibnamefont {O’Malley}},\ and\ \bibinfo {author} {\bibfnamefont {S.}~\bibnamefont {Eidenbenz}},\ }in\ \href@noop {} {\emph {\bibinfo {booktitle} {2021 IEEE International Conference on Quantum Computing and Engineering (QCE)}}}\ (\bibinfo {organization} {IEEE},\ \bibinfo {year} {2021})\ pp.\ \bibinfo {pages} {137--147}\BibitemShut {NoStop}%
\bibitem [{\citenamefont {Lee}\ \emph {et~al.}(2023)\citenamefont {Lee}, \citenamefont {Xie}, \citenamefont {Cai}, \citenamefont {Saito},\ and\ \citenamefont {Asai}}]{lee2023depth}%
  \BibitemOpen
  \bibfield  {author} {\bibinfo {author} {\bibfnamefont {X.}~\bibnamefont {Lee}}, \bibinfo {author} {\bibfnamefont {N.}~\bibnamefont {Xie}}, \bibinfo {author} {\bibfnamefont {D.}~\bibnamefont {Cai}}, \bibinfo {author} {\bibfnamefont {Y.}~\bibnamefont {Saito}},\ and\ \bibinfo {author} {\bibfnamefont {N.}~\bibnamefont {Asai}},\ }\bibfield  {title} {\emph {\bibinfo {title} {A depth-progressive initialization strategy for quantum approximate optimization algorithm}},\ }\href@noop {} {\bibfield  {journal} {\bibinfo  {journal} {Mathematics}\ }\textbf {\bibinfo {volume} {11}},\ \bibinfo {pages} {2176} (\bibinfo {year} {2023})}\BibitemShut {NoStop}%
\bibitem [{\citenamefont {Tate}\ and\ \citenamefont {Eidenbenz}(2025)}]{tate2025theoretical}%
  \BibitemOpen
  \bibfield  {author} {\bibinfo {author} {\bibfnamefont {R.}~\bibnamefont {Tate}}\ and\ \bibinfo {author} {\bibfnamefont {S.}~\bibnamefont {Eidenbenz}},\ }\bibfield  {title} {\emph {\bibinfo {title} {Theoretical approximation ratios for warm-started qaoa on 3-regular max-cut instances at depth p= 1}},\ }\href@noop {} {\bibfield  {journal} {\bibinfo  {journal} {Theoretical Computer Science}\ ,\ \bibinfo {pages} {115571}} (\bibinfo {year} {2025})}\BibitemShut {NoStop}%
\bibitem [{\citenamefont {Bezanson}\ \emph {et~al.}(2017)\citenamefont {Bezanson}, \citenamefont {Edelman}, \citenamefont {Karpinski},\ and\ \citenamefont {Shah}}]{bezanson2017julia}%
  \BibitemOpen
  \bibfield  {author} {\bibinfo {author} {\bibfnamefont {J.}~\bibnamefont {Bezanson}}, \bibinfo {author} {\bibfnamefont {A.}~\bibnamefont {Edelman}}, \bibinfo {author} {\bibfnamefont {S.}~\bibnamefont {Karpinski}},\ and\ \bibinfo {author} {\bibfnamefont {V.~B.}\ \bibnamefont {Shah}},\ }\bibfield  {title} {\emph {\bibinfo {title} {{Julia: A fresh approach to numerical computing}}},\ }\href {https://doi.org/10.1137/141000671} {\bibfield  {journal} {\bibinfo  {journal} {SIAM review}\ }\textbf {\bibinfo {volume} {59}},\ \bibinfo {pages} {65--98} (\bibinfo {year} {2017})}\BibitemShut {NoStop}%
\bibitem [{\citenamefont {Golden}\ \emph {et~al.}(2023{\natexlab{b}})\citenamefont {Golden}, \citenamefont {Baertschi}, \citenamefont {O’Malley}, \citenamefont {Pelofske},\ and\ \citenamefont {Eidenbenz}}]{JuliQAOA}%
  \BibitemOpen
  \bibfield  {author} {\bibinfo {author} {\bibfnamefont {J.}~\bibnamefont {Golden}}, \bibinfo {author} {\bibfnamefont {A.}~\bibnamefont {Baertschi}}, \bibinfo {author} {\bibfnamefont {D.}~\bibnamefont {O’Malley}}, \bibinfo {author} {\bibfnamefont {E.}~\bibnamefont {Pelofske}},\ and\ \bibinfo {author} {\bibfnamefont {S.}~\bibnamefont {Eidenbenz}},\ }in\ \href {https://doi.org/10.1145/3624062.3624220} {\emph {\bibinfo {booktitle} {Proceedings of the SC ’23 Workshops of The International Conference on High Performance Computing, Network, Storage, and Analysis}}},\ \bibinfo {series and number} {SC-W 2023}\ (\bibinfo  {publisher} {ACM},\ \bibinfo {year} {2023})\BibitemShut {NoStop}%
\bibitem [{\citenamefont {Anderson}\ \emph {et~al.}(1999)\citenamefont {Anderson}, \citenamefont {Bai}, \citenamefont {Bischof}, \citenamefont {Blackford}, \citenamefont {Demmel}, \citenamefont {Dongarra}, \citenamefont {Du~Croz}, \citenamefont {Greenbaum}, \citenamefont {Hammarling}, \citenamefont {McKenney},\ and\ \citenamefont {Sorensen}}]{lapackcite}%
  \BibitemOpen
  \bibfield  {author} {\bibinfo {author} {\bibfnamefont {E.}~\bibnamefont {Anderson}}, \bibinfo {author} {\bibfnamefont {Z.}~\bibnamefont {Bai}}, \bibinfo {author} {\bibfnamefont {C.}~\bibnamefont {Bischof}}, \bibinfo {author} {\bibfnamefont {S.}~\bibnamefont {Blackford}}, \bibinfo {author} {\bibfnamefont {J.}~\bibnamefont {Demmel}}, \bibinfo {author} {\bibfnamefont {J.}~\bibnamefont {Dongarra}}, \bibinfo {author} {\bibfnamefont {J.}~\bibnamefont {Du~Croz}}, \bibinfo {author} {\bibfnamefont {A.}~\bibnamefont {Greenbaum}}, \bibinfo {author} {\bibfnamefont {S.}~\bibnamefont {Hammarling}}, \bibinfo {author} {\bibfnamefont {A.}~\bibnamefont {McKenney}},\ and\ \bibinfo {author} {\bibfnamefont {D.}~\bibnamefont {Sorensen}},\ }\href@noop {} {\emph {\bibinfo {title} {{LAPACK} Users' Guide}}},\ \bibinfo {edition} {3rd}\ ed.\ (\bibinfo  {publisher} {Society for Industrial and Applied Mathematics},\ \bibinfo {address} {Philadelphia, PA},\ \bibinfo {year} {1999})\BibitemShut {NoStop}%
\bibitem [{\citenamefont {Harris}\ \emph {et~al.}(2020)\citenamefont {Harris}, \citenamefont {Millman}, \citenamefont {van~der Walt}, \citenamefont {Gommers}, \citenamefont {Virtanen}, \citenamefont {Cournapeau}, \citenamefont {Wieser}, \citenamefont {Taylor}, \citenamefont {Berg}, \citenamefont {Smith}, \citenamefont {Kern}, \citenamefont {Picus}, \citenamefont {Hoyer}, \citenamefont {van Kerkwijk}, \citenamefont {Brett}, \citenamefont {Haldane}, \citenamefont {del R{\'{i}}o}, \citenamefont {Wiebe}, \citenamefont {Peterson}, \citenamefont {G{\'{e}}rard-Marchant}, \citenamefont {Sheppard}, \citenamefont {Reddy}, \citenamefont {Weckesser}, \citenamefont {Abbasi}, \citenamefont {Gohlke},\ and\ \citenamefont {Oliphant}}]{harris2020array}%
  \BibitemOpen
  \bibfield  {author} {\bibinfo {author} {\bibfnamefont {C.~R.}\ \bibnamefont {Harris}}, \bibinfo {author} {\bibfnamefont {K.~J.}\ \bibnamefont {Millman}}, \bibinfo {author} {\bibfnamefont {S.~J.}\ \bibnamefont {van~der Walt}}, \bibinfo {author} {\bibfnamefont {R.}~\bibnamefont {Gommers}}, \bibinfo {author} {\bibfnamefont {P.}~\bibnamefont {Virtanen}}, \bibinfo {author} {\bibfnamefont {D.}~\bibnamefont {Cournapeau}}, \bibinfo {author} {\bibfnamefont {E.}~\bibnamefont {Wieser}}, \bibinfo {author} {\bibfnamefont {J.}~\bibnamefont {Taylor}}, \bibinfo {author} {\bibfnamefont {S.}~\bibnamefont {Berg}}, \bibinfo {author} {\bibfnamefont {N.~J.}\ \bibnamefont {Smith}}, \bibinfo {author} {\bibfnamefont {R.}~\bibnamefont {Kern}}, \bibinfo {author} {\bibfnamefont {M.}~\bibnamefont {Picus}}, \bibinfo {author} {\bibfnamefont {S.}~\bibnamefont {Hoyer}}, \bibinfo {author} {\bibfnamefont {M.~H.}\ \bibnamefont {van Kerkwijk}}, \bibinfo {author} {\bibfnamefont {M.}~\bibnamefont {Brett}}, \bibinfo {author} {\bibfnamefont
  {A.}~\bibnamefont {Haldane}}, \bibinfo {author} {\bibfnamefont {J.~F.}\ \bibnamefont {del R{\'{i}}o}}, \bibinfo {author} {\bibfnamefont {M.}~\bibnamefont {Wiebe}}, \bibinfo {author} {\bibfnamefont {P.}~\bibnamefont {Peterson}}, \bibinfo {author} {\bibfnamefont {P.}~\bibnamefont {G{\'{e}}rard-Marchant}}, \bibinfo {author} {\bibfnamefont {K.}~\bibnamefont {Sheppard}}, \bibinfo {author} {\bibfnamefont {T.}~\bibnamefont {Reddy}}, \bibinfo {author} {\bibfnamefont {W.}~\bibnamefont {Weckesser}}, \bibinfo {author} {\bibfnamefont {H.}~\bibnamefont {Abbasi}}, \bibinfo {author} {\bibfnamefont {C.}~\bibnamefont {Gohlke}},\ and\ \bibinfo {author} {\bibfnamefont {T.~E.}\ \bibnamefont {Oliphant}},\ }\bibfield  {title} {\emph {\bibinfo {title} {Array programming with {NumPy}}},\ }\href {https://doi.org/10.1038/s41586-020-2649-2} {\bibfield  {journal} {\bibinfo  {journal} {Nature}\ }\textbf {\bibinfo {volume} {585}},\ \bibinfo {pages} {357--362} (\bibinfo {year} {2020})}\BibitemShut {NoStop}%
\bibitem [{\citenamefont {Fletcher}(2000)}]{fletcher2000practical}%
  \BibitemOpen
  \bibfield  {author} {\bibinfo {author} {\bibfnamefont {R.}~\bibnamefont {Fletcher}},\ }\href@noop {} {\emph {\bibinfo {title} {Practical methods of optimization}}}\ (\bibinfo  {publisher} {John Wiley \& Sons},\ \bibinfo {year} {2000})\BibitemShut {NoStop}%
\bibitem [{\citenamefont {Oitmaa}(2020)}]{Oitmaa_2020_TF}%
  \BibitemOpen
  \bibfield  {author} {\bibinfo {author} {\bibfnamefont {J.}~\bibnamefont {Oitmaa}},\ }\bibfield  {title} {\emph {\bibinfo {title} {{Frustrated transverse-field Ising model}}},\ }\href {https://doi.org/10.1088/1751-8121/ab63e6} {\bibfield  {journal} {\bibinfo  {journal} {Journal of Physics A: Mathematical and Theoretical}\ }\textbf {\bibinfo {volume} {53}},\ \bibinfo {pages} {085001} (\bibinfo {year} {2020})}\BibitemShut {NoStop}%
\bibitem [{\citenamefont {Kellermann}\ \emph {et~al.}(2019)\citenamefont {Kellermann}, \citenamefont {Schmidt},\ and\ \citenamefont {Zimmer}}]{PhysRevE.99.012134}%
  \BibitemOpen
  \bibfield  {author} {\bibinfo {author} {\bibfnamefont {N.}~\bibnamefont {Kellermann}}, \bibinfo {author} {\bibfnamefont {M.}~\bibnamefont {Schmidt}},\ and\ \bibinfo {author} {\bibfnamefont {F.~M.}\ \bibnamefont {Zimmer}},\ }\bibfield  {title} {\emph {\bibinfo {title} {Quantum ising model on the frustrated square lattice}},\ }\href {https://doi.org/10.1103/PhysRevE.99.012134} {\bibfield  {journal} {\bibinfo  {journal} {Phys. Rev. E}\ }\textbf {\bibinfo {volume} {99}},\ \bibinfo {pages} {012134} (\bibinfo {year} {2019})}\BibitemShut {NoStop}%
\bibitem [{\citenamefont {Bob\'ak}\ \emph {et~al.}(2018)\citenamefont {Bob\'ak}, \citenamefont {Jur\ifmmode \check{c}\else \v{c}\fi{}i\ifmmode~\check{s}\else \v{s}\fi{}inov\'a}, \citenamefont {Jur\ifmmode \check{c}\else \v{c}\fi{}i\ifmmode~\check{s}\else \v{s}\fi{}in},\ and\ \citenamefont {\ifmmode \check{Z}\else \v{Z}\fi{}ukovi\ifmmode~\check{c}\else \v{c}\fi{}}}]{PhysRevE.97.022124}%
  \BibitemOpen
  \bibfield  {author} {\bibinfo {author} {\bibfnamefont {A.}~\bibnamefont {Bob\'ak}}, \bibinfo {author} {\bibfnamefont {E.}~\bibnamefont {Jur\ifmmode \check{c}\else \v{c}\fi{}i\ifmmode~\check{s}\else \v{s}\fi{}inov\'a}}, \bibinfo {author} {\bibfnamefont {M.}~\bibnamefont {Jur\ifmmode \check{c}\else \v{c}\fi{}i\ifmmode~\check{s}\else \v{s}\fi{}in}},\ and\ \bibinfo {author} {\bibfnamefont {M.}~\bibnamefont {\ifmmode \check{Z}\else \v{Z}\fi{}ukovi\ifmmode~\check{c}\else \v{c}\fi{}}},\ }\bibfield  {title} {\emph {\bibinfo {title} {Frustrated spin-$\frac{1}{2}$ ising antiferromagnet on a square lattice in a transverse field}},\ }\href {https://doi.org/10.1103/PhysRevE.97.022124} {\bibfield  {journal} {\bibinfo  {journal} {Phys. Rev. E}\ }\textbf {\bibinfo {volume} {97}},\ \bibinfo {pages} {022124} (\bibinfo {year} {2018})}\BibitemShut {NoStop}%
\bibitem [{\citenamefont {Kato}\ and\ \citenamefont {Misawa}(2015)}]{PhysRevB.92.174419}%
  \BibitemOpen
  \bibfield  {author} {\bibinfo {author} {\bibfnamefont {Y.}~\bibnamefont {Kato}}\ and\ \bibinfo {author} {\bibfnamefont {T.}~\bibnamefont {Misawa}},\ }\bibfield  {title} {\emph {\bibinfo {title} {{Quantum tricriticality in antiferromagnetic Ising model with transverse field: A quantum Monte Carlo study}}},\ }\href {https://doi.org/10.1103/PhysRevB.92.174419} {\bibfield  {journal} {\bibinfo  {journal} {Phys. Rev. B}\ }\textbf {\bibinfo {volume} {92}},\ \bibinfo {pages} {174419} (\bibinfo {year} {2015})}\BibitemShut {NoStop}%
\bibitem [{\citenamefont {Sadrzadeh}\ \emph {et~al.}(2016)\citenamefont {Sadrzadeh}, \citenamefont {Haghshenas}, \citenamefont {Jahromi},\ and\ \citenamefont {Langari}}]{PhysRevB.94.214419}%
  \BibitemOpen
  \bibfield  {author} {\bibinfo {author} {\bibfnamefont {M.}~\bibnamefont {Sadrzadeh}}, \bibinfo {author} {\bibfnamefont {R.}~\bibnamefont {Haghshenas}}, \bibinfo {author} {\bibfnamefont {S.~S.}\ \bibnamefont {Jahromi}},\ and\ \bibinfo {author} {\bibfnamefont {A.}~\bibnamefont {Langari}},\ }\bibfield  {title} {\emph {\bibinfo {title} {{Emergence of string-valence bond solid state in the frustrated J1-J2 transverse field Ising model on the square lattice}}},\ }\href {https://doi.org/10.1103/PhysRevB.94.214419} {\bibfield  {journal} {\bibinfo  {journal} {Phys. Rev. B}\ }\textbf {\bibinfo {volume} {94}},\ \bibinfo {pages} {214419} (\bibinfo {year} {2016})}\BibitemShut {NoStop}%
\bibitem [{\citenamefont {Murtazaev}\ \emph {et~al.}(2014)\citenamefont {Murtazaev}, \citenamefont {Ramazanov},\ and\ \citenamefont {Kassan-Ogly}}]{MonteCarlofrustrated}%
  \BibitemOpen
  \bibfield  {author} {\bibinfo {author} {\bibfnamefont {A.~K.}\ \bibnamefont {Murtazaev}}, \bibinfo {author} {\bibfnamefont {M.~K.}\ \bibnamefont {Ramazanov}},\ and\ \bibinfo {author} {\bibfnamefont {F.~A.}\ \bibnamefont {Kassan-Ogly}},\ }\bibfield  {title} {\emph {\bibinfo {title} {Frustrations and phase transitions in the ising model on square lattice}},\ }\href {https://doi.org/10.1088/1742-6596/510/1/012026} {\bibfield  {journal} {\bibinfo  {journal} {Journal of Physics: Conference Series}\ }\textbf {\bibinfo {volume} {510}},\ \bibinfo {pages} {012026} (\bibinfo {year} {2014})}\BibitemShut {NoStop}%
\bibitem [{\citenamefont {Zhu}\ and\ \citenamefont {White}(2014)}]{DMRG_frustrate}%
  \BibitemOpen
  \bibfield  {author} {\bibinfo {author} {\bibfnamefont {Z.}~\bibnamefont {Zhu}}\ and\ \bibinfo {author} {\bibfnamefont {S.~R.}\ \bibnamefont {White}},\ }\bibfield  {title} {\emph {\bibinfo {title} {Quantum phases of the frustrated xy models on the honeycomb lattice}},\ }\href {https://doi.org/10.1142/S0217984914300166} {\bibfield  {journal} {\bibinfo  {journal} {Modern Physics Letters B}\ }\textbf {\bibinfo {volume} {28}},\ \bibinfo {pages} {1430016} (\bibinfo {year} {2014})},\ \Eprint {https://arxiv.org/abs/https://doi.org/10.1142/S0217984914300166} {https://doi.org/10.1142/S0217984914300166} \BibitemShut {NoStop}%
\bibitem [{\citenamefont {Crosson}\ \emph {et~al.}(2014)\citenamefont {Crosson}, \citenamefont {Farhi}, \citenamefont {Lin}, \citenamefont {Lin},\ and\ \citenamefont {Shor}}]{crosson2014different}%
  \BibitemOpen
  \bibfield  {author} {\bibinfo {author} {\bibfnamefont {E.}~\bibnamefont {Crosson}}, \bibinfo {author} {\bibfnamefont {E.}~\bibnamefont {Farhi}}, \bibinfo {author} {\bibfnamefont {C.~Y.-Y.}\ \bibnamefont {Lin}}, \bibinfo {author} {\bibfnamefont {H.-H.}\ \bibnamefont {Lin}},\ and\ \bibinfo {author} {\bibfnamefont {P.}~\bibnamefont {Shor}},\ }\bibfield  {title} {\emph {\bibinfo {title} {Different strategies for optimization using the quantum adiabatic algorithm}},\ }\href@noop {} {\bibfield  {journal} {\bibinfo  {journal} {arXiv preprint arXiv:1401.7320}\ } (\bibinfo {year} {2014})}\BibitemShut {NoStop}%
\bibitem [{\citenamefont {Preskill}(2018)}]{Preskill_2018}%
  \BibitemOpen
  \bibfield  {author} {\bibinfo {author} {\bibfnamefont {J.}~\bibnamefont {Preskill}},\ }\bibfield  {title} {\emph {\bibinfo {title} {{Quantum Computing in the NISQ era and beyond}}},\ }\href {https://doi.org/10.22331/q-2018-08-06-79} {\bibfield  {journal} {\bibinfo  {journal} {Quantum}\ }\textbf {\bibinfo {volume} {2}},\ \bibinfo {pages} {79} (\bibinfo {year} {2018})}\BibitemShut {NoStop}%
\bibitem [{\citenamefont {Dorier}\ \emph {et~al.}(2005)\citenamefont {Dorier}, \citenamefont {Becca},\ and\ \citenamefont {Mila}}]{dorier2005quantum}%
  \BibitemOpen
  \bibfield  {author} {\bibinfo {author} {\bibfnamefont {J.}~\bibnamefont {Dorier}}, \bibinfo {author} {\bibfnamefont {F.}~\bibnamefont {Becca}},\ and\ \bibinfo {author} {\bibfnamefont {F.}~\bibnamefont {Mila}},\ }\bibfield  {title} {\emph {\bibinfo {title} {Quantum compass model on the square lattice}},\ }\href@noop {} {\bibfield  {journal} {\bibinfo  {journal} {Physical Review B—Condensed Matter and Materials Physics}\ }\textbf {\bibinfo {volume} {72}},\ \bibinfo {pages} {024448} (\bibinfo {year} {2005})}\BibitemShut {NoStop}%
\bibitem [{\citenamefont {Galda}\ \emph {et~al.}(2021)\citenamefont {Galda}, \citenamefont {Liu}, \citenamefont {Lykov}, \citenamefont {Alexeev},\ and\ \citenamefont {Safro}}]{galda2021transferability}%
  \BibitemOpen
  \bibfield  {author} {\bibinfo {author} {\bibfnamefont {A.}~\bibnamefont {Galda}}, \bibinfo {author} {\bibfnamefont {X.}~\bibnamefont {Liu}}, \bibinfo {author} {\bibfnamefont {D.}~\bibnamefont {Lykov}}, \bibinfo {author} {\bibfnamefont {Y.}~\bibnamefont {Alexeev}},\ and\ \bibinfo {author} {\bibfnamefont {I.}~\bibnamefont {Safro}},\ }in\ \href@noop {} {\emph {\bibinfo {booktitle} {2021 IEEE International Conference on Quantum Computing and Engineering (QCE)}}}\ (\bibinfo {organization} {IEEE},\ \bibinfo {year} {2021})\ pp.\ \bibinfo {pages} {171--180}\BibitemShut {NoStop}%
\bibitem [{\citenamefont {Shaydulin}\ \emph {et~al.}(2023)\citenamefont {Shaydulin}, \citenamefont {Lotshaw}, \citenamefont {Larson}, \citenamefont {Ostrowski},\ and\ \citenamefont {Humble}}]{shaydulin2023parameter}%
  \BibitemOpen
  \bibfield  {author} {\bibinfo {author} {\bibfnamefont {R.}~\bibnamefont {Shaydulin}}, \bibinfo {author} {\bibfnamefont {P.~C.}\ \bibnamefont {Lotshaw}}, \bibinfo {author} {\bibfnamefont {J.}~\bibnamefont {Larson}}, \bibinfo {author} {\bibfnamefont {J.}~\bibnamefont {Ostrowski}},\ and\ \bibinfo {author} {\bibfnamefont {T.~S.}\ \bibnamefont {Humble}},\ }\bibfield  {title} {\emph {\bibinfo {title} {Parameter transfer for quantum approximate optimization of weighted maxcut}},\ }\href@noop {} {\bibfield  {journal} {\bibinfo  {journal} {ACM Transactions on Quantum Computing}\ }\textbf {\bibinfo {volume} {4}},\ \bibinfo {pages} {1--15} (\bibinfo {year} {2023})}\BibitemShut {NoStop}%
\bibitem [{\citenamefont {Claes}\ and\ \citenamefont {Dam}(2021)}]{Claes2021instance}%
  \BibitemOpen
  \bibfield  {author} {\bibinfo {author} {\bibfnamefont {J.}~\bibnamefont {Claes}}\ and\ \bibinfo {author} {\bibfnamefont {W.~v.}\ \bibnamefont {Dam}},\ }\bibfield  {title} {\emph {\bibinfo {title} {Instance {I}ndependence of {S}ingle {L}ayer {Q}uantum {A}pproximate {O}ptimization {A}lgorithm on {M}ixed-{S}pin {M}odels at {I}nfinite {S}ize}},\ }\href {https://doi.org/10.22331/q-2021-09-15-542} {\bibfield  {journal} {\bibinfo  {journal} {{Quantum}}\ }\textbf {\bibinfo {volume} {5}},\ \bibinfo {pages} {542} (\bibinfo {year} {2021})}\BibitemShut {NoStop}%
\bibitem [{\citenamefont {Wurtz}\ and\ \citenamefont {Lykov}(2021)}]{wurtz2021fixedangleconjectureqaoa}%
  \BibitemOpen
  \bibfield  {author} {\bibinfo {author} {\bibfnamefont {J.}~\bibnamefont {Wurtz}}\ and\ \bibinfo {author} {\bibfnamefont {D.}~\bibnamefont {Lykov}},\ }\href {https://arxiv.org/abs/2107.00677} {\bibinfo {title} {{The fixed angle conjecture for QAOA on regular MaxCut graphs}}} (\bibinfo {year} {2021}),\ \Eprint {https://arxiv.org/abs/2107.00677} {arXiv:2107.00677 [quant-ph]} \BibitemShut {NoStop}%
\bibitem [{\citenamefont {Chernyavskiy}\ and\ \citenamefont {Bantysh}(2023)}]{chernyavskiy2023method}%
  \BibitemOpen
  \bibfield  {author} {\bibinfo {author} {\bibfnamefont {A.~Y.}\ \bibnamefont {Chernyavskiy}}\ and\ \bibinfo {author} {\bibfnamefont {B.}~\bibnamefont {Bantysh}},\ }\bibfield  {title} {\emph {\bibinfo {title} {{A Method to Compute QAOA Fixed Angles}}},\ }\href@noop {} {\bibfield  {journal} {\bibinfo  {journal} {Russian Microelectronics}\ }\textbf {\bibinfo {volume} {52}},\ \bibinfo {pages} {S352--S356} (\bibinfo {year} {2023})}\BibitemShut {NoStop}%
\bibitem [{\citenamefont {Boulebnane}\ and\ \citenamefont {Montanaro}(2021)}]{boulebnane2021predictingparametersquantumapproximate}%
  \BibitemOpen
  \bibfield  {author} {\bibinfo {author} {\bibfnamefont {S.}~\bibnamefont {Boulebnane}}\ and\ \bibinfo {author} {\bibfnamefont {A.}~\bibnamefont {Montanaro}},\ }\href {https://arxiv.org/abs/2110.10685} {\bibinfo {title} {{Predicting parameters for the Quantum Approximate Optimization Algorithm for MAX-CUT from the infinite-size limit}}} (\bibinfo {year} {2021}),\ \Eprint {https://arxiv.org/abs/2110.10685} {arXiv:2110.10685 [quant-ph]} \BibitemShut {NoStop}%
\bibitem [{\citenamefont {Akshay}\ \emph {et~al.}(2021)\citenamefont {Akshay}, \citenamefont {Rabinovich}, \citenamefont {Campos},\ and\ \citenamefont {Biamonte}}]{Akshay_2021}%
  \BibitemOpen
  \bibfield  {author} {\bibinfo {author} {\bibfnamefont {V.}~\bibnamefont {Akshay}}, \bibinfo {author} {\bibfnamefont {D.}~\bibnamefont {Rabinovich}}, \bibinfo {author} {\bibfnamefont {E.}~\bibnamefont {Campos}},\ and\ \bibinfo {author} {\bibfnamefont {J.}~\bibnamefont {Biamonte}},\ }\bibfield  {title} {\emph {\bibinfo {title} {Parameter concentrations in quantum approximate optimization}},\ }\bibfield  {journal} {\bibinfo  {journal} {Physical Review A}\ }\textbf {\bibinfo {volume} {104}},\ \href {https://doi.org/10.1103/physreva.104.l010401} {10.1103/physreva.104.l010401} (\bibinfo {year} {2021})\BibitemShut {NoStop}%
\end{thebibliography}%

\end{document}